\documentclass[twocolumn, numberedappendix, twocolappendix, apj]{openjournal}

\usepackage{graphicx}
\usepackage{dcolumn}
\usepackage{bm}
\usepackage{physics}
\usepackage{color}
\definecolor{Blue}{rgb}{0,0.08,0.65}
\definecolor{Green}{rgb}{0.2,0.55,0.35}
\definecolor{grey}{rgb}{0.75,0.75,0.75}
\definecolor{Orange}{rgb}{1.0,0.5,0.15}
\definecolor{brown}{rgb}{0.7,0.25,0.0}
\definecolor{Pink}{rgb}{1.0,0.5,0.5}
\definecolor{darkerred}{rgb}{0.8,0,0}
\definecolor{darkerblue}{rgb}{0,0,0.8}
\definecolor{darkergreen}{rgb}{0,0.5,0}
\definecolor{darkcyan}{rgb}{0,0.6,0.6}

\usepackage[normalem]{ulem}
\usepackage{placeins} 
\usepackage{float} 

\hypersetup{
colorlinks=true,
linkcolor=Blue,
filecolor=red,
citecolor=Blue,
urlcolor=Blue
}

\newcommand{\del}{\partial}

\begin{document}

\title{Making (dark matter) waves: \\ Untangling wave interference for multi-streaming dark matter}

\author[0000-0002-1524-6949]{Alex Gough$^\star$}
\thanks{$^\star$\href{mailto:a.gough2@newcastle.ac.uk}{a.gough2@newcastle.ac.uk}}
\author[0000-0001-7831-1579]{Cora Uhlemann}
\affiliation{School of Mathematics, Statistics and Physics, Newcastle University, Herschel Building, NE1 7RU Newcastle-upon-Tyne, U.K.}%

\begin{abstract}
The classical dynamics of collisionless cold dark matter, commonly described by fluid variables or a phase-space distribution, can be captured in a single semiclassical wavefunction. We illustrate how  classical multi-streaming creates wave interference in a toy model corresponding to the dynamics of the Zel'dovich approximation and link it to diffraction optics.  Wave interference dresses the classical skeleton of cold dark matter with universal features akin to the physical imprints of wavelike (or fuzzy) dark matter. We untangle this wave interference to obtain single-stream wavefunctions corresponding to the classical fluid streams, by writing the wavefunction in an integral form. Our wave decomposition captures the full phase-space information and isolates the multi-stream phenomena related to vorticity and velocity dispersion. We link the wave interference features of our system to the standard forms of diffraction catastrophe integrals, which produce bright caustics in optical fields analogous to the cold dark matter density field. Our two complementary descriptions of dark matter wave-fields present rich universal features that can unlock new ways of modelling and probing wavelike dark matter on the scales of the cosmic web.
\end{abstract}


\maketitle


\section{Introduction}

A field-level description of the large-scale structure of our Universe is one of the key challenges for upcoming galaxy surveys such as \textit{Euclid} \citep{Euclid_mission}, LSST \citep{LSST_mission}, and DESI \citep{DESI_mission}, and would be a powerful tool for parameterising the cosmology-dependence of observables. Within the standard $\Lambda$CDM model, the cold dark matter (CDM) component of the Universe dominates the dynamics of structure formation. Since running fully-fledged non-linear simulations for cosmological inference is time-consuming and costly, hybrid and analytic approaches to CDM dynamics are  a crucial step on the path to field-level descriptions of the cosmic large-scale structure traced by galaxy clustering and weak lensing. 

A variety of analytic techniques exist for predicting the dynamics of cold dark matter, most based on leveraging the fact that at early times the dark matter is well described as a perfect fluid. In standard perturbation theory \citep[SPT, e.g.][]{Bernardeau2002PhR} the density and velocity are expanded perturbatively about their background values. As it relies on the perturbative smallness of these quantities, SPT struggles to accurately describe the large densities arising from gravitational collapse. Furthermore, as cold dark matter is collisionless, fluid streams cross and create regions of (formally) infinite Eulerian density, called caustics, strongly limiting the region of applicability of this perturbative theory. A useful alternative to SPT is Lagrangian perturbation theory (LPT) \citep{ Zeldovich1970, Buchert1989A&A, Bouchet1992ApJL,  Bouchet1995A&A, Villone2017EPJH}, which takes the displacement from fluid elements' initial positions as its perturbing quantity, avoiding the singular densities from SPT.  However, updating the gravitational potential which displaces particles requires the Eulerian density, which requires mapping between the initial (Lagrangian) and final (Eulerian) positions. A similar conversion is necessary to facilitate a comparison with galaxy survey data, which naturally lives in Eulerian space. 

In this paper, we consider an alternative approach to modelling cold dark matter dynamics that combines advantages of Lagrangian and Eulerian frameworks. We introduce a wavefunction $\psi$, to play the role of the dark matter field, and rely on reproducing cold dark matter dynamics in a semiclassical limit, using the propagator formalism established in \cite{Uhlemann2019, Rampf2021MNRAS}. In particular we examine how classical CDM phenomenology is encoded in the features of the wavefunction, and how it can be extracted, using a simple toy model. This toy model makes use of a ``free-particle Schr\"odinger equation'' in 1+1 dimensions \citep{ColesSpencer2003, ShortColes2006, Uhlemann2019}, which is closely related to the classical Zel'dovich approximation (lowest order LPT). While \cite{Uhlemann2019} derives the higher order behaviour within this propagator formalism, here we focus on dissecting the wavefunction in the simplest case, which already features a rich phenomenology  and facilitates links to a well-studied classical model. Using a wavefunction in this way allows for LPT-like dynamics to be captured with direct access to Eulerian observables like the density and velocity. 

The use of wavefunctions to model dark matter arises in a variety of other contexts in cosmology, which are physically distinct but conceptually and phenomenologically closely related to the propagator formalism used here. In studying non-relativistic self-gravitating systems, the Schr\"odinger-Poisson equation is the principle equation of interest. The Schr\"odinger-Poisson equation describes the evolution of a wavefunction, describing the dark matter field, which experiences self-interaction via a gravitational potential described by a Poisson equation. 

Taken as the true dynamical equations of motion, the Schr\"odinger-Poisson equation describes a non-relativistic, scalar dark matter particle (as a limit of the Klein-Gordon equation) \citep{Hui2017} with interesting astrophysical signatures. Such dark matter candidates are often motivated from particle physics including the QCD axion \citep{Peccei1977PhRvL}, the string axiverse \citep[e.g.][]{Svrcek2006JHEP, Arvanitaki2010PhRvD}, and other axion-like particles \citep{Jaeckel2022arXiv}. In the context of cosmology, these candidates often have very small mass scales ($\sim 10^{-22} \ \rm eV$) so that their wave phenomena are present on astrophysical scales. Though not completely interchangeable terms, such dark matter candidates are referred to as ultra-light axions (ULAs), fuzzy dark matter (FDM), Bose-Einstein condensate (BEC) dark matter or simply wave dark matter ($\psi$DM) in the literature. For a recent review of such systems see e.g. \cite{Niemeyer2020, Hui2021, Ferreira2021A&ARv}.

Alternatively, one can interpret the Schr\"odinger-Poisson equation as a theoretical trick to study cold dark matter.  In the original work by Widrow and Kaiser \citep{WidrowKaiser1993},  they propose solving the Schr\"odinger-Poisson equation as an alternative to $N$-body simulations, leveraging the fact that the phase-space distribution of $\psi$ and classical CDM obey the same equation to $\order{\hbar^2}$. In such cases, we absorb the mass $m$ into the value of $\hbar$, and take $\hbar$ to parameterise how coarsely phase-space is sampled. Within this so-called Schr\"odinger method, wave interference effects were shown to emulate the multi-stream phenomena of velocity dispersion and vorticity \citep{Uhlemann2014,Kopp2017,Mocz2018} and criteria for its capability to effectively reproduce classical features were quantified  \citep{Garny2020,Eberhardt2020}.

Equations of Schr\"odinger-Poisson type can also be derived as an alternate scheme for closing the Vlasov cumulant hierarchy which appears in CDM dynamics (see e.g. \cite{Rampf2021arXiv} for a recent review of the cosmological Vlasov-Poisson equations). If one requires that the cumulant generating function is constructed only from the density and velocity (the first two cumulants) degrees of freedom, then the resulting cumulant generating function takes the form of Schr\"odinger-Poisson \citep{Uhlemann2018finitelygenerated}. This provides a self consistent closure scheme beyond truncation at the level of a perfect fluid. 

The propagator formalism used in this work is physically distinct from the Schr\"odinger method proposed by Widrow and Kaiser, as discussed in \cite{Uhlemann2019}. Morally however, both methods use a wavefunction to model cold dark matter and take advantage of formally having uniform resolution in position space\footnote{While in principle the wavefunction method has uniform resolution, practical computational choices might drive changes in resolution. If one only cares about the physics of halos this can be limiting, as computation must be spent resolving the phase in voids \citep{Schive2014} or making use of a hybrid scheme as in \cite{SchwabeNiemeyer2022}. However, wave models have had success in extracting information from underdense environments such as the Lyman-$\alpha$ forest \citep{Porqueres_2020}. For our toy model this concern is not relevant, as free evolution of wavefunctions is computationally fast and simple.}, which provides complementarity to $N$-body simulations as originally intended. While the focus of this work always implicitly interprets the wave-mechanical model in the semiclassical approach, one could instead consider the wave effects as physical phenomena in the context of a true wave dark matter. While the long-term dynamics of the free Schr\"odinger equation is different from Schr\"odinger-Poisson, manifestations of wave interference effects created by the onset of multi-streaming are universal features. Additionally, perturbative treatments based on the propagator formalism can be valuable for pushing the volumes of wave dark matter simulations from currently around $1-10$ Mpc \citep{Schive2014,Mocz2018,Mina2020,May2021,SchwabeNiemeyer2022} to truly cosmological scales.

This paper is structured as follows. In Section~\ref{sec:multi-streaming and interference} we present a 1+1D toy model for structure formation and discuss how the phenomenology of multi-streaming can be understood and in the context of both classical and quantum systems. This toy model is based on a wavefunction satisfying the free Schr\"odinger equation, and corresponds to the Zel'dovich approximation. In Section~\ref{sec:unweaving_the_wavefunction} we discuss how the classical streaming behaviour can be extracted from wave interference by way of stationary phase analysis. Section~\ref{sec:phase_properties} discusses interference effects in the wave-mechanical model and how these encode information beyond perfect fluid dynamics.  The wavefunction can be separated into an ``average'' part, which describes the bulk fluid behaviour, and a ``hidden'' part, which isolates the sources of vorticity and velocity dispersion after shell-crossing. Finally, Section~\ref{sec:catastrophe_theory} connects the wave properties of the wave-mechanical system to catastrophe theory, in particular to diffraction integrals which classify the properties of different singularity types. This provides both justification for studying a simple model, and quantitative results where the stationary phase analysis is insufficient. We conclude in Section~\ref{sec:conclusion}, where we also provide an outlook on the generalisations of our model to more realistic scenarios and potential applications of the splitting presented.

\section{Dark matter multi-streaming, caustics, and interference}\label{sec:multi-streaming and interference}

In the following we present the phenomenology associated with multi-streaming in classical cold dark matter. We introduce the propagator formalism for our wave-mechanical model, show how the classical features associated with multi-streaming and shell-crossing appear as wave interference, and how to extract Eulerian observables from the wavefunction. The classical toy model (sine wave collapse) is shown in Figure~\ref{fig:zeldo_phase_sheet} and compared to the wave analogue model in Figure~\ref{fig:free_schrodi_evol}.

\subsection{Coordinate system and units}
In the weak field regime, on scales smaller than the Hubble radius, and for non-relativistic velocities, one can use the Newtonian limit rather than the full Einstein equations to describe structure formation. We take our background cosmology to be an Einstein-de Sitter (EdS) universe, as we are interested in structure formation during the matter dominated era. The entire discussion is straightforwardly extended to full $\Lambda$CDM.

We write our equations in comoving coordinates $\bm{x} = \bm{r}/a$, where $\bm{r}$ is the physical space coordinate and $a$ is the cosmic scale factor. We take our time variable to be $a$ rather than coordinate time $t$, which allows for a well defined initialisation as $a\to 0$ \citep{Rampf2015}. We define our peculiar velocity to be $\bm{v} = \dv*{\bm{x}}{a}$, which is related to the total velocity by $\bm{U} = H\bm{r} + Ha^2 \bm{v}$. The standard conjugate momentum $\bm p$ is related to our peculiar velocity by $\bm{p}/m = a^2\dv*{\bm{x}}{t} = a^{3/2}\bm{v}$. For a fluid with density $\rho(\bm{x},a)$ we decompose into a background part $\bar{\rho}(a)$ and a density contrast $\delta(\bm{x},a)$ related by $\rho = \bar{\rho}(1+\delta)$. We also choose units such that $4\pi G\bar{\rho}_0 = 3/2$, which makes $\dv*{a}{t} = a^{-1/2}$ and simplifies the units in several equations. Unlabelled spatial derivatives are always taken to mean with respect to Eulerian coordinates e.g. $\bm{\nabla} = \bm{\nabla}_{\! \bm{x}}$, but we will occasionally explicitly label Eulerian derivatives to emphasise the space we work in or to avoid confusion.

While the focus of this paper is on the 1-dimensional toy model presented in the following section, here results would hold in more dimensions they are written in vector notation, to indicate where the discussion readily generalises to more realistic systems. 

\subsection{The Zel'dovich Approximation}
Lagrangian perturbation theory (LPT) describes how particle (or fluid element) positions in the initial field, $\bm{q}$, map to positions in the final field via a displacement field, $\bm{\xi}(\bm{q})$. That is, the final position of a fluid element, $\bm{x}$, is given by the mapping
\begin{equation}\label{eqn:lagrangian_mapping}
    \bm{x}(\bm{q},a) = \bm{q} + \bm{\xi}(\bm{q},a)\,.
\end{equation}
In this view, we ``flow along with'' the fluid parcels, rather than taking a static set of coordinates to view the fluid as in the Eulerian picture.  Lagrangian perturbation theory then solves for the displacement field $\bm{\xi}$ perturbatively. At fixed order, LPT performs better than perturbing in the fluid variables directly, as there is a way of capturing non-local information due to flowing along with the fluid.

The mapping $\bm{x}(\bm{q},a)$ can become non-injective, taking particles from different initial positions to the same final position, a phenomenon called multi-streaming. The first occurrence of multi-streaming is called (first) shell-crossing. Assuming our fluid describes a collisionless set of particles (as in CDM), these streams of particles flow through each other. The single and multi-stream regions of space are separated from each other by regions of (formally) infinite Eulerian density called caustics  \citep[see e.g. Figure 3 in][]{Rampf2021arXiv}.

On large scales, the Zel'dovich approximation \citep{Zeldovich1970} corresponding to first order LPT, performs very well at predicting the formation of structures in the cosmic web. In 1-dimensional collapse, the Zel'dovich approximation is exact before shell crossing, while in higher dimensional systems it receives corrections due to tidal fields, encoded in higher order LPT. The Zel'dovich approximation in an EdS cosmology corresponds to the displacement field
\begin{equation}\label{eqn:zeldo_displacement_field}
    \vb*{\xi}^{\rm ZA}(\bm{q},a) = -a\bm{\nabla}_{\! \bm{q}} \varphi_g^{(\text{ini})}(\bm{q})\,,
\end{equation}
where $\varphi_g^{\rm (ini)}$ is the initial gravitational potential. This form of the displacement field corresponds to constant velocity motion, provided $a$ is taken as the time variable. With this displacement field, the final position of fluid elements is simply given by ballistic motion with a constant velocity
\begin{equation}
\bm{v}^{\rm ZA}(\bm{q})=-\bm{\nabla}_{\! \bm{q}} \varphi_g^{(\text{ini})}(\bm{q})\,,
\end{equation}
set by the initial gravitational field.

\subsection{Multi-streaming}
In the following we illustrate the onset of shell-crossing and multi-streaming in a simple 1-dimensional example in the Zel'dovich approximation.

Formally linearising the fluid variables (discussed in more detail in Section~\ref{sec:observables}) about their background values and evaluating the linearised fluid equations at arbitrarily early times $a\to 0$, it is found that analytic solutions at $a=0$ require that the boundary conditions on the initial density contrast, $\delta^{\rm (ini)}$, and the initial velocity potential, $\phi_v^{\rm (ini)}$, (which determines the velocity through $\bm{v} = \bm{\nabla} \phi_v$) must be tethered to each other \citep{Brenier2003} 
\begin{equation}\label{eqn:ZA_boundary_conditions}
    \delta^{\rm (ini)} = 0\,, \quad \phi_v^{\rm (ini)} = - \varphi_g^{\rm (ini)}\,.
\end{equation}
These boundary conditions select for the growing-mode solutions and are vorticity free, in accordance with our requirement of a potential velocity \citep{Rampf2015}.

\begin{figure}[h!t]
    \centering
    \includegraphics[width=\columnwidth]{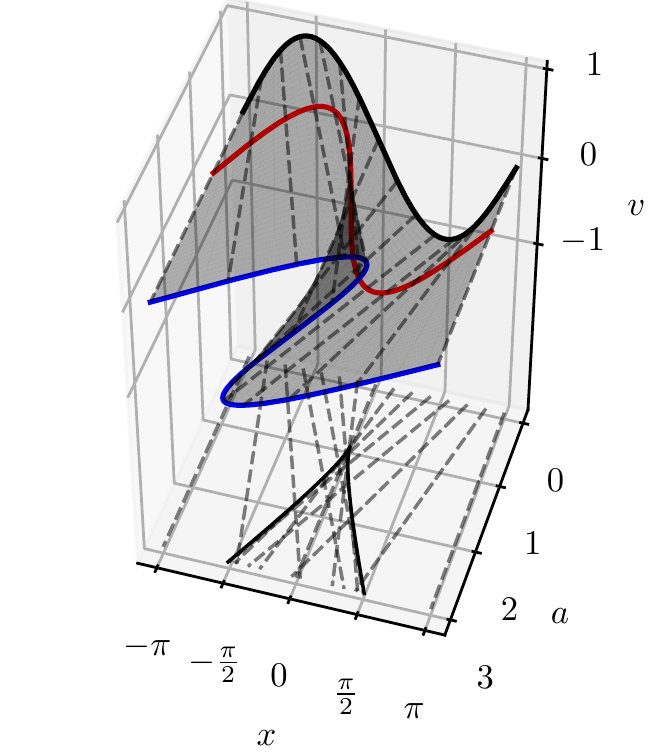}
    \caption{Phase-space sheet describing the evolution of a Fourier mode under the Zel'dovich approximation described by the displacement mapping $x(q,a) = q - a \sin(q)$\footnote{An animated version of this figure and its spacetime projection can be found in the auxilliary files or on \href{https://commons.wikimedia.org/wiki/File:Sine_wave_collapse_in_phase_space.gif}{wikimedia commons}.}. For times $a>1$, the phase-space sheet becomes triple valued in velocity. Projecting this phase-space sheet onto the $(x,a)$ plane results in a cusped line (the caustic line) which separates the single and triple stream regions of spacetime (given by equation~\eqref{eqn:shell_cross_region}). The dashed lines show the trajectories of a few individual particles through phase-space. The black, red, and blue curves show the phase-sheet at fixed times $a=0,1,3$. At $a=1$ (red) the phase-sheet becomes vertical, and thereafter it is multi-valued.}
    \label{fig:zeldo_phase_sheet}
\end{figure}

Because of these boundary conditions, we can study how a single Fourier mode in the velocity potential evolves under the Zel'dovich approximation. Figure~\ref{fig:zeldo_phase_sheet} shows the evolution of one such mode, corresponding to an initially uniformly dense fluid with initial velocity potential $\phi_v^{\rm (ini)}(q) = \cos(q)$ so that the initial velocity is given by $v^{\rm ZA} = \nabla_{\! q} \phi_v^{\rm (ini)}(q)$. Note that the velocity variable $v$ used in the figure is a phase-space variable, related to the usual canonical momentum $p$ by $v = p/a^{3/2}$ in our units, and should not be confused for the Eulerian velocity field.

While the fluid remains single valued, the density $\rho$ can be obtained by the mapping between Lagrangian (initial) positions and Eulerian positions via conservation of mass $\rho \dd[n]{\bm{x}} = \bar{\rho} \dd[n]{\bm{q}}$. Until the first shell-crossing, this can be exactly integrated, giving
\begin{equation}
    \rho = 1/J\,,
\end{equation}
where $J = \det (\partial\bm{x}/\partial \bm{q})$ is the Jacobian for the Lagrangian mapping. For a  1-dimensional system with initial velocity potential $\phi_v^{\rm (ini)}(q)$, this gives a density of
\begin{equation}\label{eqn:zeldo_density}
    \rho(x,a) = \frac{1}{\abs{1+a \dv[2]{}{q}\phi_v^{\rm (ini)}(q(x))}}\,.
\end{equation}
For the choice $\phi_v^{\rm (ini)}(q) = \cos(q)$ we see that this density diverges at $a=1$, corresponding to the time of shell-crossing. After shell-crossing there is a region of final positions, $x$, which are triple valued in velocity. Projecting the phase-space sheet onto the $(x,a)$ plane, the region separating the single-stream and the multi-stream regions of spacetime forms a cusp shape given by 
\begin{equation}
    (x(q),a(q)) = \left(q-\tan(q), \frac{1}{\cos(q)}\right)\,, \quad q \in \left(-\frac{\pi}{2}, \frac{\pi}{2}\right)\,.
\end{equation}
Locations on the caustic correspond to two initial points which cross streams, and to divergences in the density field. The $q$ in this parametric form corresponds to the initial position which maps to $x(q,a)$ via equation~\eqref{eqn:lagrangian_mapping}, and is closest to the origin, which is always in the central region between $\pm \frac{\pi}{2}$. For a fixed time, the region which has shell-crossed is given by
\begin{equation}
    \abs{x} < \sqrt{a^2-1}-\arccos(1/a)\,,
    \label{eqn:shell_cross_region}
\end{equation}
for $a>1$. 

We could instead study this system in full phase-space, to avoid the problem of multi-valuedness. In Lagrangian coordinates the distribution function in phase-space is given by (in $n$ dimensions)
\begin{equation}\label{eqn:lagrangian_phase_space_dist} 
    f_{\rm L}(\bm{x},\bm{p}) = \int \dd[n]{\bm{q}} \delta_{\rm D}(\bm{x}-\bm{q}-\bm{\xi}(\bm{q})) \delta_{\rm D}\left[\frac{\bm{p}}{a^{3/2}}-\bm{v}^{\rm L}(\bm{q})\right]\,,
\end{equation}
where $\bm{\xi}(\bm{q})$ is the Zel'dovich displacement field in equation~\eqref{eqn:zeldo_displacement_field} and $\bm{v}^{\rm L}(\bm{q}) = \bm{v}(\bm{x}(\bm{q},a),a)$ is the Lagrangian representation of the velocity evaluated at the Eulerian position $\bm{x}(\bm{q},a)$.

\subsection{From multiple streams to wave interference}

Rather than considering a set of classical collisionless particles or a classical fluid, we represent our dark matter field using a complex wavefunction $\psi$. 

Following in the spirit of \cite{Uhlemann2019}, we take the classical action of particles moving with constant velocity to build a propagator for our wavefunction. The classical action for a free fluid particle beginning at position $\bm{q}$ and evolving to position $\bm{x}$ at time $a$ is
\begin{equation}\label{eqn:free_action}
    S_0(\bm{q};\bm{x},a) = \frac{(\bm{q}-\bm{x})^2}{2a}\,.
\end{equation}
The associated transition amplitude for this action is the standard exponential of this action
\begin{equation}
    K_0(\bm{q};\bm{x},a) = \mathcal{N}\exp\left(\frac{i}{\hbar}S_0(\bm{q};\bm{x},a)\right)\,,
\end{equation}
where $\mathcal{N} = (2\pi i \hbar a)^{-n/2}$ is a normalisation factor for $n$ dimensions to ensure that the propagation returns a Dirac delta distribution for time $a=0$. Here we consider $\hbar$ to be a parameter of our model, building a quantum system which should produce classical behaviour in the $\hbar \to 0$ limit (subject to subtleties discussed in Section~\ref{sec:observables}). 

In the context of quantum mechanics, the transition amplitude can be used as a propagator moving initial wavefunctions to solutions at later time
\begin{equation}\label{eqn:psi propagated}
    \psi_0(\bm{x},a) = \int \dd[n]{\bm{q}} K_0(\bm{q};\bm{x},a) \psi_0^{\rm (ini)}(\bm{q})\,,
\end{equation}
where $\psi_0^{\rm (ini)}(\bm{q}) = \psi_0(\bm{q}, a=0)$ is the initial wavefunction. Both the propagator $K_0$ and the associated wavefunction, $\psi_0$, satisfy the potential free Schrödinger equation
\begin{equation}
\label{eq:freeSchroedi}
    i\hbar \del_a \psi_0 = -\frac{\hbar^2}{2}\nabla^2\psi_0\,.
\end{equation}
In this paper we will only be considering this free Schrödinger evolution, using the free classical action in equation~\eqref{eqn:free_action}, and therefore will drop the 0 subscripts on quantities from here on. We note that this wavefunction model is the same as the ``free particle approximation'' introduced in \cite{ColesSpencer2003}, compared to linearised fluid in \cite{ShortColes2006} and applied to cosmic voids in \cite{Gallagher_2022_SPvoids}. For details on how this propagator formalism can be extended to a perturbative theory relying on a Schr\"odinger equation with a time-independent Hamiltonian including an external effective potential  
\begin{equation}\label{eqn:schrodinger_with_potential}
    i\hbar \del_a \psi = \mathcal{H} \psi = -\frac{\hbar^2}{2}\nabla^2\psi + V_{\rm eff} \psi\,,
\end{equation}
see \cite{Uhlemann2019}. 

Using the polar representation of a wavefunction, we can decompose it into $\psi(\bm{x},a) = \sqrt{\rho(\bm{x},a)}e^{i\phi_v(\bm{x},a)/\hbar}$, which is called the Madelung representation \citep{Madelung1927}. Under this decomposition we see that the boundary conditions \eqref{eqn:ZA_boundary_conditions} require an initial wavefunction with Zel'dovich-like dynamics of the form  $\psi^{\rm (ini)}(\bm{q}) = \exp(i\phi_v^{\rm (ini)}(\bm{q})/\hbar)$. The propagator equation \eqref{eqn:psi propagated} can then be written
\begin{align} 
    \psi(\bm{x},a) &= \mathcal{N}\int \dd[n]{\bm{q}} \exp\left[{\frac{i}{\hbar}S(\bm{q};\bm{x},a)}\right]\exp\left[{\frac{i}{\hbar}\phi_v^{\rm (ini)}(\bm{q})}\right] \nonumber \\
    &= \mathcal{N} \int \dd[n]{\bm{q}} \exp\left[{\frac{i}{\hbar}\zeta(\bm{q};\bm{x},a)}\right]\,,
    \label{eqn:psi_integral_with_zeta}
\end{align}
where $\zeta= S + \phi_v^{\rm (ini)}$ includes both the propagation of the wavefunction and its initial conditions. We note here that we take the opposite sign choice for the velocity potential to \cite{Uhlemann2019}, so that $\bm{v} = \bm{\nabla} \phi_v$ instead of $\bm{v} = - \bm{\nabla} \phi_v$.

This free wavefunction can be solved for using a variety of analytic or numerical techniques. For free evolution, solving the free Schrödinger equation with Fourier transforms at each time slice is significantly faster than computing the wavefunction at late time using the highly oscillatory integral \eqref{eqn:psi_integral_with_zeta}\footnote{For a wavefunction obeying the Schrödinger equation with a potential, the time evolution can be obtained by using a symplectic integration scheme, splitting the Hamiltonian into a kinetic and potential part (see e.g. Appendix D of \cite{Uhlemann2019} or Chapter 3 of \cite{BinneyTremaine2008}).}.  However, as we will see in Sections~\ref{sec:unweaving_the_wavefunction} and \ref{sec:catastrophe_theory}, being able to directly analyse these highly oscillatory integrals of the form~\eqref{eqn:psi_integral_with_zeta} will be of physical interest, so it is useful to develop techniques to approximate them. 

In this paper we focus on $\zeta(q;x,a)$ which depend on 1-dimensional $q$ and $x$, and are analytic in $q$. In this case, we can replace the integration along the real line with \emph{any} contour in the complex plane which has the same end points without changing the value of the integral via Cauchy's theorem. In such cases, a simple way to accelerate the numerical integration is to simply step the integration contour in the direction which increases $\Im(\zeta)$ (decreasing the magnitude of the integrand). While not well suited to very complex forms of $\zeta$, and requiring some amount of tuning for the size of the displacement to take, any shift which reduces the amount of oscillation is helpful in obtaining accurate numerical evaluations of highly oscillatory integrals.

\begin{figure}[h!t]
    \centering
    \includegraphics[width=\columnwidth]{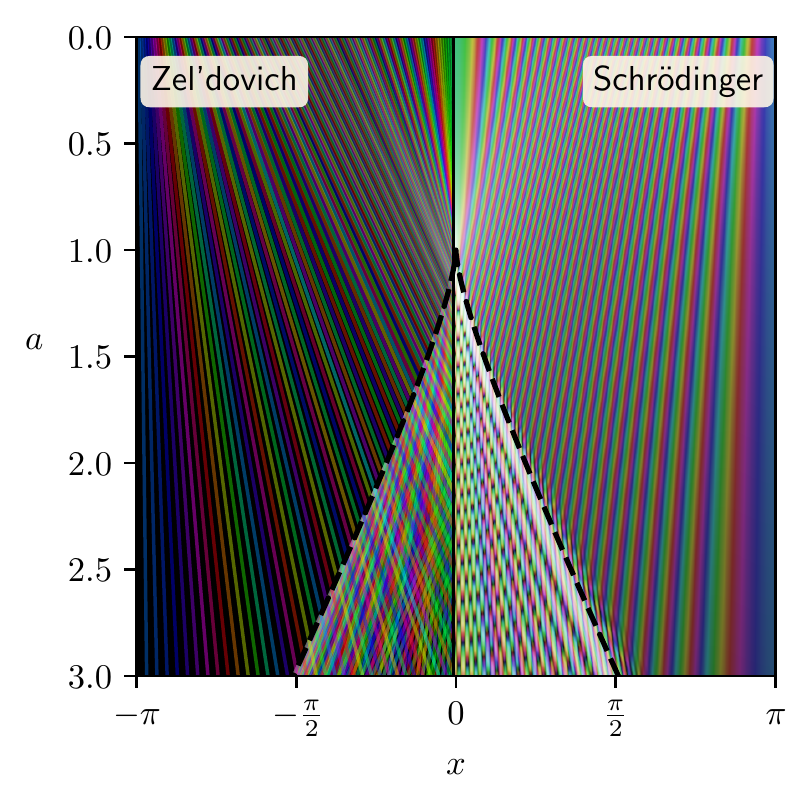}
    \caption{Comparison of the wavefunction evolution and classical Zel'dovich evolution with the same initial conditions. The wavefunction initial conditions are $\psi^{\rm (ini)}(q) = \exp((i/\hbar)\cos(q))$ with $\hbar = 0.01$ evolved on a grid of $(1024)^2$ (spacetime) cells. The complex value of the wavefunction is encoded through domain colouring, with brightness corresponding to amplitude, and hue corresponding to phase (the exact colouring scheme is in Appendix~\ref{app:domain_colouring}). Hence, lines of constant colour in this image correspond paths of constant phase. The Zel'dovich trajectories are coloured according to their initial phase $\cos(q)/\hbar$. For times $a>1$, interference patterns arise in the wavefunction, characterised by rapid oscillations in space and across time, corresponding to classical multi-streaming. The black dashed curve separating the classically shell-crossed and single-stream regions is given by equation~\eqref{eqn:shell_cross_region}.}
    \label{fig:free_schrodi_evol}
\end{figure}

Figure~\ref{fig:free_schrodi_evol} shows the evolution of a wavefunction corresponding the same Zel'dovich initial conditions as in Figure~\ref{fig:zeldo_phase_sheet}, compared to particle trajectories under Zel'dovich evolution. This wavefunction is defined by the initial data $\psi_0^{\rm (ini)}(q) = \exp((i/\hbar)\phi_v(q))$ with $\phi_v(q) = \cos(q)$. This wavefunction is the principal object of study in this paper, in particular understanding how the classical steam information is encoded in the wave phenomena. The figure uses a domain colouring technique to show the complex value of the wavefunction at all spacetime points. The brightness corresponds to the magnitude of the $\psi$, while the hue corresponds to the phase. The Zel'dovich trajectories are coloured according to their initial phase $\cos(q)/\hbar$, which they carry along their trajectory. The wavefunction is evaluated using the contour shifting technique, and reproduces the wavefunction solved via Fourier methods in Figure 1 of \cite{Uhlemann2019}.

\begin{figure}[h!t]
    \centering
    \includegraphics[width=\columnwidth]{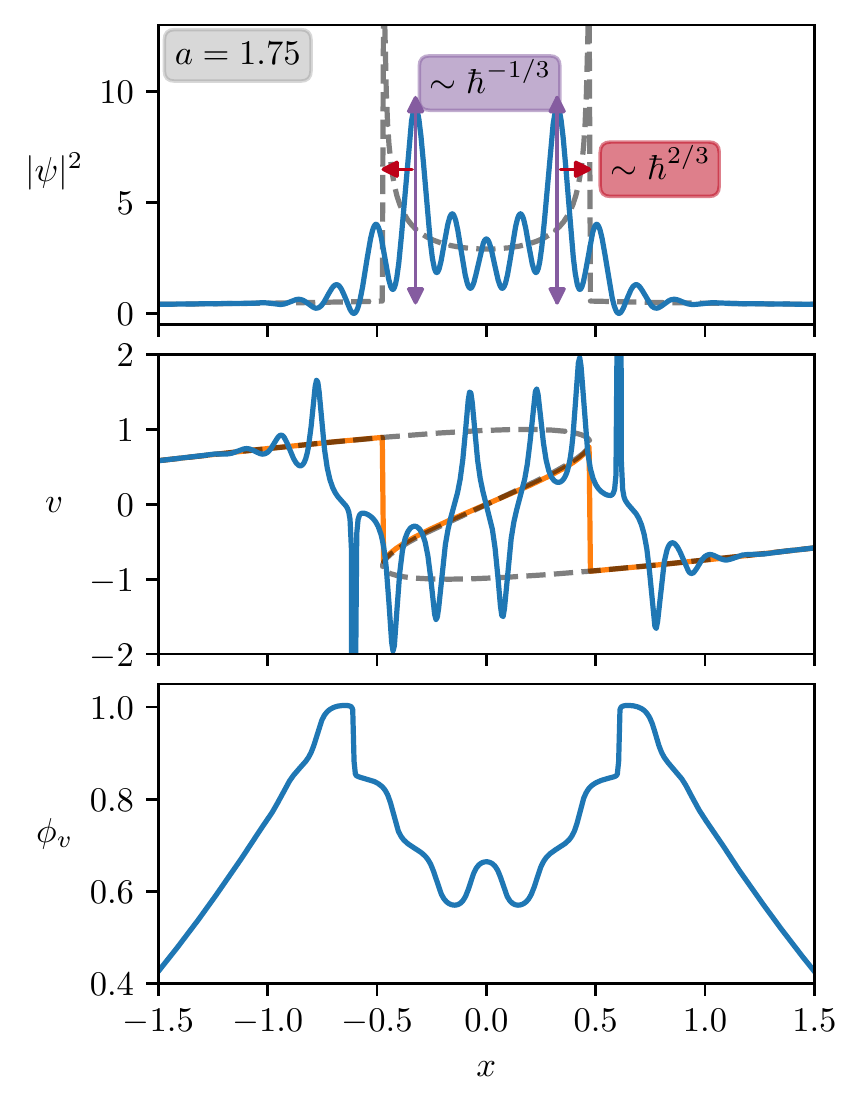}
    \caption{Features of the wavefunction $\psi$ (with $\hbar = 0.05$) with Zel'dovich initial conditions at time $a=1.75$.
    (Upper panel) The density of the wavefunction $\psi$. The dashed line corresponds to the classical Zel'dovich density at the same time. The wavefunction follows this classical density, dressing it in wave interference. The parameter $\hbar$ sets certain scaling properties, including the characteristic width and peak height of the density. The exact indices in this case will be explained in Section~\ref{sec:catastrophe_theory}. (Middle panel) The velocity associated with the wavefunction, given as the gradient of the wavefunction phase, or as the probability current divided by the density as in equation~\eqref{eqn:wavefunction_velocity_moment}. The full Zel'dovich phase-space sheet (dashed) and the classical stream-weighted average velocity (orange) are shown for comparison. Where the phase-space sheet is multi-valued, the wavefunction rapidly oscillates between the classical values. (Bottom panel) The phase function of the wavefunction, which produces the velocity field in a Madelung decomposition as $v = \nabla \phi_v$. At $x\approx \pm 0.6$ the phase develops a discontinuous jump, leading to ill-defined gradient. The impact of these phase jumps will be discussed in greater detail in Section~\ref{sec:phase_properties}. 
    }
    \label{fig:fold_annotated}
\end{figure}

The features in the wavefunction system shown in Figure~\ref{fig:free_schrodi_evol} are similar to those seen in the evolution of the classical Zel'dovich mode. We see the formation of a bright (dense), cusp shaped caustic line beginning at $a=1$. The interior region of this caustic exhibits interference, characterised by spatio-temporal oscillations, which must correspond to the classical multi-stream regime. Of notable difference however, the caustic region no longer corresponds to infinite density, it has been regularised by the wavelength associated with the size of $\hbar$. It also now corresponds to a finite sized region, rather than being infinitely thin. These interference phenomena are shown at a constant time in Figure~\ref{fig:fold_annotated}, in comparison to the classical Zel'dovich density and stream velocities. The particular scalings of the peak density and the fringe spacing are given by certain powers of $\hbar$. These scalings are obtained by catastrophe theory, discussed in Section~\ref{sec:catastrophe_theory}. We see that this wave-mechanical model ``dresses'' the classical observables in this interference phenomena.

\subsection{Dark matter observables in phase- and real space}\label{sec:observables}

\subsubsection{Fluid equations from a wavefunction}

The Madelung representation  $\psi(\bm{x},a) = \sqrt{\rho(\bm{x},a)}e^{i\phi_v(\bm{x},a)/\hbar}$, makes connection between the wavefunction and fluid-like variables. Note that at the level of the evolution equations, this decomposition is only meaningful where the density field and phase are sufficiently smooth. If this wavefunction satisfies the free Schrödinger equation \eqref{eq:freeSchroedi}, under this Madelung decomposition the real and imaginary parts of the Schrödinger equation and become a continuity and a Bernoulli equation,
\begin{subequations}\label{eqn:quantum_fluids}
\begin{align}
    &\del_a \rho + \bm{\nabla} \cdot (\rho \bm{\nabla} \phi_v) = 0\,, \\
    &\del_a \phi_v + \frac{1}{2} \abs{\bm{\nabla}\phi_v}^2 + Q = 0\,.\label{eqn:quantum_bernoulli}
\end{align}
The final term in the Bernoulli equation, 
\begin{equation}\label{eqn:quantum_pressure}
    Q = -\frac{\hbar^2}{2}\frac{\nabla^2\sqrt{\rho}}{\sqrt{\rho}}\,,
\end{equation}
\end{subequations}
is traditionally called the ``quantum pressure,'' even though strictly it does not appear as a pressure but rather a velocity dispersion term in the associated Euler equation. If the Schr\"odinger equation contained a potential term, this would also appear in the Bernoulli equation~\eqref{eqn:quantum_bernoulli}. In the classical limit this term vanishes and we obtain standard (potential free) fluid equations in $a$-time, assuming a potential velocity field determined by $\bm{v} = \bm{\nabla}\phi_v$. 

We note that the quantum pressure term in equation~\eqref{eqn:quantum_fluids} is important for precision cosmology, and that wave techniques which lack the quantum pressure term can differ substantially from full wave simulations. Compared to $N$-body simulations, simulations including quantum pressure cause up to $5$ -- $10\%$ suppression in the matter power spectrum at low redshift, particularly near the quantum Jeans scale \citep{Veltmaat2016PhRvD, Nori_2018_AX-GADGET}. This suppression is further enhanced when the difference in initial conditions between fuzzy and cold dark matter is taken into account. While Madelung based solvers such as AX-GADGET \citep{Nori_2018_AX-GADGET} can capture the role the quantum pressure plays in suppressing structure, solving the Madelung equations~\eqref{eqn:quantum_fluids} does not well capture the details of interference patterns compared to full Schr\"odinger solvers, particularly the points where $|\psi|$ is close to zero where the Madelung variables develop discontinuities while $\psi$ remains well behaved throughout its evolution \citep{Mocz_2015_SP_SPH,Veltmaat2016PhRvD,  Hopkins_2019_numerics_FDM, SchwabeNiemeyer2022}. The interplay between these factors further motivates advances in analytic techniques for these systems, and care to be taken in running simulations.

\subsubsection{Phase-space distributions for quantum systems}

To more completely characterise a wavefunction beyond the Madelung split, we consider the dynamics of a quantum state in phase-space. For a pure quantum state $\psi$ in $n$-dimensions, the Wigner  function \citep{Wigner1932}
\begin{equation}\label{eqn:wigner_dist}
    f_W(\bm{x},\bm{p})= \int \frac{\dd[n]{\tilde{\bm{x}}}}{(\pi\hbar)^n} \exp\left[\frac{2i\bm{p}\cdot \tilde{\bm{x}}}{ a^{3/2}\hbar}\right] \psi\left(\bm{x}-\tilde{\bm{x}}\right) \psi^*\left(\bm{x}+\tilde{\bm{x}}\right)\,,
\end{equation}
represents a quasi-probability distribution in phase-space (here $\psi^*$ denotes complex conjugation). However, the Wigner function does display some unphysical properties for a phase-space distribution function, namely that it can by negative in small localised regions (bounded to a minimum value of $-4\pi/\hbar$ for pure states). It is for this reason that the naïve limit $\hbar \to 0$ does not derive the classical solution in a continuous way, and the classical limit needs to be treated with care \citep{Takahashi1989}. A theorem due to Hudson demonstrates that the only non-negative Wigner functions those which correspond to Gaussian wavefunctions, so for generic states these negative regions necessarily exist \citep{Hudson1974}.

To account for this, one can smooth the Wigner distribution using Gaussian filters in both position and momentum space, resulting in the coarse-grained Wigner function, or the Husimi distribution \citep{Husimi1940}. If the smoothing scales $\sigma_x, \sigma_p$ are chosen to satisfy the uncertainty principle
\begin{equation}\label{eqn:uncertainty_principle}
    \sigma_x \sigma_p \geq \frac{\hbar}{2}\,,
\end{equation} 
the resulting distribution is guaranteed to be non-negative. Under such a smoothing, the semiclassical $\hbar \to 0$ limit can be approached in a mathematically sound way \citep{Lions1993,Gerard1997,Zhang2002,Athanassoulis2009,Athanassoulis2018}. Features on scales of $\order{\hbar}$ do not survive the classical limit after this smoothing. For example, wave dark matter systems can exhibit non-trivial self-similar solutions which vanish identically in the classical limit \citep{Galazo2022}.

From this phase-space distribution, we can construct observables as momentum weighted averages (moments) of this distribution. For example, the first two moments of the Wigner distribution are the density and momentum-flux
\begin{align}
    &\int \dd[n]{\bm{p}} f_W(\bm{x},\bm{p}) = \rho(\bm{x}) = \abs{\psi}^2\,, \\ 
    &\int \dd[n]{\bm{p}} \frac{\bm{p}}{a^{3/2}}f_W(\bm{x},\bm{p}) = \bm{j}(\bm{x}) = \hbar \Im(\psi^*\bm{\nabla}\psi)\,. \label{eqn:wavefunction_velocity_moment}
\end{align}
Using these kinetic moments, we can define a velocity field $\bm{v}=\bm{j}/\rho$ everywhere, even where the phase is not smooth. If one wishes to work with fluid variables even in regions where $\phi_v$ is not smooth, they should replace the quantum continuity and Bernoulli equations~\eqref{eqn:quantum_fluids} with equivalent equations using the momentum flux $\bm{j}$ \citep{Uhlemann2014}.

We often prefer to work with the connected and independent parts of these moments (those which are not simply products of lower order moments), called the cumulants. For example, the second moment of $f_W$ is
\begin{equation}
\int\dd[n]{\bm{p}} \frac{p_i p_j}{a^{3}}f_W(\bm{x},\bm{p}) = \frac{j_i(\bm{x}) j_j(\bm{x})}{\rho(\bm{x})} + \rho(\bm{x})\sigma_{ij}(\bm{x})\,,
\end{equation}
which contains a product of the first moment, and a connected part determined by the velocity dispersion $\sigma_{ij}$. These cumulants directly correspond to (Eulerian) observables: the logarithmic matter density $\ln\rho$, velocity $v_i$, and the velocity dispersion $\sigma_{ij}$. Written in terms of the density, the velocity dispersion associated with a wavefunction is
\begin{align}
    \sigma_{ij} &= \frac{\hbar^2}{4}\left(\frac{\nabla_{\! i} \rho \nabla_{\! j} \rho}{\rho^2} - \frac{\nabla_{\! i}\nabla_{\! j} \rho}{\rho}\right)= -\frac{\hbar^2}{4}\nabla_{\! i}\nabla_{\! j} \ln \rho\,, \label{eqn:sigma_ij}
\end{align}
and is directly related to the quantum ``pressure'' in equation~\eqref{eqn:quantum_bernoulli}. In this formula we can see explicitly how the wave-based model generates higher-order cumulants from derivatives of the wavefunction building blocks.

Under a Madelung split the velocity associated with a wavefunction appears potential, as $\bm{v} = \bm{\nabla} \phi_v$, however, the phase can develop discontinuities (jumps) which source vorticity. These phase jumps only occur at places where the density vanishes and the phase becomes ill-defined. The curvature in the density near the zeros of the wavefunction can then source velocity dispersion via equation~\eqref{eqn:sigma_ij}.

All the higher-order cumulants for the system can be obtained as moments of the Wigner distribution in this way, and will be non-zero after shell-crossing, even when starting from initial conditions described by a perfect fluid. The procedure of extracting moments and cumulants from the wavefunction is exactly the same as extracting moments/cumulants from a Schr\"odinger-Poisson system, although the underlying evolution equation is different from the propagator formalism laid out here.

To determine the full behaviour of a system in phase-space, in principle one would have to solve the evolution equations associated with each cumulant, which becomes difficult as the evolution of the $n^{\rm th}$ cumulant depends on the $(n+1)^{\rm th}$ cumulant. The only consistent truncation to the hierarchy is the perfect fluid model, which corresponds to (in Eulerian coordinates)
\begin{equation}\label{eqn:eulerian_fluid_phase_space}
    f_{\rm fluid}(\bm{x},\bm{p}) = \frac{\rho(\bm{x})}{\bar{\rho}} \delta^{(3)}_{\rm D}\left(\frac{\bm{p}}{a^{3/2}}-\bm{\nabla}_{\! \bm{x}}\phi_v(\bm{x}) \right),
\end{equation}
where all cumulants of order 2 and higher are set to 0 (and remain 0). However, once shell-crossing occurs and the system develops multiple streams, this perfect fluid description is no longer adequate to describe the dynamics of the system as a whole. At this point, the phase-space distribution cannot take the form in equation~\eqref{eqn:eulerian_fluid_phase_space}, and all higher cumulants are sourced dynamically.

The Wigner function's time evolution, induced by the time evolution of $\psi$ through the appropriate Schr\"odinger equation, naturally captures these dynamically sourced higher cumulants. This guarantees that all phase-space information is encoded in the wavefunction, even beyond shell-crossing. This is in the same spirit of looking at the phase-space distribution for a fluid in Lagrangian space  (c.f. Figure~\ref{fig:zeldo_phase_sheet}). A Lagrangian fluid as defined in equation~\eqref{eqn:lagrangian_phase_space_dist} can still develop an infinite hierarchy of cumulants if $x(q)$ is a multiple-to-one mapping describing a mixture of fluid streams in Eulerian space. However, in the wavefunction model we get the same encoded information, but a more direct way to access Eulerian observables, without having to consider mapping between Lagrangian and Eulerian space.

\section{Unweaving the wavefunction}\label{sec:unweaving_the_wavefunction}

\subsection{Obtaining single-stream wavefunctions from interference}

For the free evolution of a wavefunction with Zel'dovich-like initial conditions, the multi-stream region is replaced with small scale wave interference. In the following we demonstrate that the shell-crossed wavefunction can be decomposed into a sum of single-stream, non-interfering wavefunctions, each corresponding to a classical Zel'dovich stream.

\subsubsection{Stationary phase decomposition}

Taking a 1-dimensional wavefunction with Zel'dovich initial conditions, we write the wavefunction over spacetime as an oscillatory integral as in equation~\eqref{eqn:psi_integral_with_zeta}
\begin{equation}
    \psi(x,a) = \mathcal{N} \int \dd{q} \exp\left[{\frac{i}{\hbar}\zeta(q;x,a)}\right], \label{eqn:psi_oscillatory}
\end{equation}
where $\zeta=S + \phi_v^{\rm (ini)}$. The integrand of equation~\eqref{eqn:psi_oscillatory} is highly oscillatory and therefore the dominant contribution to the integral will come from points where $\zeta$ is stationary with respect to the integration variable $q$, in a technique known as the stationary phase approximation (Appendix~\ref{app:SPA}). It is worth noting that the \emph{phase} in the stationary phase approximation (SPA) refers to the function $\zeta$, not the phase of the wavefunction $\phi_v$. The approximation is dominated by the stationary points of $\zeta$, not $\phi_v$.

For a stationary point $q_*$ satisfying $\nabla_{\! q}\zeta(q_*;x,a) = 0$, if $\nabla^2_{\! q}\zeta(q_*;x,a) \neq 0$, the stationary phase contribution from that point reads (suppressing the $x,a$ arguments of $\zeta$)
\begin{equation}\label{eqn:psi_spa}
    \psi_{q_*}^{\rm SPA}(x,a) =  \frac{\exp(\frac{i\pi}{4}[\operatorname{sgn} (\nabla^2_{\! q}\zeta(q_*))-1])}{\sqrt{a\abs{\nabla^2_{\! q}\zeta(q_*)}}} \exp\left(\frac{i}{\hbar}\zeta(q_*)\right).
\end{equation}
The full approximation is then given by a sum over all of these stationary points,
\begin{equation} \label{eqn:psi_spa_full}
    \psi^{\rm SPA}(x,a) =\sum_{q_*} \psi_{q_*}^{\rm SPA}(x,a)\,,
\end{equation}
which in the semiclassical $\hbar \to 0$ limit obeys the asymptotic relation $\psi(x,a) \to \psi^{\rm SPA}(x,a)$.

For the choice $\phi_v^{\rm (ini)}(q) = \cos(q)$, number of (real) stationary points is determined by the same caustic condition as equation~\eqref{eqn:shell_cross_region}, with three real roots in the shell-crossed region, two real roots on the caustic line, and a single real root in the single-stream regime. Indeed this is not surprising, as for the Zel'dovich-like initial conditions we have
\begin{equation}\label{eqn:zeta}
    \zeta(q;x,a) = \frac{(x-q)^2}{2a}+\phi_v(q)\,,
\end{equation}
so the condition for $\zeta$ to be stationary with respect to $q$ is 
\begin{align}\label{eqn:stationary_condition}
    \nabla_{\! q}\zeta(q;x,a) = -\frac{x}{a} + \frac{q}{a} + \nabla_{\! q}\phi_v(q) = 0\,,
\end{align}
which is equivalent to the Zel'dovich displacement mapping $x = q + a\nabla_{\! q} \phi_v(q)$.  That is, the quantum propagation integral is dominated by the $q$ which satisfy the classical displacement mapping.

\begin{figure}[h!t]
    \centering
    \includegraphics[width=\columnwidth]{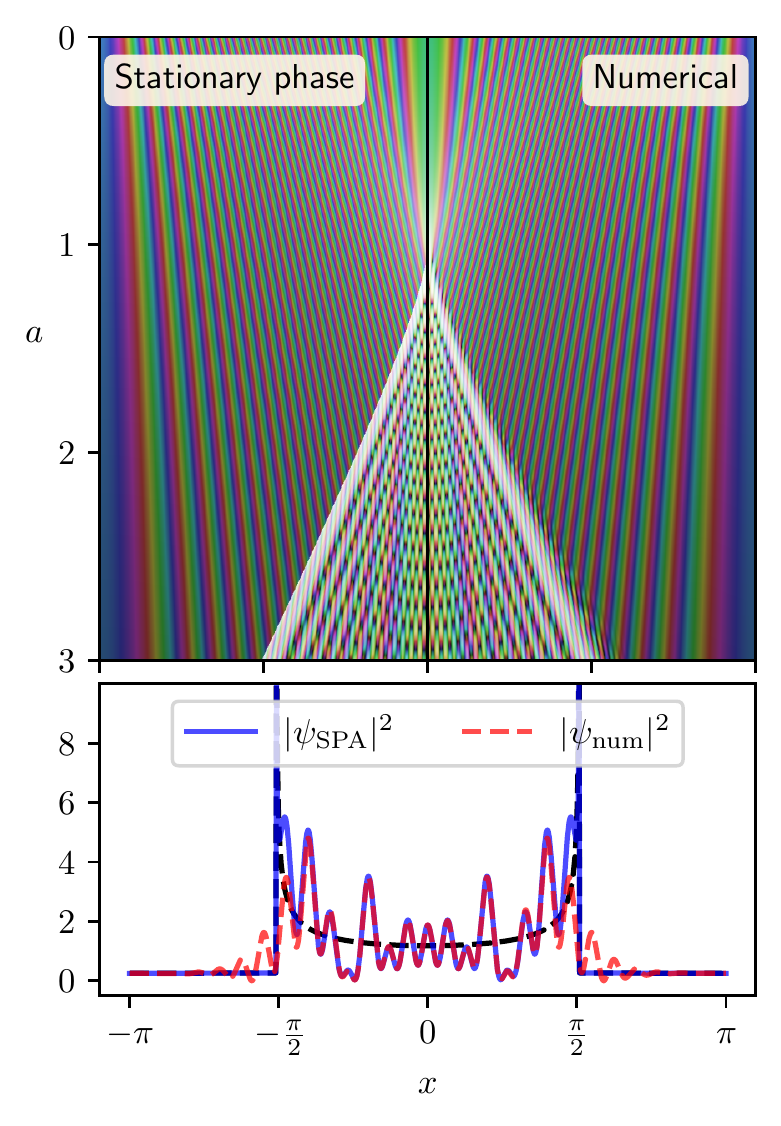}
    \caption{Comparison of the stationary phase approximation (SPA) and the numerical evolution of $\psi^{\rm (ini)}(q)=\exp(i\cos(q)/\hbar)$. (Upper panel) The full spacetime evolution of this wavefunction, with $\hbar=0.01$. (Lower panel) The density profile post shell-crossing of this wavefunction, with $\hbar = 0.05$. The black dashed line shows the classical Zel'dovich density, or equivalently the sum of the densities of the individual stationary phase streams. From these figures we see that the stationary phase approximation well approximates the numerical solution, except very close to the caustic boundary. This indicates that well inside the single or multi-stream regime that the interference can be resolved into a sum over classical trajectories. Near the caustic line, stationary phase analysis fails as the stationary points are no longer well separated in the complex plane. Behaviour close to the caustic should be studied using the normal forms of diffraction catastrophe integrals, discussed in Section~\ref{sec:catastrophe_theory}.}
    \label{fig:compare_SPA_to_numeric_joint}
\end{figure}

Figure~\ref{fig:compare_SPA_to_numeric_joint} compares the full stationary phase analysis of the Fourier mode to the numeric integration of the wavefunction. The stationary phase approximation provides an excellent description of the wavefunction except near to the caustic lines themselves. However, well into the single or multi-stream regime, far away from high density caustics, this stationary phase decomposition provides a good description of the full wavefunction.

\begin{figure*}[t]
    \centering
    \includegraphics[width=2\columnwidth]{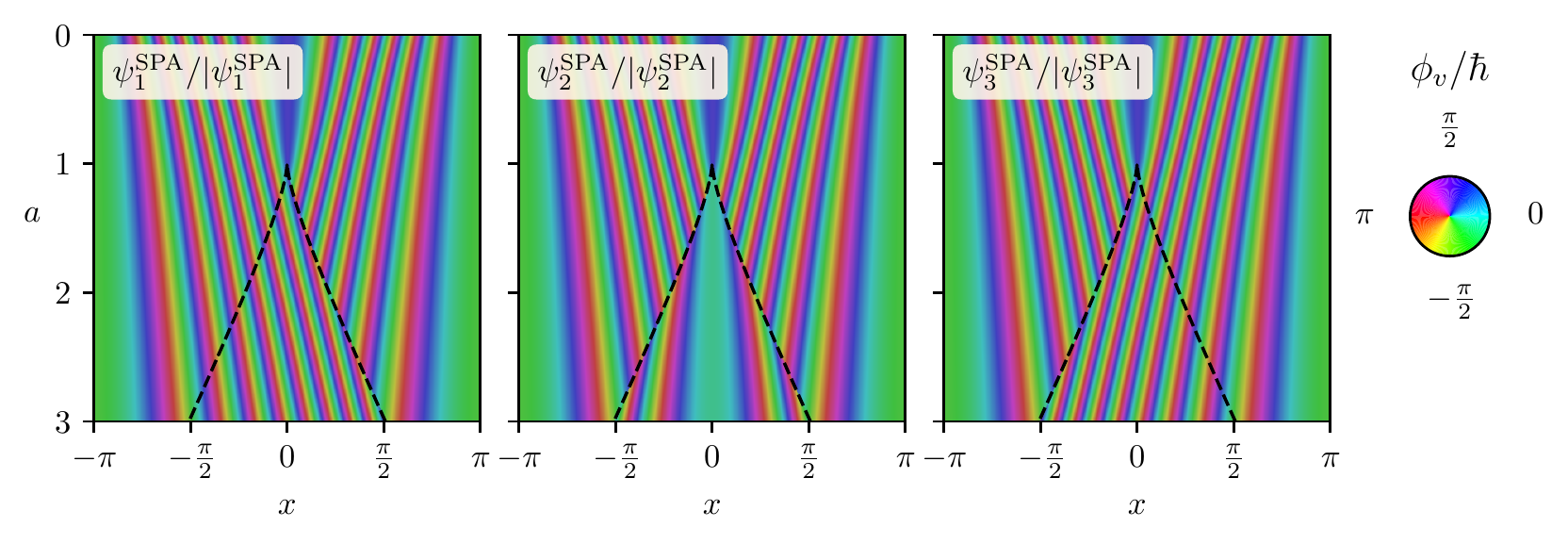}
    \caption{The polar angle of the normalised wavefunctions ($\hbar = 0.05$) split according to each stationary point of the function $\zeta$ in equation~\eqref{eqn:zeta}. Each of these stationary points corresponds to one of the classical Zel'dovich trajectories. Outside the cusp there is a single stationary point, corresponding to the single stream (shown in all three panels), while inside the classical cusp three wavefunctions exist, corresponding to the three classical streams illustrated in Figure~\ref{fig:zeldo_phase_sheet}.}
    \label{fig:stream_splitting_phase_hbar0.05}
\end{figure*}

Figure~\ref{fig:stream_splitting_phase_hbar0.05} shows the individual terms in equation~\eqref{eqn:psi_spa_full} as separate wavefunctions. We see that indeed these are single-stream wavefunctions which are non-interfering and resemble individual Zel'dovich streams. Outside the classical cusp there is a single-stream wavefunction, and inside the cusp there are three stream wavefunctions, corresponding to the three sheets of the Zel'dovich phase-space sheet in Figure~\ref{fig:zeldo_phase_sheet}. This is precisely analogous to taking the geometric optics limit for an optical wave field, resolving wave phenomena as sums over classical light rays.

\begin{figure}[h!t]
    \centering
    \includegraphics[width=\columnwidth]{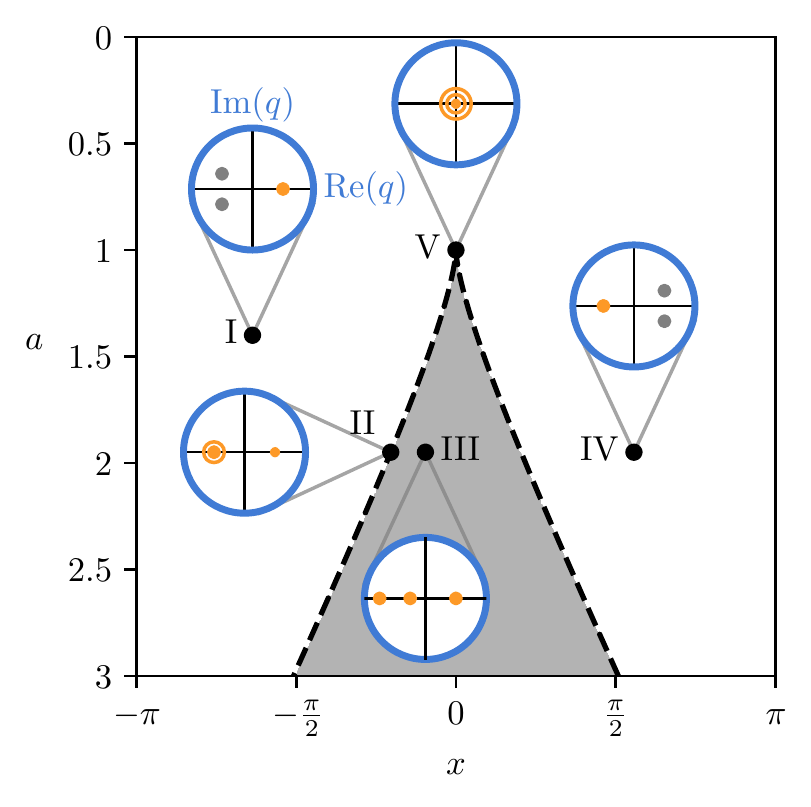}
    \caption{The structure of stationary points $q_*$ which solve $\nabla_q \zeta(q_*;x,a) = 0$ (equation~\eqref{eqn:stationary_condition}) for different values of $x$ and $a$. The blue circles show the distribution of $q_*$ in the complex-$q$ plane. Beginning at point (I) in the single-stream (white) region, two solutions are complex, with only one real root. Approaching the caustic (dashed black) line, as with point (II), the two complex roots merge into a real double root. Points within the multi-stream (shaded) region, such as (III), have all three roots are real. Crossing the $x=0$ line causes the sign of the real part of the roots to flip, as can be seen comparing points (I) and (IV). At the cusp point (V), all three roots coalesce into a single real solution. For our stationary point analysis, we only use contributions from real stationary points (coloured in yellow in this diagram). }
    \label{fig:root_structure}
\end{figure}

Figure~\ref{fig:root_structure} illustrates how the stationary points are distributed in the complex-$q$ plane for different values of $x$ and $a$. By moving from the single-stream region across the caustic line, the complex stationary points coalesce into a double root, before becoming distinct real roots in the shell-crossed region. This merging and splitting of roots at the caustic is why stationary phase contributions of the form \eqref{eqn:psi_spa} no longer well approximate the integral in equation~\eqref{eqn:psi_oscillatory}. On the caustic defined by $\nabla_{\! q} \zeta = \nabla_{\! q}^2 \zeta = 0$ the SPA wavefunction form~\eqref{eqn:psi_spa} does not apply, and near the caustic the stationary points are not sufficiently separated (with respect to the size of $\hbar$) in the complex plane to be treated independently. 

A more accurate asymptotic approximation can be made by considering the contributions from the complex roots along a proper steepest descent contour of the integrand, which are the same as contours of constant $\mathrm{Im}(i\zeta/\hbar)$ and as such remove the oscillatory part of the integral \citep[see e.g.][for general discussion of this method]{BenderandOrszag}.  The asymptotic analysis of such oscillatory integrals in the complex plane in general is difficult, as the geometry of the steepest descent contours, and the number of complex stationary points which are relevant can change suddenly in different regions of the parameter space (Stokes' phenomena). The Stokes' lines, where the steepest descent contour changes how many stationary points it intersects by one, for a cusp integral are similar to the case considered here and calculated analytically in \cite{Wright1980}. While we restrict ourselves to simple stationary phase analysis of real roots, \cite{Feldbrugge2019} present some techniques for automatically accounting for the changes in the nature of the stationary points and integration contours in a more rigorous manner. Figure 14 from \cite{Feldbrugge2019} is similar to our Figure~\ref{fig:root_structure}, showing the roots, steepest descent contours, and Stokes' lines for the canonical cusp integral discussed in Section~\ref{sec:catastrophe_theory}. While the specific function these stationary points arise from is different than the function $\zeta(q;x,a)$ corresponding to our toy model, the qualitative features are the same. We focus here only on the real stationary points as they have clear physical interpretation as classical trajectories, and recover classical CDM dynamics.

The principal qualitative feature which cannot be reproduced by stationary phase decomposition is the existence of zeros in the density field \emph{outside} the shell-crossed region. These zeros are of physical interest, as they result in discontinuities in the phase of the wavefunction which will be discussed in more detail in Section~\ref{sec:phase_properties}. For an optical system, zeros outside the cusp correspond to interference between a single real ray and a complex ray. Since the amplitude of the complex ray decays exponentially, it only has enough amplitude to fully destructively interfere with the real ray once outside the caustic \citep{Wright1980}. For the free wavefunction, the interference which produces these exterior zeros occurs between the infalling part of the wavefunction (the ``real ray'', corresponding to a single stream wavefunction from the SPA decomposition), and a partial reflection of the wavefunction due to the quantum pressure term in the Bernoulli equation~\eqref{eqn:quantum_bernoulli} (the ``complex ray''). As the quantum pressure disappears in the classical limit, only density field zeros within in the multi-stream region remain.

\subsection{Phase-space distributions for individual streams}

To further establish the correspondence between these stationary phase contributions and classical streams, we can look at the entire phase-space distribution associated with these wavefunctions. While the individual $\psi^{\rm SPA}_{q_*}$ can be constructed without constructing a quantum phase-space distribution, it is worth looking at the correspondence in the full phase-space analysis.

\begin{figure}[h!t]
    \centering
    \includegraphics[width=\columnwidth]{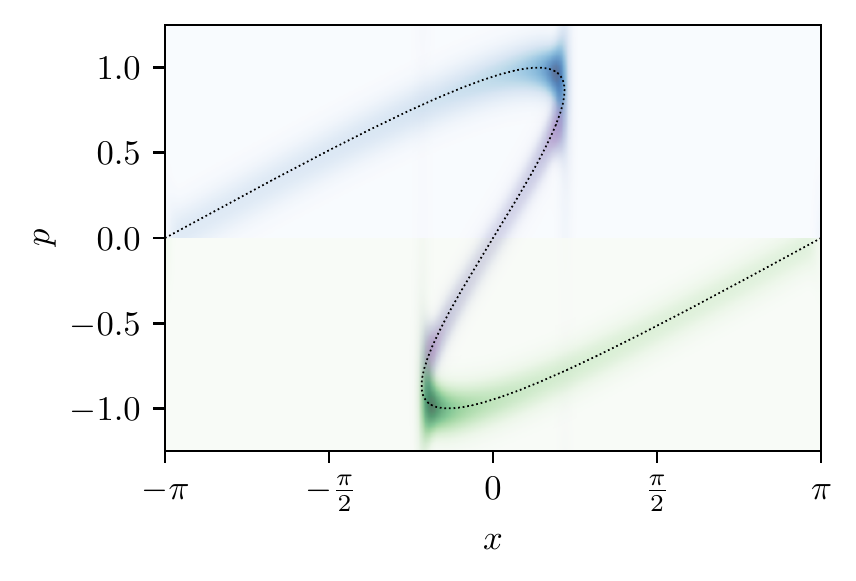}
    \caption{The Husimi phase-space distributions for each of the individual wavefunctions associated with stationary points of $\zeta$ post shell-crossing ($\hbar=0.01$). We see that this stationary phase decomposition naturally dissects the classical Zel'dovich phase-space sheet (dotted line). At the points where the phase-space sheet turns over, the Husimi functions blur in momentum space, due to the sharp localisation of the wavefunction in position space. Here we use a Gaussian filter to coarse-grain the Wigner distribution with $\sigma_x = 0.049$ and $\sigma_p= {\hbar}/({2\sigma_x})$ that satisfy the uncertainty principle \eqref{eqn:uncertainty_principle}. }
    \label{fig:stream_split_wigner_overplot}
\end{figure}

The Husimi phase-space distributions associated with each term in the stationary phase approximation are shown in Figure~\ref{fig:stream_split_wigner_overplot}. We see that the stationary points naturally dissect the phase-space sheet into its three layers, as anticipated from looking at the phase plots in Figure~\ref{fig:stream_splitting_phase_hbar0.05}. We see also that the quantum phase-space sheet traces the characteristic S-shaped curve expected for Zel'dovich dynamics, with a characteristic width around the classical phase-space sheet owing to the uncertainty principle associated with the size of $\hbar$. The Husimi phase-space distributions are calculated using an adapted version of the \texttt{CHiMES}\footnote{\href{https://github.com/andillio/CHiMES}{https://github.com/andillio/CHiMES}} code \citep{Eberhardt_2021_CHiMES}. We choose to plot Husimi distributions instead of Wigner distributions to emphasise the classical phase-space structure and to avoid the presence of aliasing in discrete Wigner distributions \citep{Claasen1983, Chassande-Mottin2005}. 

The Husimi phase-space distributions of the individual stationary phase terms present an accurate tracing of the Zel'dovich phase-space sheet, except near the points where the phase-space sheet turns over. This is expected, as the points where the phase-space sheet turns over are precisely the classical caustic line, where the number of terms in the stationary phase approximation changes from 1 to 3. Near these points, the wavefunction streams develop a sharp discontinuity, localising on one or both sides of the stream, leading to a blurring in the momentum as required by uncertainty. This is expected however, as we know that near the classical caustic line the sum of the SPA terms does not accurately approximate the full wavefunction as discussed in the previous section. This is encouraging, as it demonstrates that this splitting does not just accurately reproduce low order cumulants, but the full dynamics of the system.

\subsection{Comparing SPA streams to Zel'dovich properties}
\label{subsec:SPAstreamsZeldo}

To make the relationship between this stationary phase decomposition and the classical trajectories more explicit, we can read off the density and velocity associated with each stationary point. From equation~\eqref{eqn:psi_spa}, we see that the density of the $i^{\rm th}$ wavefunction is given by
\begin{align}
    \rho_i(x,a) &= \frac{1}{\abs{a\nabla^2_{\! q}\zeta(q_i;x,a)}} = \frac{1}{\abs{1-a\nabla_{\! q}^2 \phi_v^{\rm (ini)}(q_i(x,a))}}\,, \label{eqn:spa_density_to_zeldo}\\
    &= \frac{1}{\abs{1-a\cos(q_i(x,a))}}\,, 
\end{align}
where the final equality holds for the specific case $\phi_v^{\rm (ini)}(q)=\cos(q)$ considered for our Zel'dovich mode. This is precisely the same density profile one predicts from the Zel'dovich approximation (c.f. equation~\eqref{eqn:zeldo_density}).

\begin{figure}[h!t]
    \centering
    \includegraphics[width=\columnwidth]{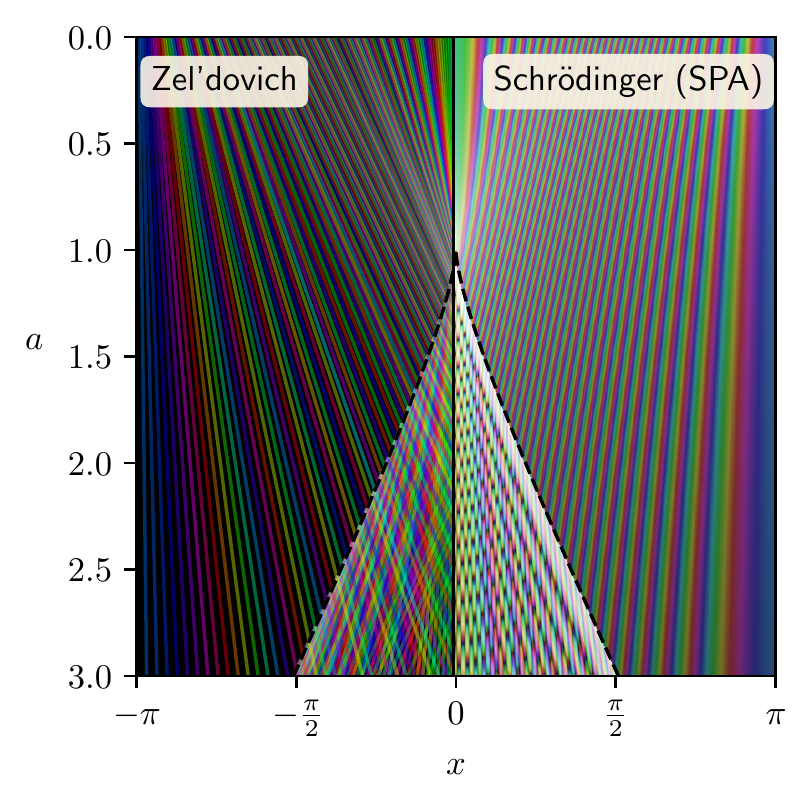}
    \caption{Comparison of classical Zel'dovich trajectories to the stationary phase approximation for a  wavefunction (with $\hbar=0.01$) evolving under the free Schr\"odinger equation. The stationary phase wavefunction is the proper analogue to this classical system, having the exact same density and velocity profiles, and forming the same shell-crossed region.}
    \label{fig:zeldo_vs_SPA}
\end{figure}

The velocity field associated with each stream can be extracted by taking gradients (with respect to $x$) of the phase in equation~\eqref{eqn:psi_spa}. This recovers the velocity of the $i^{\rm th}$ stream at position $x$ and time $a$,
\begin{equation}
    v_i(x, a) = \nabla_{\! x} \zeta(q_i,x,a)\eval_{q_i = \rm const} = \frac{x-q_i}{a}\,,
\end{equation}
which is the constant velocity motion expected from the Zel'dovich approximation.\footnote{We note that the velocity can also be obtained as $\bm{j}/\rho$ using the expression for $\psi$ given in equation~\eqref{eqn:psi_spa_full}, as is done in Appendix C of \cite{Uhlemann2019}.} This is completely generic, not requiring specific initial conditions. We thus see that the density and velocity of the streams extracted from stationary phase \emph{exactly} match those of a Zel'dovich system with the same initial conditions. Because of this exact correspondence, the stationary phase decomposition is the correct way to upgrade individual Zel'dovich streams into wavefunctions. Figure~\ref{fig:zeldo_vs_SPA} shows these two analogue systems side by side. Both of these systems have precisely the same extent of the shell-crossed region, with the SPA wavefunction not presenting the same finite width to the caustic line that the full numerically evolved wavefunction does.

As a note of caution however, we point out that the phase of the individual stationary phase streams in equation~\eqref{eqn:psi_spa} is \emph{not} simply the initial phase transported to the final position via the displacement mapping. The phase of $\psi^{\rm SPA}_{q_i}$ also contains time propagation terms responsible for converting the Lagrangian derivative determining velocity into a Eulerian one. Starting from the link between Eulerian and Lagrangian velocity potentials,
\begin{subequations}
\label{eq:velpot_Eulerian}
\begin{equation}
    \nabla_{\! x}\phi_v^{\rm E}(x) = \pdv{q}{x}\nabla_{\! q}\phi^{\rm (ini)}_v(q) = \frac{1}{J}\nabla_{\! q}\phi_v^{\rm (ini)}(q)\,,
\end{equation}
with the Jacobian $J=\partial x/\partial q$, we can integrate to get the Eulerian velocity potential $\phi_v^{\rm E}$,
\begin{align}
    \phi_v^{\rm E}(x(q,a)) &= \int \dd{q} \nabla_{\! q}\phi_v^{\rm (ini)}(q) J \\
    &= \cos(q) - \frac{a}{2} \cos^2(q) + C\,,
\end{align}
\end{subequations}
where the second line holds for $\phi_v^{\rm (ini)}(q) = \cos(q)$ and $C$ is an integration constant.

``Zel'dovich wavefunctions,'' constructed from the Zel'dovich density and the Eulerian phase, 
$
\psi^{\rm (Z)}_i(x,a) = \sqrt{\rho_i(x,a)} \exp(\frac{i}{\hbar}\phi_v^{\rm E}(x(q_i,a)) )
$
reproduce the wavefunctions of the individual SPA streams. So while the colouring scheme used for the particle trajectories in Figures~\ref{fig:free_schrodi_evol} and \ref{fig:zeldo_vs_SPA} provides intuition for the similarities between the dynamics, the appropriate phase for one of the component wavefunctions at spacetime position $(x,a)$ is not simply the phase $\phi^{\rm (ini)}(q_i(x,a))$ as if carried by particles, but the Eulerian phase $\phi^{\rm E}_v(x(q_i,a))$. The explicit difference between these phases can be seen in the left panel of Figure~\ref{fig:zeldo_loop_avg_phase}, with the Lagrangian velocity potential evaluated at a Eulerian position shown in blue and the proper Eulerian velocity potential (or SPA wave phase) in orange. 

\subsection{Comparison to other wave techniques}
Here we briefly summarise how the stationary phase decomposition presented here relates to similar looking models used in the literature. 

\cite{Veltmaat_2018_FDM_halos} and \cite{SchwabeNiemeyer2022} interface $N$-body simulations and full Schr\"odinger-Poisson solvers to study individual halos. Both of these methods rely on translating $N$-body particles into wavepackets to build a wavefunction which sets the initial and boundary conditions for the Schr\"odinger-Poisson solver on small scales. They use classical wavefunctions \citep{Wyatt_2005_quantum_trajectories} for individual particle wavepackets to construct an overall wavefunction. In \cite{Veltmaat_2018_FDM_halos} wavepackets in position space are used to obtain a phase from the combined wavefunction, while the amplitude is determined by the classical density in order to erase the interference of overlapping wavepackets which is present even in the single-stream regime. This problem could be avoided by using wavepacket `beams' localised in phase space, for which \cite{SchwabeNiemeyer2022} present an approximation for obtaining a phenomenological wavefunction including interference from collapsed halos. Our SPA split for the wavefunction corresponds to assigning classical wavefunctions to the classical fluid streams instead of individual particles.

\cite{Lague_2021_FDM_LPT} approach modelling mixed dark matter containing FDM and CDM by adapting LPT. Their approach is to use the CDM LPT displacement but adjust for the difference in the linear growth using a scale-dependent transfer function for fuzzy dark matter. This incorporates the quantum pressure at the linear level while neglecting nonlinear effects in the displacement described in Appendix B of \cite{Uhlemann2014}. Our propagator approach transports a dark matter wave function thus incorporating the wave nature in the evolution, which is lost when treating FDM as particles moved by a modified displacement field.

\section{Wave interference effects}\label{sec:phase_properties}

We now turn our attention to another piece of interesting phenomenology associated with multi-streaming: the dynamical production of vorticity and higher order cumulants like velocity dispersion. In the wave-mechanical model, these features arise from interference and are concentrated in regions where the wavefunction vanishes and topological defects form. In Section~\ref{sec:hidden_features} we show that the  interference features can be isolated in a ``hidden'' part of the wavefunction, which decorates the ``average'' part of the wavefunction describing the Zel'dovich fluid behaviour.

\subsection{Phase jumps and vorticity}

\begin{figure}[h!t]
    \centering
    \includegraphics[width=\columnwidth]{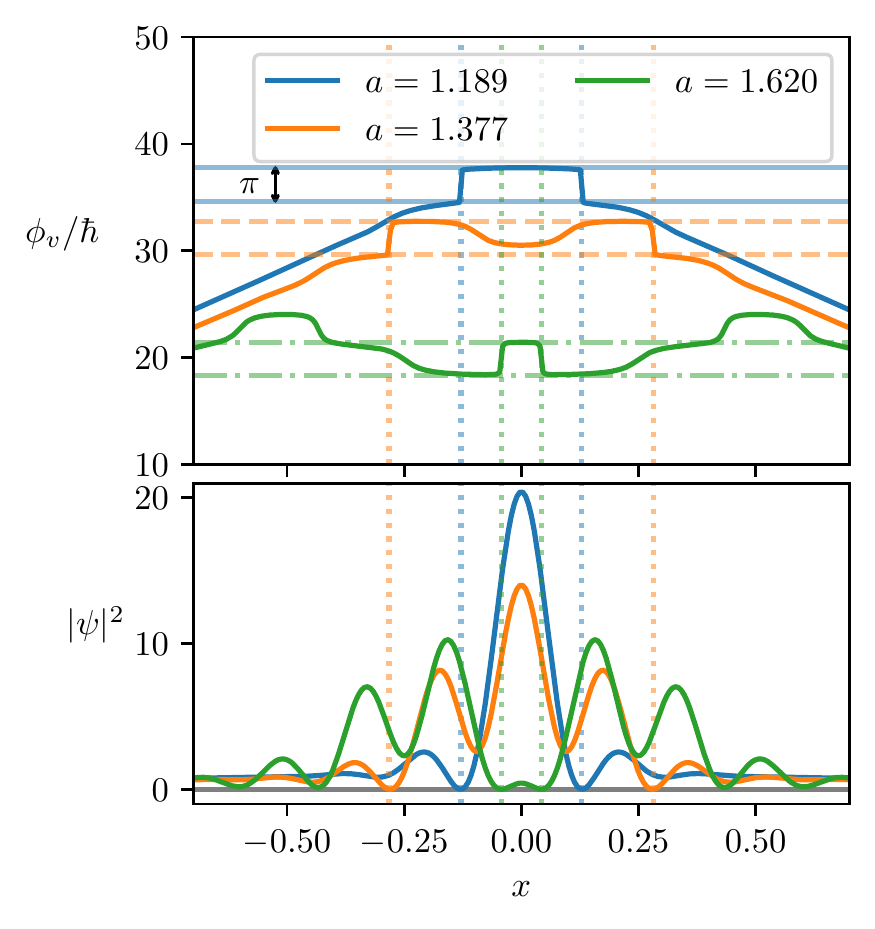}
    \caption{The unwrapped phase (upper panel) and density (lower panel) of $\psi$ at times past shell-crossing when defects occur (with $\hbar=0.05$). The spatial phase discontinuity is always $\pm \pi$, as indicated by the horizontal lines, which are spaced apart by $\pi$. These phase discontinuities source the non-potential part of the velocity, and occur at points where the density vanishes (as seen in the lower panel). The narrow regions around the zeros in density source velocity dispersion through equation~\eqref{eqn:sigma_ij}. The times were chosen to show the first two phase jumps occurring outside the cusp (as seen in Figure~\ref{fig:bp_located}),  and the first interior phase jump. }
    \label{fig:phase_jumps_1d}
\end{figure}

As discussed in Section~\ref{sec:observables}, post shell-crossing, all cumulants are sourced dynamically. In the wave-mechanical model, these higher order cumulants are  hidden in the oscillations of the wavefunctions and can be extracted from moments of the Wigner distribution function \eqref{eqn:wigner_dist}. 
In addition to these higher order cumulants, such as velocity dispersion, shell-crossing sources vorticity, even for initially irrotational velocity fields. In the classical system this vorticity is due to multi-stream averaging, which results in a non-potential velocity field. In the wave-mechanical model, the non-potential parts of the velocity field occur at isolated points where the density vanishes. At these so-called branch points, the  wavefunction can develop phase jumps that cause infinite spatial gradients. As shown in Figure~\ref{fig:phase_jumps_1d}, the spatial dislocation of the phase is a jump of $\pm \pi$ as the complex value of the wavefunction passes through zero \citep{Hui2021}. Since the wavefunction is single valued, the winding of $\phi_v/\hbar$ around a closed loop $\gamma$ must be an integer multiple of $2\pi$ leading to a quantisation of $\phi_v$ \citep{Feynman1958, Gross1961}. This can be expressed as the integral quantisation condition
\begin{equation}\label{eqn:quantised_vorticity}
    \Gamma =  \oint_{\gamma} \bm{\nabla} \phi_v \cdot \dd{\bm{x}} = 2\pi\hbar n\,, \quad n\in \mathbb{Z}\,,
\end{equation}
related to vorticity via Stokes' theorem $\Gamma = \int_\Sigma (\bm{\nabla} \times \bm{v}) \cdot \hat{\bm{n}} \dd{A}$, where $\Sigma$ is the surface bounded by the curve $\gamma$, which has normal vector $\hat{\bm{n}}$ and area element $\dd{A}$. In 2-dimensional systems these vortices are points and in 3-dimensional systems closed vortex loops trace the defects in the phase of the wavefunction. The association of these branch points and vanishing of the wavefunction with the sourcing of vorticity and higher order cumulants is precisely why the Madelung formalism breaks down post shell-crossing.

These vortex cores are of phenomenological interest for wave dark matter  signatures in collapsed structures, for example the way that such vortex lines interact with the presence of solitonic cores in fuzzy dark matter halos \citep{hallock_vortex_2011, Rindler-Daller2012MNRAS, Hui2021JCAP, Hui2021,   Schobesberger2021MNRAS},   or in their role in cosmic filament spins \citep{Alexander2021arXiv}. Near the vortex cores, the velocity scales as $v\sim 1/r_\perp$, owing to equation~\eqref{eqn:quantised_vorticity}, while the density profile near a branch point scales as $\abs{\psi}^2 \sim r_{\perp}^2$, where $r_\perp$ is the perpendicular distance from the vortex core/line \citep{Hui2021JCAP}.

For the 1+1D toy model, we do not develop true vorticity, as in one spatial dimension the condition~\eqref{eqn:quantised_vorticity} is trivially satisfied. Nevertheless, the localised phase jumps create a non-potential velocity field and the associated vanishing of the density induces velocity dispersion, both of which are classically produced by multi-stream averaging. The derivatives of the density about these zeros produces velocity dispersion via equation~\eqref{eqn:sigma_ij}.

\subsubsection{Locating the phase jumps}

\begin{figure}[h!t]
    \centering
    \includegraphics[width=\columnwidth]{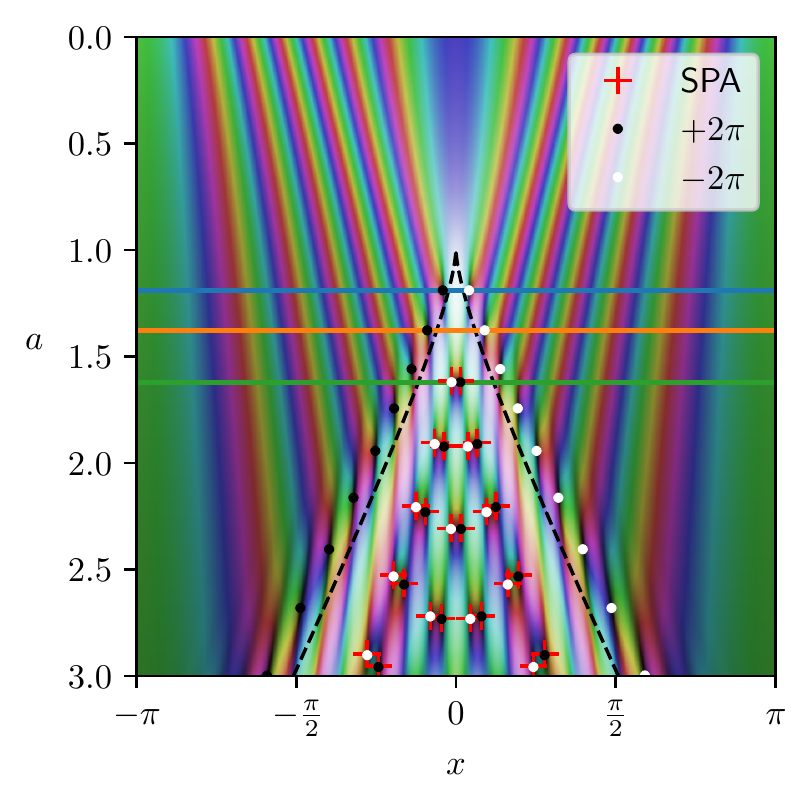}
    \caption{Location of branch points in the evolution of the free wavefunction ($\hbar=0.05$). The branch points predicted by the stationary phase approximation (SPA) are shown as red crosses, and accurately capture the position of the interior branch points from the full wavefunction. Points are coloured according to their (spacetime) circulation value in the diagram via equation~\eqref{eqn:quantised_vorticity}. The horizontal lines indicate the times chosen for the spatial phase profiles shown in Figure~\ref{fig:phase_jumps_1d}.}
    \label{fig:bp_located}
\end{figure}

Figure~\ref{fig:bp_located} shows the evolution of the wavefunction, with the spacetime positions of the phase jumps located, and identified with their sign. While the figure shows the time-evolution of a 1-dimensional system, we make use of the 2D spacetime diagram to find these branch points. Numerically, we located these branch point by masking the phase to only include regions of low density (since branch points only occur where $\abs{\psi} = 0$). Then the circulation (in spacetime) $\Gamma$ can be calculated by discrete line integral, and the branch points are taken to be the centroids of the connected regions where the circulation is $\pm 2 \pi$. The same procedure could instead be applied to constant time slices of a 2-dimensional wavefunction, locating vortex cores.

Figure~\ref{fig:bp_located} demonstrates that the phase jumps predicted from the SPA decomposition accurately locate the branch points inside the classical caustic. The first branch points to form are a pair outside the classical caustic line, owing to interference from the partial reflection of the wavefunction due to the quantum pressure term in equation~\eqref{eqn:quantum_bernoulli}. We will first focus on the interior branch points and their relation to multi-streaming and return to the exterior branch points in Section~\ref{sec:catastrophe_theory}.

\subsubsection{Conserved quantities associated with vorticity}
In a classical fluid system, the Kelvin-Helmholtz theorem \citep{Helmholtz1858,Kelvin1869} requires that the circulation of the velocity field,
\begin{equation}
    \Gamma = \oint_{\gamma(a)}\bm{v}\cdot \dd{\bm{x}} = \int_{\Sigma(a)} (\bm{\nabla} \times \bm{v}) \cdot \hat{\bm{n}}\dd{A}\,,
\end{equation}
is conserved. Here $\gamma(a)$ is a loop, bounding a surface $\Sigma(a)$ at time $a$ with associated normal vector $\hat{\bm{n}}$ and area element $\dd{A}$. The integration surface is taken to be transported by the fluid flow from an initial surface $\Sigma^{\rm (ini)}$. In particular, this means that an initially irrotational flow will conserve its initial value,
\begin{equation}
    \Gamma = \int_{\Sigma^{\rm (ini)}}(\bm{\nabla} \times \bm{v}^{\rm (ini)}) \cdot \hat{\bm{n}}\dd{A} = 0\,.
\end{equation}
This requires that any vorticity produced in the multi-stream regime arises only from stream averaging. 

Under certain circumstances the conservation of circulation also applies to quantum and semiclassical systems \citep{Damski2003}. For sufficiently smooth initial conditions one can apply the same definition of $\Gamma$, using the velocity defined by $\bm{j}/\rho$ and ensuring that the integral contour only goes through regions where the velocity is well defined, as we did in equation~\eqref{eqn:quantised_vorticity}. However, for a loop $\gamma(a)$, the conservation of circulation only holds if such a loop initially only goes through points where the velocity is well defined, and will evolve only through such points \citep{Damski2003}, thus excluding the branch point locations, where a relaxed quantisation condition~\eqref{eqn:quantised_vorticity} applies.
For an initially irrotational system with $\Gamma=0$ these dislocations must occur in pairs (called rotons \citep{Savchenko1999}) such that
\begin{equation}
    \int (\bm{\nabla} \times \bm{v}) \cdot \hat{\bm{n}} \dd{A} = 2\pi (n_+ - n_-) \hbar = 0\,, \quad n_\pm \in \mathbb{N}\,,
\end{equation}
if the surface is taken to be the entire space to ensure global conservation. The 1+1-dimensional system shown in Figure~\ref{fig:bp_located} resembles this pair creation property, with the spacetime circulation being globally conserved throughout the evolution.

For the 1+1D toy model, the circulation is trivially satisfied on an interval and the appropriate conserved quantity is the Poincar\'e-Cartan invariant from Hamiltonian dynamics (e.g. Chapter 9 of \cite{Arnold1978_classicalmechanics} or Chapter 3 of \cite{heller_semiclassical_2018}). As our semiclassical model has a natural Hamiltonian structure (see e.g. equation~\eqref{eqn:schrodinger_with_potential}), this conserved quantity remains valid for higher dimensions. The Poincar\'e-Cartan invariant is formulated by extending the phase-space, relying on the time $a$ as an extra variable conjugate to (minus) the Hamiltonian $\mathcal{H}$. The quantity
\begin{equation}\label{eqn:PC_invariant}
    \tilde{\Gamma}(\gamma) = \oint_\gamma \bm{v} \cdot \dd{\bm{x}} - \mathcal{H}\dd{a}\,,
\end{equation}
is then conserved along the Hamiltonian flow and generalises the circulation theorem from fluid dynamics.

\begin{figure}[h!t]
    \centering
    \includegraphics[width=\columnwidth]{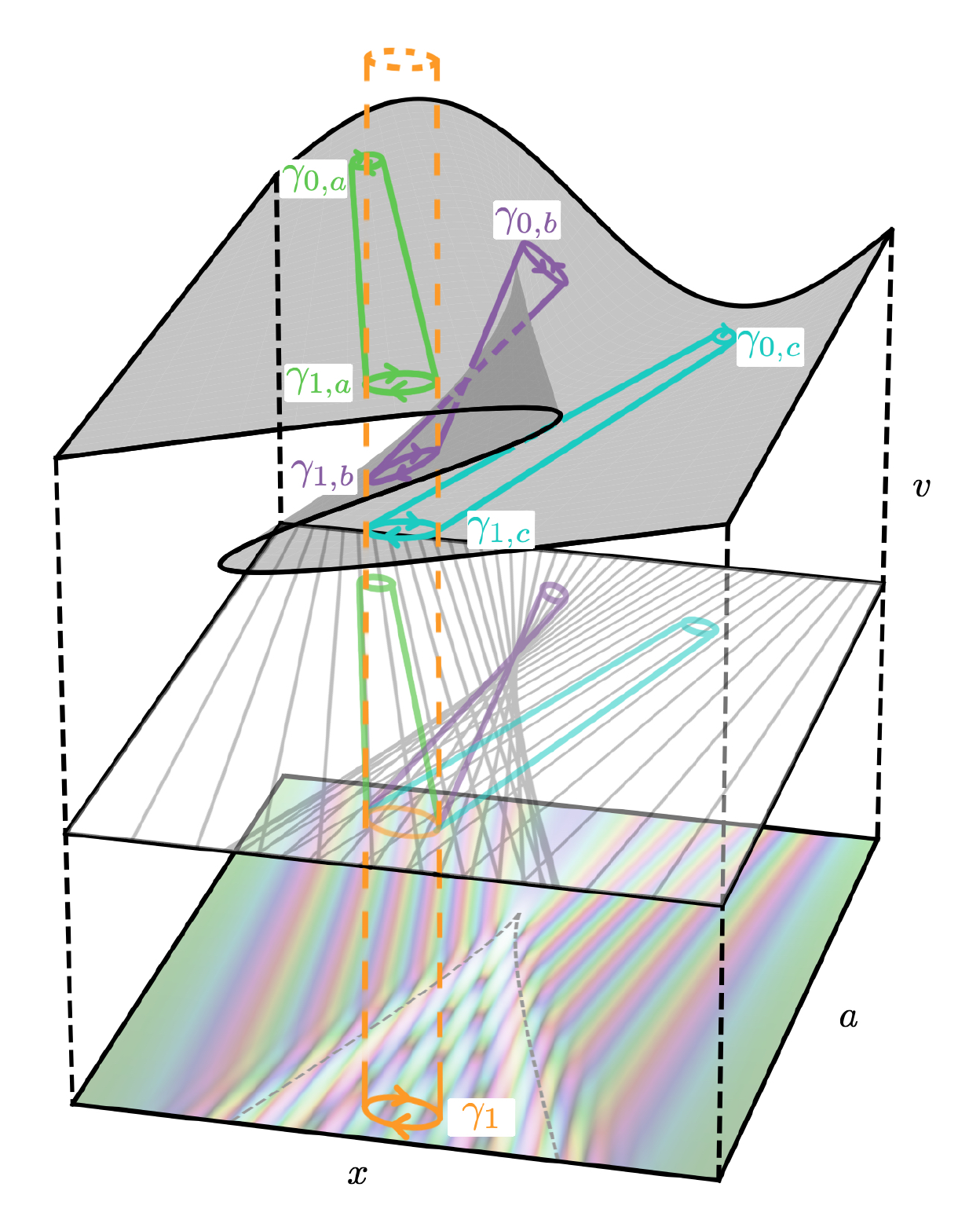}
    \caption{Illustration of an inadmissible spacetime loop for the Poincar\'e-Cartan invariant. Shown are the phase-space sheet for the system (top, as in Figure~\ref{fig:zeldo_phase_sheet}), together with projections onto spacetime for both classical trajectories (middle), and the wave mechanical model (bottom). The orange contour $\gamma_1$ drawn in spacetime is not appropriate to calculate the invariant, as it corresponds to three different loops in phase space, labelled $\gamma_{1,i}, i\in\{a,b,c\}$. As such, the orange spacetime loop cannot be evolved uniquely by Hamiltonian flow to relate it to another loop via the invariant. Since the $\gamma_{1,i}$ loops are local in phase-space, each has a unique time evolution (shown on both the phase-space sheet and in projection), and can be evolved backwards by Hamiltonian flow to loops $\gamma_{0,i}$. The invariant is conserved along each of these loops individually, e.g. $\tilde{\Gamma}(\gamma_{0,i})=\tilde{\Gamma}(\gamma_{1,i})$.  Admissible loops drawn in spacetime are must stay in the singe-stream region (classically), and must not include branch points (in the quantum case), which guarantees unique evolution when projected to phase space. }
    \label{fig:PC_invariant}
\end{figure}
Figure~\ref{fig:PC_invariant} shows a schematic diagram of loops in phase-space, demonstrating that not all loops in spacetime can be used to calculate the Poincar\'e-Cartan invariant. The orange contour in spacetime corresponds to three different loops in phase space, and cannot be evolved through Hamiltonian flow. This essentially amounts to asking ``which $\bm{v}$'' to chose in equation~\eqref{eqn:PC_invariant}. Loops in spacetime which are entirely outside the classically shell-crossed region (and stay away from the outside branch points) are uniquely mapped to loops on the phase-space sheet, and therefore can be related to their initial values. This can be seen by considering a loop in the bottom projection of Figure~\ref{fig:PC_invariant} in spacetime fully outside the caustic and projecting it upwards. Such a loop intersects the phase-sheet exactly once, which then allows unique time evolution via the Hamiltonian flow. Such loops conserve the value of the Poincar\'e-Cartan invariant $\tilde{\Gamma}$ until the Hamiltonian flow makes them intersect the classical caustic.

\subsection{Hidden features beyond a perfect fluid}\label{sec:hidden_features}

We now show that it is possible to decouple the presence of these phase jumps and the associated oscillating density from an ``average'' wavefunction capturing the stream-averaged fluid density and velocity. 

\subsubsection{The average wavefunction from SPA decomposition}

The stationary phase decomposition provides a way to construct the ``average wavefunction'' which captures the bulk fluid dynamics. We can construct the density weighted mean velocity in the multi-stream region, and require the phase of the ``average wavefunction'' produces this mean velocity.
As seen in Section~\ref{subsec:SPAstreamsZeldo}, the densities and velocities of the individual stationary phase wavefunctions are exactly those expected classically. Thus, we can straightforwardly construct the density-weighted mean velocity from the stationary phase streams 
\begin{subequations}
\label{eq:average_phase}
\begin{equation}
    \bar{v}(x,a) = \frac{\sum\limits_i \rho_i(x,a) v_i(x,a)}{\sum\limits_j \rho_j(x,a)}\,,
\end{equation}
where these sums run over the appropriate number of streams at position $(x,a)$. We define an ``average velocity potential'' for this mean velocity by $\bar{v} = \nabla_{\! x}\phi_{\rm avg}$ and calculate this in Fourier space,
\begin{equation}
    \phi_{\rm avg} = \mathcal{F}^{-1}\left[\frac{1}{ik^2}k\cdot \mathcal{F}[\bar{v}]\right],
\end{equation}
\end{subequations}
where $\mathcal{F}[\cdot]$ is a fast Fourier transform. From this smooth potential we construct an ``average wavefunction'' 
\begin{equation}
\label{eq:psi_avg}
    \psi_{\rm avg} = \sqrt{\sum_i \rho_i} \exp(\frac{i}{\hbar}\phi_{\rm avg}),
\end{equation}
representing a hypothetical perfect fluid with the classical density and mean velocity. 

\begin{figure*}[h!t]
    \centering
    \includegraphics[width=2\columnwidth]{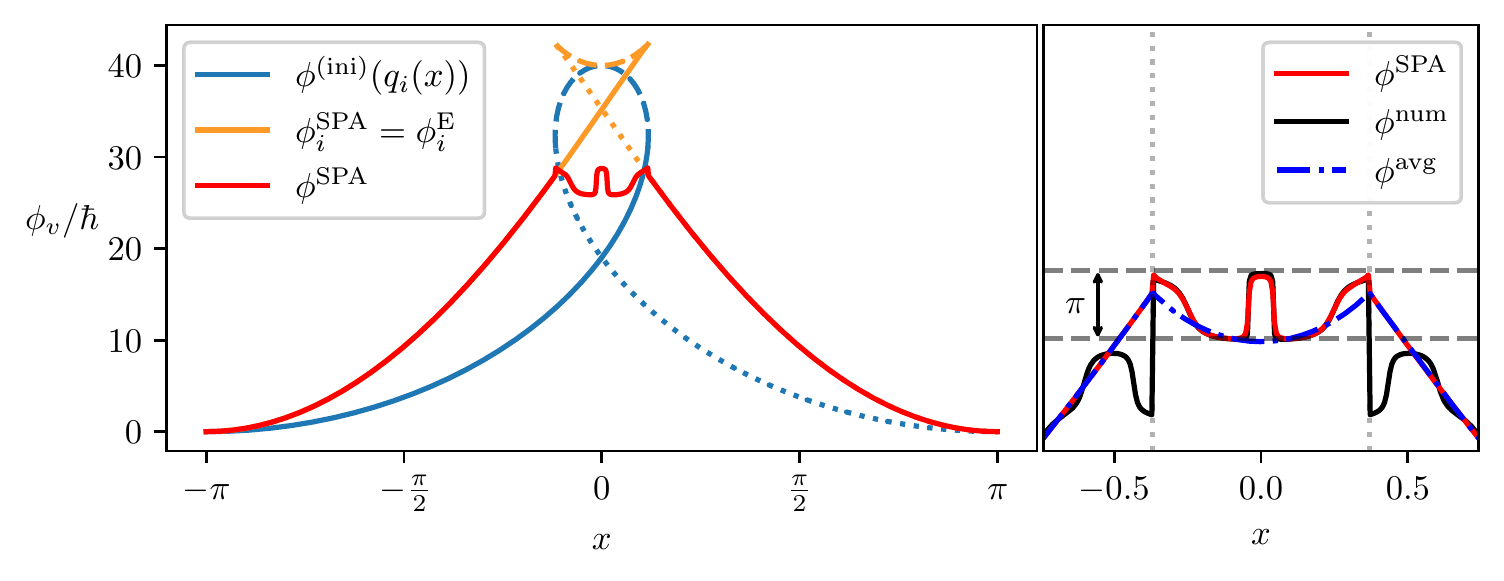}
    \caption{(Left) The phases of individual streams of the wavefunction model (with $\hbar=0.05$) past shell-crossing, at the time of the first interior phase jump. We show the phase of the individual SPA stream wavefunctions, $\phi^{\rm SPA}_i$ (which are equal to the Eulerian velocity potentials from equation~\eqref{eq:velpot_Eulerian}) in comparison to the Lagrangian phase $\phi^{\rm (ini)}(q_i)$ plotted against the Eulerian position $x(q_i)$. In the single-stream region, the Eulerian and SPA phases are equal (the red line completely overlaps the orange). The linestyles indicate the portions corresponding to different initial positions/stationary points $q_i(x,a)$. The phase of the full SPA wavefunction is also shown. (Right) A zoomed in view of the phase in the shell-crossed region. The phase of the numerical wavefunction, $\phi^{\rm num}$, is shown in comparison to the stationary phase approximation. Note that the discontinuity in $\phi^{\rm num}$ at the caustic boundary is not physical, the phase has simply been shifted up by $2\pi$ to align the profile with $\phi^{\rm SPA}$. The phase associated with the mean fluid velocity, $\phi_{\rm avg}$, well describes the overall profile, allowing the phase jumps to be isolated in $\phi_{\rm hid} = \phi_v-\phi_{\rm avg}$. }
    \label{fig:zeldo_loop_avg_phase} 
\end{figure*}

Figure~\ref{fig:zeldo_loop_avg_phase} shows the individual phases of the SPA wavefunctions, together with the phase of the full SPA at the time of an interior phase jump. Additionally, we show the Lagrangian velocity potential evaluated at a Eulerian position $\phi^{\rm (ini)}_i = \cos(q_i(x,a))/\hbar$ in blue and the proper Eulerian velocity potential, which agrees with the phases of the individual SPA streams $\phi_i^{\rm (SPA)}$ in orange. The right panel zooms into the shell-crossed region, comparing the full numerical phase of the wavefunction (black) and the stationary phase approximation to the phase (red), which match extremely well. The phase associated with the mean velocity (blue dot-dashed) well describes the overall profile of the phase.

We will show that the hidden features of the SPA wavefunction describing the wave-mechanical system decorate the Zel'dovich density and velocity, encoded in an average wavefunction from equation~\eqref{eq:psi_avg}, with wave interference effects. Those interference effects can be captured in a ``hidden wavefunction''
\begin{equation}
    \label{eq:psi_hidden}
    \psi_{\rm hid}=\frac{\psi_{\rm SPA}}{\psi_{\rm avg}}=\sqrt{\rho^{\rm hid}
}\exp\left(\frac{i}{\hbar}\phi_{\rm hid}\right),
\end{equation}
which produces multi-stream phenomena of a non-potential velocity from the phase jumps and a velocity dispersion from the oscillating density.

\subsubsection{Hidden phase associated with non-potential velocity}

We seek a decomposition of the phase of the wavefunction into a smooth and a ``hidden''  part, where the smooth part of the phase produces the potential part of the velocity field. The remaining ``hidden'' phase then contains the discontinuities which, in 2- or 3-dimensional systems would source vorticity. Such a phase decomposition has already been put forward for 2-dimensional optical systems in \cite{Fried1998}, as reviewed in Appendix~\ref{app:hid_phase_fried}. That form of the hidden phase is not appropriate for our 1-dimensional toy model, but could be applicable to 2-dimensional extensions.

We take the smooth part of our phase to be the phase of the average wavefunction, defined in equation~\eqref{eq:average_phase}. We then define our hidden phase as the difference between the full phase of the wavefunction (either predicted by stationary phase or calculated numerically) and this smooth phase to isolate the phase jumps,
\begin{equation}
    \phi_{\rm hid} = \phi_v - \phi_{\rm avg}\,.
\end{equation}

\begin{figure*}[h!t]
    \centering
    \includegraphics[width=2\columnwidth]{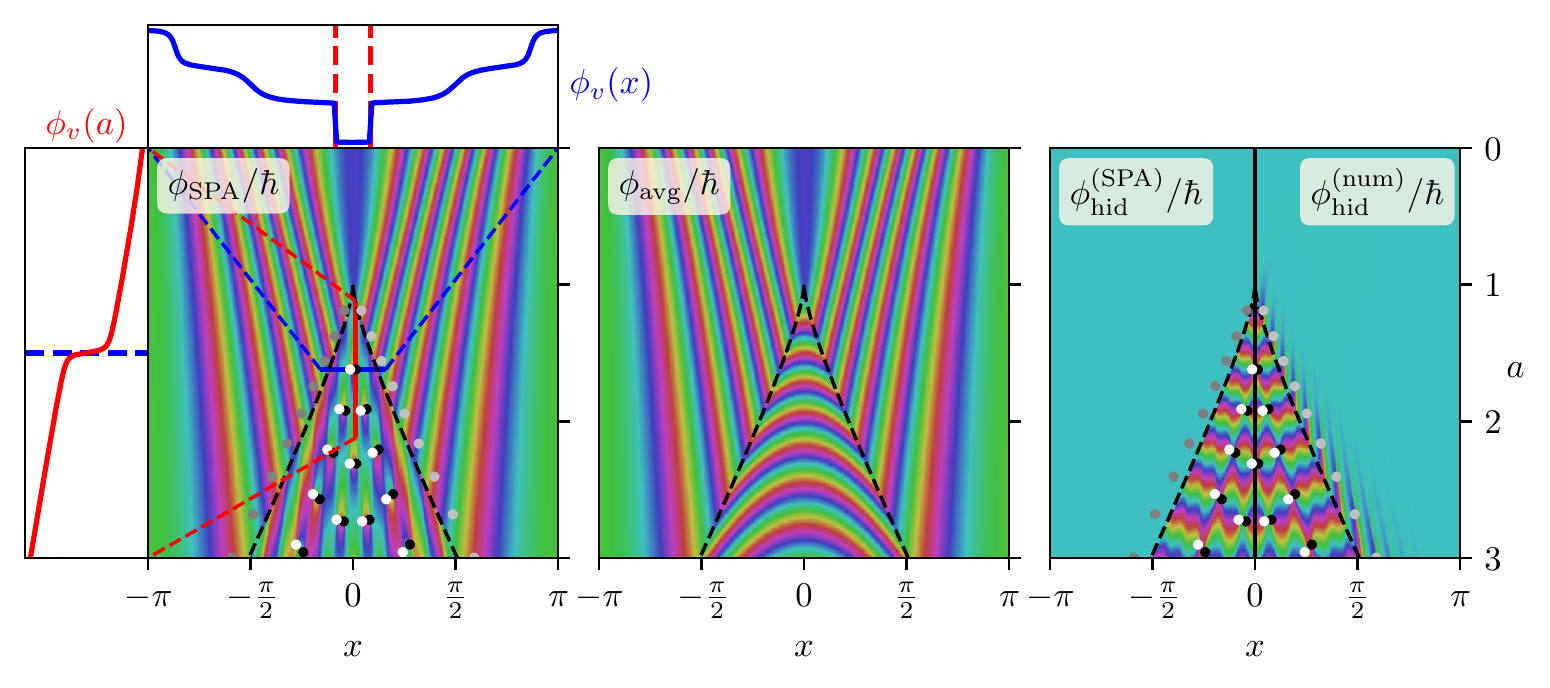}
    \caption{The spacetime evolution of the polar angle of the full stationary phase wavefunction (with $\hbar=0.05$, (left)), the ``average phase'' associated with the mean velocity (middle), and the discontinuous ``hidden phase'' (right). The left panel also shows the spatial and temporal phase profiles through a branch point (both of which jump by $\pm\hbar\pi$ upon crossing the branch point). The hidden phase in the right panel is calculated as using the stationary phase approximation, $\phi_{\rm hid}^{\rm (SPA)}=\phi_{\rm SPA}-\phi_{\rm avg}$ (left side), and numerically, $\phi_{\rm hid}^{\rm (num)}=\phi_{\rm num}-\phi_{\rm avg}$ (right side). The black and white points shown in the outer panels are the numerically located branch points of the full wavefunction, coloured according to their spacetime circulation as in Figure~\ref{fig:bp_located}. The hidden phase in the numerical case does not identically vanish outside the cusp, due to the presence of branch points outside the cusp which stationary phase analysis cannot produce (which are coloured light and dark grey to emphasise this point). }
    \label{fig:phase_profiles_and_hid}
\end{figure*}

Figure~\ref{fig:phase_profiles_and_hid} shows both the smooth average phase $\phi_{\rm avg}$ and the jumpy hidden phase $\phi_{\rm hid}$ over the entire spacetime region. We see that if the hidden phase is constructed as $\phi_{\rm hid}^{\rm (SPA)}=\phi_{\rm SPA} - \phi_{\rm avg}$, then it vanishes identically in the single-stream region. This is expected as the average velocity before shell-crossing is identical to the single-stream velocity. If we instead construct the hidden phase using the phase of the numerically solved wavefunction, $\phi_{\rm hid}^{\rm (num)}=\phi_{\rm num}-\phi_{\rm avg}$, then presence of branch points outside the classical caustic causes a nonzero hidden phase  in the single-stream region, which the stationary phase analysis cannot capture (as seen in the final panel of Figure~\ref{fig:phase_profiles_and_hid}).

\subsubsection{Hidden density associated with velocity dispersion}

Analogous to the hidden phase, we can write a ``hidden density'' as the part of the density which dresses the Zel'dovich dynamics in wave-phenomena. The hidden density $\rho^{\rm hid } = \abs{\psi^{\rm SPA}}^2/\sum_i \rho_i$, describes the oscillations which trace the Zel'dovich density.

\begin{figure}[h!t]
    \centering
    \includegraphics[width=\columnwidth]{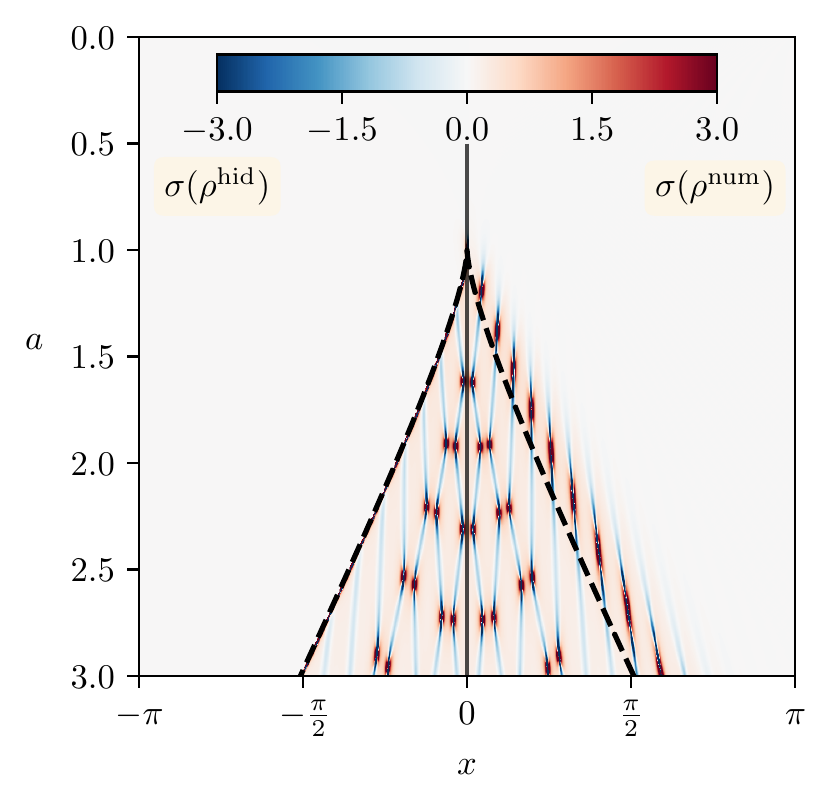}
    \caption{The scalar velocity dispersion $\sigma$ associated with the wavefunction $\psi$ (for $\hbar = 0.05$) calculated by equation~\eqref{eqn:sigma_ij} using the hidden density $\rho^{\rm hid} = \abs{\psi^{\rm SPA}}^2/\sum_i \rho_i$ (left) and the full (numerically solved) density $\rho^{\rm num} = \abs{\psi^{\rm num}}^2$ (right). The curvature near the zeros in the density field (which correspond to the branch points) sources velocity dispersion.}
    \label{fig:velocity_dispersion}
\end{figure}

The velocity dispersion associated with a wavefunction is defined in equation~\eqref{eqn:sigma_ij}, and is generally a tensor. However, in one spatial dimension, the velocity dispersion is simply a scalar quantity, sourced by the curvature in the (log) density,
\begin{equation}\label{eqn:scalar_vel_disp}
    \sigma = \frac{\hbar^2}{4}\nabla_{\! x}^2 \ln{\rho}\,.
\end{equation}
Figure~\ref{fig:velocity_dispersion} shows this scalar velocity dispersion using both the full numerically evolved density $\rho^{\rm num}$, and the hidden density $\rho^{\rm hid}$. We see that the velocity dispersion is well captured by the hidden density in the multi-stream region, and that a positive velocity dispersion is sourced by the positive curvature in the density at the locations of the branch points. 

Since the velocity dispersion is proportional to the curvature of the density, the minima of $\sigma$ seen in Figure~\ref{fig:velocity_dispersion} correspond to the maxima of the density field, while the maxima of $\sigma$, are sourced by the minima (the zero points) of the oscillating density. When coarse grained over, these small regions of negative velocity dispersion will be compensated by the positive velocity dispersion at the branch point, and will reproduce the velocity dispersion seen in the Zel'dovich approximation, for example as seen in Figure 1 of \cite{BuehlmannHahn2019}.

The fact that the velocity dispersion can become negative in the wave-mechanical case owes to the fact that is properly thought of as a modification to the Newtonian potential, or an stress-energy term, in the fluid equations~\eqref{eqn:quantum_fluids}. In particular, the Euler equation associated with equation~\eqref{eqn:quantum_bernoulli} reads
\begin{equation}
    \partial_a v_i + (\bm{v}\cdot\bm{\nabla})v_i + \underbrace{\nabla_{\! i}\left(-\frac{\hbar^2}{2}\frac{\nabla^2\sqrt{\rho}}{\sqrt{\rho}}\right)}_{\frac{1}{\rho}\nabla_{\! j}\left(\rho\sigma_{ij}\right) } = 0\,. 
\end{equation}
Interpreted in this way, as an additional source of stress-energy, $\sigma_{ij}$ can be negative in the wave-mechanical model.

To make this connection to the internal energy of the system more explicit, consider the expectation value of the quantum pressure $Q$ defined in equation~\eqref{eqn:quantum_pressure}. This can be written as the integral (in $n$ spatial dimensions)
\begin{subequations}
\begin{equation}
    \langle Q \rangle = \int \dd[n]{\bm{x}} \rho(\bm{x}) Q(\bm{x})
    =\int \dd[n]{\bm{x}} \rho \cdot \frac{\hbar^2}{2}\frac{(\bm{\nabla}\sqrt{\rho})^2}{\rho}\,,
\end{equation}
making use of $\rho=|\psi|^2$ and integration by parts  in the second step. We recognise this new part of the integrand as the internal energy of a wavefunction \citep{Yahalom2018MolPh}
\begin{equation}
    \varepsilon_q = \frac{\hbar^2}{2}\frac{(\bm{\nabla}\sqrt{\rho})^2}{\rho} = \frac{\hbar^2}{8}(\bm{\nabla}\ln\rho)^2\,,
\end{equation}
such that $\langle Q \rangle = \langle \varepsilon_q \rangle$. In addition to this global relation between the quantum pressure and the internal energy, we can locally relate them to the trace of the velocity dispersion
\begin{equation}
    \sigma_{ii} = Q + \varepsilon_q\,.
\end{equation}
\end{subequations}

For our 1+1D model, the scalar velocity dispersion $\sigma$ is determined by the hidden wavefunction (as seen in Figure~\ref{fig:velocity_dispersion}), which gives rise to the internal energy via the 1-dimensional versions of these relations between $\sigma_{ij}$, $Q$, and  $\varepsilon_q$.

\bigskip

In this section we directly examined the non-potential velocity and velocity dispersion, to illustrate the principal effects of the phase jumps and the oscillatory density. Higher order cumulants are determined similarly, through higher derivatives of the wavefunction, which will be largely driven by the ``hidden wavefunction'' introduced here. We have seen that the hidden wavefunction can accurately locate the interior branch points and reproduce the velocity dispersion associated with the vanishing density near those points. These interior branch points are the only ones relevant in the classical limit as they correspond to the classically shell-crossed region with multi-streaming phenomenology. In the next section we examine the exterior branch points, and other features which the stationary phase approximation alone cannot capture.

\section{Catastrophe theory}\label{sec:catastrophe_theory}

The caustic region formed by our wavefunction is very analogous to caustics formed by optical wave fields undergoing focusing. Leveraging this mathematical similarity, we turn to the theory of diffraction catastrophes, which classify the types of stable caustics and their properties, to understand the regions near to the caustic, precisely where the stationary phase decomposition fails.

\subsection{Classical catastrophe theory}

Classical catastrophe theory describes the singularities of differentiable mappings. For example, for a Zel'dovich Fourier mode, the phase-space to position space mapping $(v,x,a) \mapsto (x,a)$ develops singularities where the sheet folds over itself, as well as at the singular point where the sheet starts to twist. Importantly however, catastrophe theory is concerned with \emph{stable} singularities, where perturbing the mapping only slightly does not remove the character of the singular point. It does this by determining a set of standard \emph{generating polynomials}, $\zeta(\bm{s};\bm{C})$, which generate differentiable mappings such that all stable singularities look locally like one of these standard forms. The parameters $\bm{C} = (C_1, \dots, C_m)$ are called the \emph{control parameters}, and $\bm{s} = (s_1, \dots, s_n)$ are called the \emph{state parameters}. The control parameters are quantities on which the classical trajectories/rays depend, such as spatial position or describing the surface initial conditions are provided on. The state parameters are internal, and define the gradient mapping which produces the singularity. Roughly, it is the stationary points of $\zeta$ with respect to $\bm{s}$ which determine the number of classical trajectories/rays which meet at the singularity.

The elementary catastrophes were originally classified and named in \cite{Thom1994}, but a more systematic and complete classification is due to Arnol'd \citep{Arnold1973, Arnold1975}, which provided each generating polynomial a symbol related to the Coxeter reflection groups. In 1+1 dimensions, the only stable caustics are (using Thom's naming system) the fold and the cusp corresponding to the generating polynomials
\begin{align}
    \zeta_{\rm fold}(s; C_1) &= \frac{s^3}{3}+ C_1 s\,, \label{eqn:zeta_fold}\\
    \zeta_{\rm cusp}(s; C_1, C_2) &= \frac{s^4}{4} + C_2 \frac{s^2}{2} + C_1 s\,. \label{eqn:zeta_cusp}
\end{align}

Both of these singularities are present in the Zel'dovich phase-space sheet in Figure~\ref{fig:zeldo_phase_sheet}. The fold point corresponds to the point at shell-crossing, where the sheet begins to twist, corresponding to three trajectories intersecting. This fold point exists for a single moment in time, before splitting into a pair of fold lines which move apart from each other (called the ``unfolding'' of the catastrophe). Each point on the fold is characterised by the intersection of two trajectories. In this way, we have the heuristic identification of the spacetime parameters $(x,a)$ relating to the control parameters $(C_1, C_2)$, and the velocity phase-space variable playing the role of the state parameter $s$. Equally, one could consider the singularity in the $(q,x,a)\mapsto (x,a)$ mapping, which has the same structure as the phase-space sheet in Figure~\ref{fig:zeldo_phase_sheet}, treating the initial position $q$ as the state parameter. This second view will prove to be more directly useful in relating the integral form of our wavefunction to the standard catastrophe forms. We will refine this correspondence in the following sections.

This classical catastrophe theory has been applied to the cosmic web skeleton of caustics under the Zel'dovich approximation \citep{ArnoldShandarinZeldovich1982, Hidding2014, Feldbrugge2014, Feldbrugge2018}. In these, the classical Zel'dovich divergences and their statistics are classified and identified by examining the eigenvalues and eigenvectors of the deformation tensor, the gradient of the Lagrangian displacement field $\bm{\xi}$, with respect to Lagrangian coordinates. In the Zel'dovich approximation, this deformation tensor directly encodes the tidal information of the initial gravitational potential.

\subsection{Diffraction catastrophe theory}

Instead of simply defining smooth surfaces and differentiable mappings, the generating polynomials $\zeta$ can instead be used to generate oscillatory integrals. This allows the classification of classical caustics to be extended into the classification of caustics associated with wave phenomena. We refer to such integrals as diffraction catastrophe integrals. These integrals retain the underlying skeleton from the classical catastrophes they are built from, but now dressed with wave interference. Such systems were studied series of optical experiments in the 1970s \citep{Berry1977_focusing, Berry1977_finestructure,BerryNyeWright1979, BerryUpstill1980}.

The optical field, $u$, for a monochromatic source with frequency $\nu$ will act as an analogue for the wavefunction $\psi$ with ``quantumness'' $\hbar$ (such that $\hbar$ acts like an inverse frequency). The eikonal ($\nu \to \infty$) limit plays the role of the semiclassical ($\hbar\to 0$) limit. The intensity of the optical field is given by $\abs{u}^2$, and interference in the optical field is produced by mixing of the complex phase. A standard diffraction catastrophe integral is of the form\footnote{The prefactor of $\nu^{n/2}$ in this integral is necessary compared to the dimensionless forms listed in e.g. \cite{BerryUpstill1980} so that far from the catastrophe, the stationary phase approximation removes the $\nu$ dependence of the amplitude.} (notice the resemblance to equation~\eqref{eqn:psi_oscillatory})
\begin{equation}
    u(\bm{C};\nu) = \left(\frac{\nu}{2\pi}\right)^{n/2}\int \dd[n]{\bm{s}} \exp\left[i\nu\zeta(\bm{s};\bm{C})\right]\,.
    \label{eqn:diffraction_integral}
\end{equation}

As we have already seen, these oscillatory integrals will be mainly dominated by their stationary points, except very close to the singular point. Since we are always interested in the stationary points relative to the state parameters, by $\zeta'(\bm{s};\bm{C})$ we will always mean $\bm{\nabla}_{\! \bm{s}}\zeta(\bm{s};\bm{C})$. This is precisely how the universality of catastrophe theory arises, as for any function $\zeta$ which has a singularity, the integral will be dominated by the form of $\zeta$ near the singular point in the $\nu \to \infty$ limit. Therefore, $\zeta$ will be dominated by the leading order terms in its Taylor expansion about that point. For any system which includes a given catastrophe, a (smooth) coordinate transform bringing the expansion into one of the standard generating polynomials is guaranteed to exist \citep{Arnold2012_catastrophebook}.

\subsubsection{Properties of diffraction integrals}

Diffraction catastrophe theory also classifies certain topological properties of these integrals, which are invariant under diffeomorphism. This means that while the initial conditions in our wavefunction do not exactly give rise to any of the standard generating polynomials $\zeta(\bm{s};\bm{C})$ with their propagation and initial conditions, certain properties of the wavefunction are preserved in the coordinate change to bring them into standard form.

The relevant topological quantities for our interests are the indices which describe the maximum intensity of the caustic (which corresponds to the amount of regularisation the Zel'dovich divergences receive), and the fringe spacing in different directions. These scalings can be recovered by changing coordinates within the generating polynomial $\zeta(\bm{s};\bm{C})$ to remove the frequency dependence from the integrand. Doing so results in a relation of the following form
\begin{equation}
    u(\bm{C};\nu) = \nu^{\beta}u(\nu^{\sigma_1}C_1,\dots,\nu^{\sigma_m}C_m;\nu=1)\,,
\end{equation}
where now the quantities $\tilde{C}_i = \nu^{\sigma_i}C_i$ appearing on the right hand side are control parameters with dimensions of $\nu$ restored (note that the associated power of $\nu$ for each control parameter will be different).

The index $\beta$, called the \emph{singularity index} \citep[introduced in][]{Arnold1975}, describes the maximum intensity at the catastrophe, since the intensity of the light scales as
\begin{equation}
    \abs{u}^2 \propto \nu^{2\beta}\,.
\end{equation}
The indices $\sigma_i$ are called the \emph{fringe exponents},  \citep[introduced in][]{Berry1977_focusing}, and control the spacing of the fringes in the $i^{\rm th}$ control parameter direction. Note that since the dimensional coordinates scale as $\tilde{C}_i\sim \nu^{\sigma_i}C_i$, the fringe spacing in the $i^{\rm th}$ direction is $\order{\nu^{-\sigma_i}}$ in physical coordinates as $\nu\to\infty$. The sum of the fringe exponents, $\gamma = \sum_i \sigma_i$, is called the \emph{fringe index} and describes the (hyper-)volume scaling of the size of the singularity in control space. All the indices $\beta$, $\sigma_i$, and $\gamma$ are invariant under diffeomorphism.

\subsubsection{The fold catastrophe}

The fold catastrophe is the simplest non-trivial catastrophe, with just a single control parameter. The diffraction integral of the fold is given by using the fold generating polynomial~\eqref{eqn:zeta_fold} in the diffraction catastrophe integral~\eqref{eqn:diffraction_integral}
\begin{equation}
    u_{\rm fold}(C_1;\nu) = \sqrt{\frac{\nu}{2\pi}} \int_{-\infty}^{\infty} \dd{s} \exp\left[i\nu\left(\frac{s^3}{3}+C_1 s\right)\right].
\end{equation}
As an illustrative example, we can obtain  the singularity and fringe indices by changing the integration variable to $s=\nu^{-1/3}t$ to remove the $\nu$ dependence from the highest order term in the integrand. This leaves
\begin{subequations}
\begin{align}
    u_{\rm fold}(C_1;\nu) &= \frac{\nu^{1/6}}{\sqrt{2\pi}}\int_{-\infty}^\infty \dd{t}\exp\left[i\left(\frac{t^3}{3}+\nu^{2/3}C_1 t\right)\right], \nonumber \\
    &= \nu^{1/6} u_{\rm fold}(\nu^{2/3}C_1;\nu=1)\,, \\
    &= \sqrt{2\pi} \nu^{1/6} \operatorname{Ai}\left[\nu^{2/3}C_1\right],
\end{align}
\end{subequations}
which recovers a singularity index $\beta=1/6$ and the fringe exponent (and index) $\sigma_1 = \gamma = 2/3$. We also note that the fold catastrophe is simply the well known Airy function (up to appropriate normalisation).

\subsubsection{The cusp catastrophe}

The cusp catastrophe has two control parameters, and is given by the using the cusp generating polynomial~\eqref{eqn:zeta_cusp} in the diffraction catastrophe integral~\eqref{eqn:diffraction_integral}
\begin{subequations}
\label{eq:int_cusp}
\begin{align}
    u_{\rm cusp}(C_1, C_2;\nu) &= \sqrt{\frac{\nu}{2\pi}}\int_{-\infty}^{\infty} \dd{s} e^{i\nu\zeta_{\rm cusp}(s;C_1,C_2)}\,,  \\
    \zeta_{\rm cusp}(s;C_1,C_2) &= \frac{s^4}{4} + C_2\frac{s^2}{2} + C_1 s\,.
\end{align}
\end{subequations}

By changing the integration variable $s=\nu^{-1/4}t$, we can remove the $\nu$ dependence in the same way as with the fold and recover the catastrophe indices. This produces
\begin{subequations}
\begin{align}
    u_{\rm cusp}(\bm{C}; \nu) &= \nu^{1/4}u_{\rm cusp}(\nu^{3/4}C_1, \nu^{1/2}C_2;\nu=1)\,, \\
    &= \frac{(4\nu)^{1/4}}{\sqrt{2\pi}}\operatorname{Pe}(4^{1/4}\nu^{3/4}C_1, \nu^{1/2}C_2)\,,
\end{align}
\end{subequations}
where $\operatorname{Pe}(A, B)$ is the Pearcey integral \citep{Pearcey1946},
\begin{equation}
    \operatorname{Pe}(A,B) = \int_{-\infty}^\infty \dd{y} \exp(i(y^4 + Ay^2 + By))\,.
\end{equation}
This analysis gives us a singularity index $\beta = 1/4$, and the fringe exponents $\sigma_1 = 3/4, \sigma_2=1/2$, resulting in a fringe index $\gamma = 5/4$ for the cusp catastrophe. 

\begin{figure}[h!t]
    \centering
    \includegraphics[width=\columnwidth]{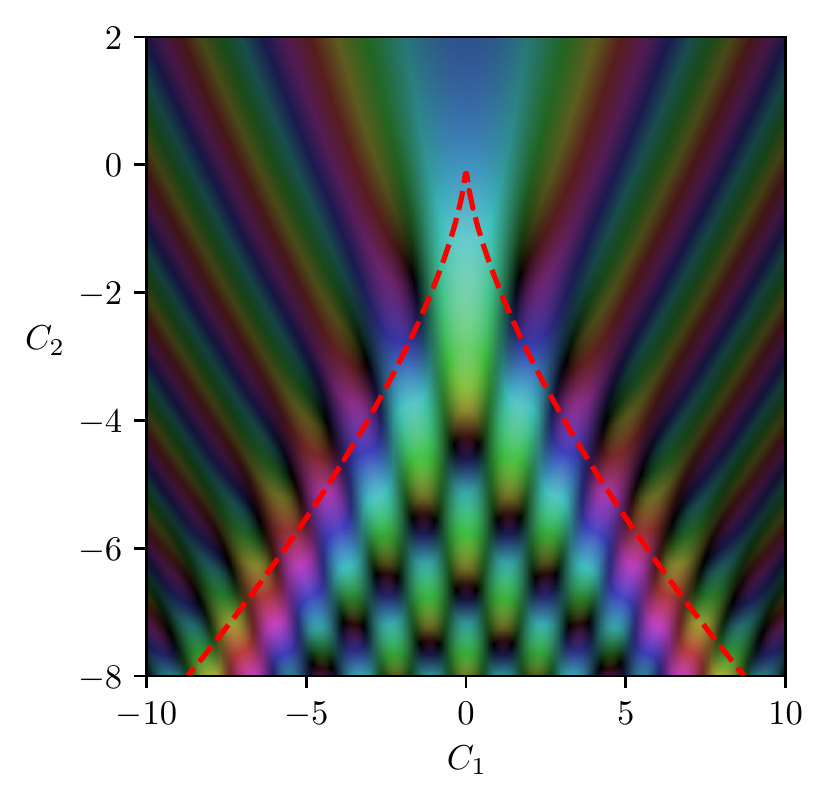}
    \caption{The cusp catastrophe $u_{\rm cusp}(C_1, C_2;\nu=1)$ from equation~\eqref{eq:int_cusp} in control parameter space, coloured according to the same domain colouring as the wavefunction. This is numerically calculated using the contour shifting technique described in Section~\ref{sec:multi-streaming and interference}.}
    \label{fig:cusp_integral_numeric}
\end{figure}

This cusp catastrophe integral, shown in Figure~\ref{fig:cusp_integral_numeric}, appears visually very similar to the spacetime evolution of the wavefunction with Zel'dovich initial conditions, with a cusped line separating a three ray region from a single ray region.

The equation of the caustic line is determined through the joint conditions
\begin{equation}
    \zeta'(s;\bm{C}) = 0\,, \quad \zeta''(s;\bm{C}) = 0\,.
\end{equation}
where the first condition says that you lie on the phase-space sheet (a cubic equation in this case) and the second condition is where the projection mapping becomes singular. In control parameters (eliminating $s$), the classical caustic line is given by
\begin{equation}\label{eqn:diff_cusp_caustic_condition}
    4C_2^3 + 27 C_1^2 = 0\,,
\end{equation}
a semi-cubic parabola.

\subsection{Mapping the wavefunction to standard catastrophes}\label{sec:fringe_properties_of_wavefunction}

We now wish to map the full wavefunction $\psi(x,a)$ of our single Fourier mode to these standard caustic forms. The role of the wavelength (inverse frequency) from the diffraction catastrophes is now played by the size of $\hbar$.

\begin{figure}[h!]
    \centering
    \includegraphics[width=\columnwidth]{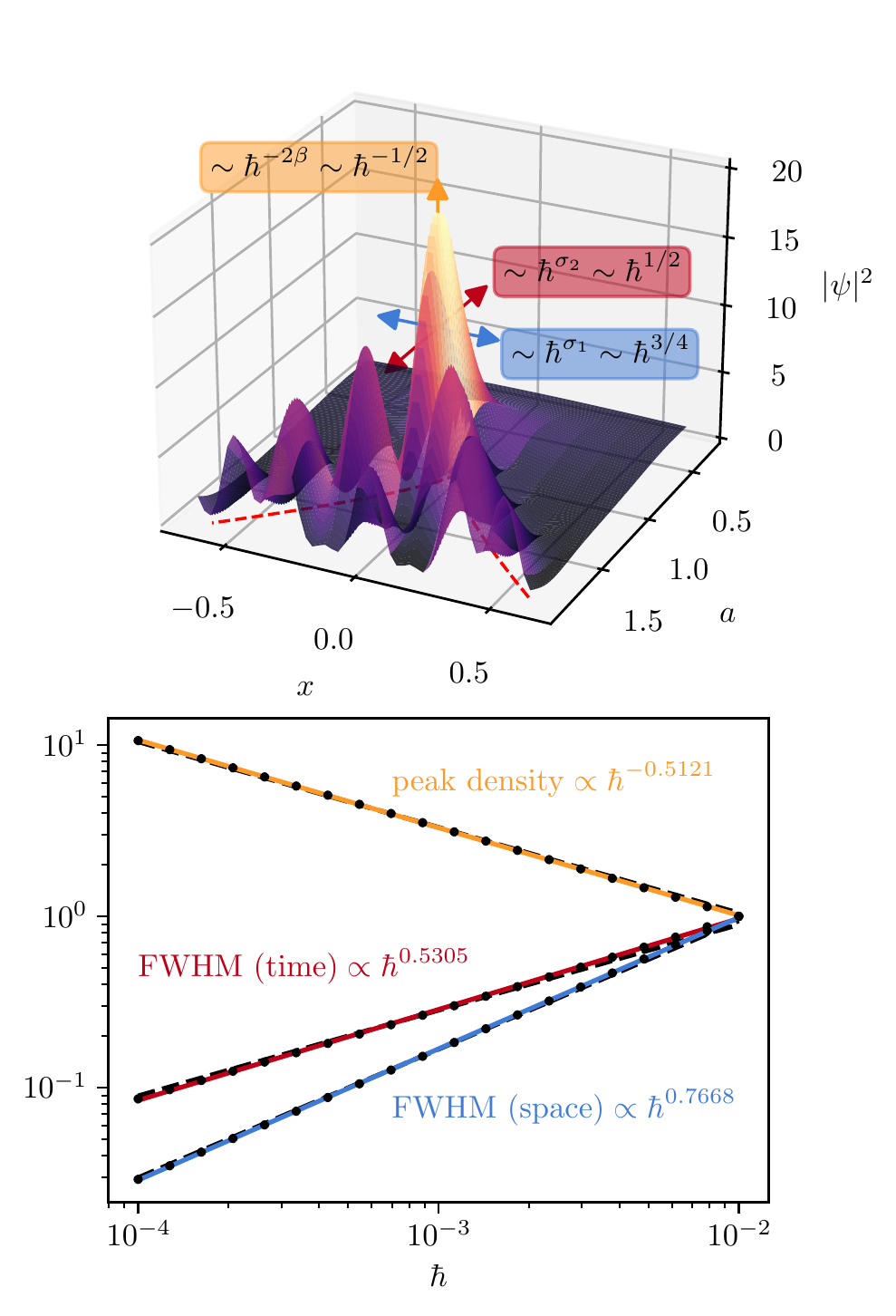}
    \caption{(Upper panel) Density of the wavefunction with Zel'dovich initial conditions for $\hbar = 0.05$. This demonstrates how the peak density, as well as the delay of shell-crossing and the spatial width of the wavefunction at the cusp point scale with $\hbar$, related to the singularity and fringe indices for the cusp catastrophe. (Lower panel) The peak height and characteristic widths around the cusp point for different values of $\hbar$, relative to their values at $\hbar=0.01$. To characterise the temporal and spatial widths we take the full width half maximum on constant space or time surfaces through the point of maximum density. The fit $\hbar$ scalings for each of these properties are shown as the solid coloured lines, while lines with the slopes anticipated from the catastrophe indices (made to go through a central data point) are shown as black dashed lines, which are mostly covered by the coloured lines. The measured $\hbar$ exponents agree well with the catastrophe theory prediction over two orders of magnitude in $\hbar$.  }
    \label{fig:3d_cusp_annotated}
\end{figure}

We write the full wavefunction using the free propagator as before
\begin{equation}\label{eqn:psi_for_catastrophe_expansion}
    \psi(x,a) = \mathcal{N}\int \dd{q} \exp\left(\frac{i}{\hbar}\left[\frac{(q-x)^2}{2a} + \cos(q)\right]\right).
\end{equation}
By Taylor expanding the cosine from the initial conditions to fourth order, we can recover a term quartic in $q$, which resembles the quartic generating function $\zeta_{\rm cusp}(s,\bm{C})$ (higher order terms in the Taylor expansion are suppressed by powers of $\hbar$, providing the universality promised by catastrophe theory). Scaling our coordinates so this leading order term is the same as the standard form, we can read off the mapping between the (dimensional) control parameters and our spacetime parameters
\begin{subequations}\label{eqn:control_of_spacetime}
\begin{align}
    \tilde{C}_1(x,a) &= -6^{1/4}\hbar^{-3/4}\frac{x}{a}\,, \\
    \tilde{C}_2(x,a) &= 6^{1/2}\hbar^{-1/2}\frac{1-a}{a}\,,
\end{align}
\end{subequations}
or their inversion
\begin{subequations}
\begin{align}
    a(\tilde{C}_1,\tilde{C}_2) &= \frac{1}{1+\sqrt{\hbar/6}\tilde{C}_2}\sim 1-\sqrt{\frac{\hbar}{6}}\tilde{C}_2\,, \label{eqn:c2_of_spacetime} \\
    x(\tilde{C}_1, \tilde{C}_2) &= -\frac{\tilde{C}_1}{\tilde{C}_2}\frac{\hbar^{3/4}}{6^{1/4}}\frac{1}{1+\sqrt{\hbar/6}}\sim -\frac{\tilde{C}_1}{\tilde{C}_2}\frac{\hbar^{3/4}}{6^{1/4}}\,.
\end{align}
\end{subequations}
The coordinate relations~\eqref{eqn:control_of_spacetime} recover the cusp fringe exponents, which we can see by expanding about the cusp point $(x,a)=(0,1)$. Writing $x = \delta x$ and $a = 1+\delta a$, and the coordinate relation becomes $\tilde{C}_1 \sim \hbar^{-3/4}\delta x$ and $\tilde{C}_2 \sim \hbar^{-1/2}\delta a$, leading to $\sigma_1 = 3/4, \sigma_2 = 1/2$ as expected from the cusp. The required coordinate transforms in equation \eqref{eqn:control_of_spacetime} are not simple rescalings, both because the cusp catastrophe happens at $a=1$ rather than 0, and a natural mixing due to the propagator containing $x/a$ terms. Applying this coordinate transformation, we obtain the following integral relationship between our wavefunction and the standard cusp integral,
\begin{align}
    \psi(x,a) \approx \exp&\left(i\left[\frac{x^2+2a}{2\hbar a}-\frac{\pi}{4}\right]\right)\frac{6^{1/4}}{a^{1/2} \hbar^{1/4}} \times \nonumber \\ 
    &u_{\rm cusp}(\tilde{C}_1(x,a),\tilde{C}_2(x,a);\nu=1)\,,
\end{align}
which correctly recovers the singularity index of $\beta = 1/4$. A schematic diagram showing the density from this wavefunction is shown in Figure~\ref{fig:3d_cusp_annotated}, together with the catastrophe scalings. In the lower panel of Figure~\ref{fig:3d_cusp_annotated}, we compare the numerical peak height, temporal width, and spatial width of the cusp peak over two orders of magnitude in $\hbar$. We use the full width half maximum (FWHM) on constant time/space surfaces through the maximum peak as a proxy for the characteristic widths. The best fit scaling exponents are in good agreement with the scalings anticipated by catastrophe theory, demonstrating that this analysis does extract the correct scaling behaviour.

In principle one could perform a similar set of coordinate transformations along each time slice after shell-crossing, bringing the form of $\psi(x,a=\text{const}>1)$ into an integral relationship with $u_{\rm fold}$. Providing the details of the coordinate transforms required is tedious and provides little physical insight into the system beyond doing this for the cusp integral. Appendix~\ref{app:fold_coord_change} demonstrates how to perform these coordinate changes on the standard cusp integral to recover the fold contribution. While one could explicitly recover the relevant coordinate transformations for the wavefunction by similar methods, we can instead rely on the existence of such coordinate changes and simply write down the relevant scalings. The fold indices $\beta_{\rm fold} = 1/6$ and $\sigma_{\rm fold}=2/3$ now finally explain the scalings presented in Figure~\ref{fig:fold_annotated} back in Section~\ref{sec:multi-streaming and interference}.

The coordinate relation equation~\eqref{eqn:c2_of_spacetime} provides the delay in the onset of shell-crossing from $a=1$, which scales as $\hbar^{1/2}$. They also provide the fact that the peak density reached for the wavefunction scales as $\hbar^{-1/2}$ (owing to $\beta_{\rm cusp} = 1/4$). From looking at constant time slices of the wavefunction post shell-crossing, we see the peaks of the wavefunction are offset from the Zel'dovich caustic positions by $\hbar^{2/3}$ (the ``finite thickness'' of the caustic), and the peak intensity of the wavefunction at fixed time ($a>1$) scales as $\hbar^{-1/3}$ (as shown in Figure~\ref{fig:fold_annotated}).  Importantly, the fringe indices provide the characteristic widths which can provide zeros in the density outside the classically shell-crossed region. This demonstrates that the exterior branch points occur on length scales $\order{\hbar^{\sigma_i}}$ from the classical caustic, for the appropriate fringe exponent. As $\sigma_i > 0$ for all the elementary catastrophes (see e.g. \cite{BerryUpstill1980, Feldbrugge2019} for the $\sigma_i$ of catastrophes in higher dimensions), these exterior branch points do not survive the classical limit. This is an important recognition, as the position of these exterior dislocations cannot be predicted by a stationary phase analysis of the real roots alone, although they could be estimated by steepest descents including contributions of one complex root as in \cite{BerryNyeWright1979}.

The (spatial) fringe spacings  in cylindrical collapse simulations performed by \cite{Lague_2021_FDM_LPT} have been found to be well approximated by half the de Broglie wavelength, taking the velocity as the root-mean-square Zel'dovich velocity at shell-crossing. Viewed as a scaling in $\hbar$, this result corresponds to $\sigma = 1$ in the language of catastrophe theory. This is larger than any of the fringe indices of the elementary catastrophes, which all have $\sigma < 1$. However, cylindrical collapse is a highly symmetric scenario, which can lead to very different scaling behaviour catastrophe theory cannot predict (in the same way a spherical lens can achieve much stronger focusing which breaks when the lens is perturbed \citep{BerryUpstill1980}). We suspect that perturbing the collapse from cylindrical would result in the focus splitting into separate caustics, each of which has scalings predicted by catastrophe theory.  

With these wave properties near the caustics, and the stationary phase decomposition described in Section~\ref{sec:unweaving_the_wavefunction}, we have a powerful dictionary characterising our wavefunction over all regions of spacetime.

\section{Conclusions}\label{sec:conclusion}

\textit{A. Summary} We have provided a detailed analysis of a single Fourier mode evolving under the Zel'dovich approximation making use of a wavefunction obeying the free Schr\"odinger equation. We demonstrate that the interfering region of the wavefunction, corresponding to classical multi-streaming, can be resolved into a sum over three stream wavefunctions, each corresponding to a classical Zel'dovich stream as shown in Figure~\ref{fig:stream_splitting_phase_hbar0.05}. These stream wavefunctions are recovered as stationary points of the function $\zeta$ from equation~\eqref{eqn:zeta}, which encodes the dynamics of the propagation of the wavefunction, and its initial conditions. By visualising phase-space in Figure~\ref{fig:stream_split_wigner_overplot}, we show that this wavefunction decomposition naturally dissects the phase-space distribution into the classical streams, without having to explicitly construct a quantum phase-space. Such a decomposition validates the ``golden rule of semiclassics'' described in \cite{heller_semiclassical_2018}: ``quantum amplitudes are to be approximated as the sum of square roots of classical probabilities, with phases given by classical actions.''

We analyse the interference properties of this Fourier mode, and provide a decomposition of the phase into a smooth average potential, which encodes the stream-averaged velocity of the fluid, and a discontinuous ``hidden phase'' which would source vorticity in a 2- or 3-dimensional system. These phase jumps accompany the zeros in the density field, which source velocity dispersion in the interference region and can be captured in a ``hidden density''.

In the regions near the high density caustics, where this separation into Zel'dovich streams fails, we have made explicit connections between our semiclassical wave model and standard ``diffraction catastrophes'', which  classify the wave phenomena arising around caustics. In particular, this predicts the peak density and the fringe spacing about the Zel'dovich divergences illustrated in Figure~\ref{fig:fold_annotated} and \ref{fig:3d_cusp_annotated}.  While the exact fringe spacings and density profiles will not follow these catastrophe scalings in the fully non-linear dynamics, we expect that their qualitative features should persist on mildly-nonlinear scales. The full classification of diffraction integrals provides a universality to these results, at least in 1+1 dimensions.

\textit{B. Outlook} This paper has focused on a simple toy model in 1+1 dimensions, under the Zel'dovich approximation. While very simple, the results here naturally extend to more realistic cases, and display certain universal features. The unweaving of a free wavefunction into the Zel'dovich streams based on stationary points of $\zeta(\bm{q};\bm{x},a)$ is not restricted to 1+1 dimensions, and readily generalises to any number of spatial dimensions. Near the caustics, in higher dimensions, one has to consider a larger set of diffraction catastrophes than considered here, but these are fully classified and their normal forms can be found in e.g. \cite{BerryUpstill1980, Feldbrugge2019}. In this way, the cosmic caustic web in two or three dimensions, as examined in \cite{ArnoldShandarinZeldovich1982, Hidding2014, Feldbrugge2014, Feldbrugge2018}, could be dressed in wave phenomena via diffraction catastrophe integrals in equation~\eqref{eqn:diffraction_integral}. Using the universal profiles of these caustics has been suggested for probing the mass of a fuzzy dark matter particle with tidal streams on galactic scales \citep{Dalal2021JCAP}.

There is potential to extend this beyond free wavefunctions using Propagator Perturbation Theory (PPT) from \cite{Uhlemann2019}. By replacing the free propagator $K_0 = e^{i S_0/\hbar}$ with a higher-order propagator, one can build a higher-order function $\zeta_n$, whose stationary points determine the classical behaviour. This perturbatively encodes the effect of tidal effects relevant in a higher dimensional system. For example, using the next to leading order propagator, the stationary phase approximation would return the results of 2LPT.

The smooth phase recovered from the stationary phase approximation as detailed in Section~\ref{sec:phase_properties} potentially holds use in extending PPT beyond shell-crossing. The perturbing element of PPT is an effective potential $V_{\rm eff}$, which in the single-stream region is the (using our sign convention) sum of the gravitational and velocity potentials. At lowest order, this vanishes (hence why we study free Schr\"odinger), and at next to leading order it is time independent. However, past shell-crossing, the velocity potential develops  discontinuities, making it an unattractive perturbation variable. The smooth profile here could be used in the interference region, allowing for the effective potential to continue to have meaning past shell-crossing in the spirit of a post-collapse perturbation theory that has been formulated for CDM \citep{Colombi2015,TaruyaColombi2017,Pietroni2018,BuehlmannHahn2019, Saga_2022_LPT_Vlasov}. It could also allow to bridge from the perturbative cosmic web regime \citep[at early times and on large scales considered here and in ][]{Uhlemann2019} to the asymptotic dynamics in the solitonic core regime \citep[at late times and on small scales as described by][]{Zimmermann2021PhRvD,Zagorac2022PhRvD, Taruya_2022_corehalo}.

These wave-mechanical models also present an attractive application complementing the power of $N$-body simulations for understanding CDM dynamics. The Husimi phase-space distributions shown in Figure~\ref{fig:stream_split_wigner_overplot} demonstrate that the wave-mechanical system samples phase-space in a complementary way to $N$-body simulations. While $N$-body simulations sample discrete points along a cold phase-space sheet and trace their evolution, the wave-mechanical model follows a uniform density sheet with finite thickness (determined by $\hbar$). This is particularly useful in underdense regions, where large-scale $N$-body simulations effectively lose resolution (see e.g. \cite{AnguloHahn2022} for a recent review). The propagator formalism adopted here has been applied to field-level inference in the Lyman-$\alpha$ forest \citep{Porqueres_2020} and extended to two components \citep{Rampf2021MNRAS} for setting initial conditions for Eulerian simulations including dark matter and baryons \citep{Hahn_2021_ICs_2fluid}.

\section*{Acknowledgements} 
We thank the anonymous referee  for valuable suggestions that helped to improve the presentation of our results. We thank Gary Liu, Andrew Baggaley, and George Stagg for useful conversations about numerically locating the phase wrappings and numerically calculating Wigner functions, and Oliver Hahn for sharing the domain colouring and FFT codes. We thank Oliver Hahn and Cornelius Rampf for useful conversations about conserved quantities and Nick Kaiser for making us aware of the work of \cite{Fried1998}.

The figures in this work were created with \textsc{matplotlib} \citep{matplotlib} making use of the \textsc{numpy} \citep{numpy}, \textsc{scipy} \citep{2020SciPy-NMeth}, and \textsc{sci-kit image} \citep{scikit-image} Python libraries.

AG is supported by an EPSRC studentship under Project 2441314 from UK Research \& Innovation. CU was partially supported by a grant from the Simons Foundation and the hospitality of the Aspen Center for Physics in the final stages of the project.

\section*{Data access statement}
No data was generated as part of this work. For the purpose of open access, the authors have applied a `Creative Commons Attribution' (CC BY) licence to this paper.
\appendix

\section{Domain colouring}\label{app:domain_colouring}
We represent complex numbers, $z = \abs{z}e^{i\theta}$, with the argument, $\theta\in [0,2\pi)$, as points in hue, lightness, saturation (hls) colour space via the mapping
\begin{equation}
    \begin{pmatrix}
    h \\ l \\ s
    \end{pmatrix} = \begin{pmatrix}
    \theta/(2\pi) \\ 1-0.5^\abs{z} \\ 0.5
    \end{pmatrix}.
\end{equation}
Each of $\{h,l,s\}$ lies in the range $0$ to $1$. These hls values can then be mapped to standard RGB values (or any other colour space) in the usual way (c.f. \cite{Wegert2012}).

\section{The stationary phase approximation}\label{app:SPA}

The stationary phase approximation (SPA) is a method for deriving an asymptotic approximation to integrals of the form
\begin{equation}\label{eqn:spa_integral_form}
    I(\nu) = \int \dd{t} f(t) \exp(i\nu g(t))\,,
\end{equation}
in the limit $\nu \to \infty$. The integration range is along the real line and the functions $f$ and $g$ are real functions. It is a standard technique in asymptotic analysis, and further detailed discussion of its applications and limits can be found in e.g. \cite{BenderandOrszag}. For the context current paper this is relevant with $\nu = 1/\hbar$ in the classical limit ($\hbar \to 0$) of the wavefunction system, or the geometric optics limit ($\nu\to \infty$) in the diffraction catastrophes. Because the integrand is highly oscillatory, the dominant contribution to this integral comes from the stationary points of $g$, denoted $t=t_*$. The contribution from one of these stationary points can be calculated by
\begin{enumerate}
    \item Restricting the integration range to be isolated around each stationary point (provided the stationary points are sufficiently separated compared to $\nu^{-1}$).
    \item Taylor expanding $f$ and $g$ about $t=t_*$, then further Taylor expanding the exponential term so that only quadratic or lower terms are left in the exponent.
    \item Replacing the limits of the integral by $\pm\infty$, and performing the resulting Gaussian integrals.
\end{enumerate}

The total contribution from one stationary point is
\begin{align}\label{eqn:SPA_1d}
    I_{t_*}^{\rm SPA}(\nu) =& \sqrt{\frac{2\pi}{\nu\abs{g''(t_*)}}}f(t_*) \times \nonumber \\
    &\quad \exp\left( i\nu g(t_*) + \frac{i\pi}{4} \operatorname{sgn}(g''(t_*))\right).
\end{align}

We note for completeness two useful direct extensions to equation~\eqref{eqn:SPA_1d}. Firstly, if the first $(p-1)$ derivatives of $g$ are all 0 at a stationary point, but $g^{(p)}(t_*)\neq 0$, the contribution receives a modification \citep{BenderandOrszag}
\begin{align}
    I_{t_*}^{\rm SPA}(\nu) = &f(t_*) \left[\frac{p!}{\nu \abs{g^{(p)}(t_*)}}\right]^{\frac{1}{p}} \frac{\Gamma(\frac{1}{p})}{p} \times \nonumber \\
    &\quad\exp\left(i\nu g(t_*) + \frac{i\pi}{4}\operatorname{sgn}(g^{(p)}(t_*))\right).
\end{align}

Secondly, we can extend this formalism to integrals over $n$-dimensional space. In this case
\begin{align}
    I_{t_*}^{\rm SPA}(\nu) = \frac{f(\bm{t}_*)}{\abs{\det\mathsf{H}(\bm{t}_*)}^{1/2}}\exp\left(i\nu g(\bm{t}_*) + \frac{i\pi}{4}\operatorname{sgn}(\mathsf{H}(\bm{t}_*))\right),
\end{align}
where $\mathsf{H}_{ij} = \pdv*{g}{t_i}{t_j}$ is the Hessian matrix (provided $\mathsf{H}(\bm{t}_*)\neq 0$). The signature of the Hessian is the difference between the number of negative and positive eigenvalues.

As a note of caution, the stationary phase approximation is not well suited to obtaining terms higher than this leading contribution in $\nu$. It performs worse than its equivalent for non-oscillatory integrals (where $g$ in equation~\eqref{eqn:spa_integral_form} is pure imaginary), called Laplace's method or the saddle-point approximation. This is because in Laplace's method, the errors introduced by steps such as pushing the integration range to $\infty$ introduce exponentially small corrections, while in stationary phase, the corrections are generally only algebraically small \citep[see Chapter 6 of][]{BenderandOrszag}. Thus higher order terms depend both on further terms in the Taylor expansion about the stationary points, \emph{and} on contributions from non-stationary points. For this reason, obtaining next order (in $\nu$) corrections should be done either by asymptotic matching, or by considering the full steepest descent contour. 

\section{Derivation of the hidden phase (following Fried)} \label{app:hid_phase_fried}

\cite{Fried1998} derives an explicit expression for the hidden phase based on the location of the branch points in an optical field (equation~\eqref{eqn:hidden_phase}). Such a phase profile could be applicable to studying constant time snapshots for a 2-dimensional system, where now the branch points act as vortex cores \citep[as shown in][]{Uhlemann2019}. Below we present the derivation of this hidden phase from this work.

The wavefunction $\psi$ plays the role of the optical field, while the velocity field $\bm{v}$ velocity field plays the role of the ``gradient field''  of the principal valued phase of the optical field, $\bm{g}$.

Consider a 2-dimensional velocity field, $\bm{v}$, with a single branch point located at  $\bm{r}_{\rm BP}$ with a circulation of $\pm 2\pi$ as defined by equation~\eqref{eqn:quantised_vorticity}. We guarantee that the circulation is always $\pm 2 \pi$ and not some other integer multiple by constructing it from the ``principal valued'' phase function. By introducing an auxiliary third dimension, $z$, we can write the circulation integral as the condition,
\begin{equation}
    \hat{\bm{e}}_z \cdot \bm{\nabla} \times \bm{v}(\bm{r}) = \pm 2\pi \delta_{\rm D}(\bm{r}-\bm{r}_{\rm BP})\,.
    \label{eqn:curl_v_BP}
\end{equation}
Performing a Helmholtz decomposition on the velocity field, we obtain the incompressible and irrotational parts of the velocity
\begin{equation}
    \bm{v} = \bm{\nabla} \phi_s + \bm{\nabla} \times \bm{H}\,.
\end{equation}
where now $\phi_s$ and $\bm{H}$ are smooth functions, in contrast to the raw phase which describes this (optical) field, which contains jumps from the branch points.

The scalar potential, $\phi_s$, determines the divergence of the velocity field,
\begin{equation}
    \bm{\nabla} \cdot \bm{v} = \nabla^2 \phi_s\,.
\end{equation}
For our 2-dimensional system, we can write the vector potential as a scalar potential in the auxiliary dimension,
\begin{equation}
    \bm{H}(\bm{r}) = h(\bm{r}) \hat{\bm{e}}_z\,. 
\end{equation}
Since the divergence of $\bm{H}$ vanishes (as it is solinoidal), we can rewrite the condition in equation~\eqref{eqn:curl_v_BP} as
\begin{equation}
    \nabla^2 h(\bm{r}) = \mp 2\pi \delta_D(\bm{r}-\bm{r}_{\rm BP})\,,
\end{equation}
which is solved by 
\begin{equation}
    h(\bm{r}) = \mp \log(\abs{\bm{r}-\bm{r}_{\rm BP}})\,.
    \label{eqn:hertz_function}
\end{equation}

To then write velocity field as the gradient of some (principal valued) velocity potential $\bm{v} = \bm{\nabla} \phi$ with some split into a ``smooth'' and a ``hidden'' part, $\phi = \phi_{\rm sm} + \phi_{\rm hid}$, we make the choice that $\phi_{\rm sm}$ should be equal to the scalar potential $\phi_{s}$ from the Helmholtz decomposition. Then, the equation for the hidden phase
\begin{equation}
    \bm{\nabla} \phi_{\rm hid} = \bm{\nabla} \times \bm{H}\,,
\end{equation}
becomes the pair of equations
\begin{subequations}
\begin{align}
    \del_x \phi_{\rm hid}(x,y) &= -\del_y [-h(x,y)]\,, \\
    \del_y \phi_{\rm hid}(x,y) &= \del_x [-h(x,y)]\,,
\end{align}
\end{subequations}
which can be interpreted as the Cauchy-Riemann conditions for a single complex analytic function
\begin{equation}
    f(z = x+iy) = -h(x,y) + i \phi_{\rm hid}(x,y)\,,
\end{equation}
with real part $-h$ and imaginary part $\phi_{\rm hid}$. Using the expression for $h(\bm{r})$ in equation~\eqref{eqn:hertz_function}, the hidden phase for a single branch point is
\begin{equation}
    \phi_{\rm hid}(\bm{r}) = \Im\{\pm\log\left[(x-x_{\rm BP} + i (y-y_{\rm BP})\right]\}.
\end{equation}

If instead the system contains multiple branch points, each will contribute a term of this form to the overall hidden phase. Since we are concerned with irrotational systems, these branch points must be created in pairs such that the number of positive and negative branch points is equal. Consider $N$ pairs of branch points, with the $i^{\rm th}$ point associated with circulation $\pm 2 \pi$ located at $\bm{r}_i^\pm = (x_i^\pm, y_i^\pm)$. The total hidden phase from these branch points is then
\begin{equation}
    \phi_{\rm hid}(\bm{r}) = \Im\left\{\log\left[\frac{\prod\limits_{n=1}^{N}(x-x_n^+) + i (y-y_n^+)}{\prod\limits_{m=1}^{N}(x-x_m^-) + i (y-y_m^-)} \right]\right\}.
    \label{eqn:hidden_phase}
\end{equation}

\section{Contribution of the fold to the cusp diffraction integral} \label{app:fold_coord_change}

Away from the singular cusp point, the cusp catastrophe unfolds into a pair of fold lines. On lines of constant $C_2<0$, near to the classical caustic line, the integral is dominated by the contribution from the folded part of the phase-space sheet. This section demonstrates how to obtain the contribution from the fold to the cusp integral on these constant $C_2$ lines.
Start with the cusp diffraction integral
\begin{subequations}
\begin{align}
        u_{\rm cusp}(C_1,C_2) &= \sqrt{\frac{\nu}{2\pi}} \int \dd{s} \exp(i\nu\zeta_{\rm cusp}(s;C_1,C_2))\,, \\
        \zeta_{\rm cusp}(s;C_1,C_2) &= \frac{s^4}{4} + C_2\frac{s^2}{2} + C_1 s\,.
\end{align}
\end{subequations}
Consider a horizontal slice through this integral at constant $C_2<0$. We want to evaluate the contribution along this line near the caustic line described by equation~\eqref{eqn:diff_cusp_caustic_condition}. On the actual caustic line, there are 2 stationary points of $\zeta$, one single and one double root. We are interested in the double root, which produces the fold. The stationary points along the caustic line satisfy:
\begin{equation}
    s_*^3 + C_2 s_* + \left(\frac{4}{27}\abs{C_2}^{3}\right)^{1/2} = 0\,,
\end{equation}
where we have written $C_1 = C_1(C_2)$ using equation~\eqref{eqn:diff_cusp_caustic_condition}. This has two solutions in $s$, 
\begin{equation}\label{eqn:s1_s2_appendix}
    s_* = \sqrt{\frac{-C_2}{3}}, -2\sqrt{\frac{-C_2}{3}}\,.
\end{equation}
The positive stationary point is the double root if $C_1>0$, and the other is if $C_1<0$. Generically, call the double root $s_2$. Change coordinates in the integral $t=s-s_2$, so that the fold occurs at $t=0$. Under this transformation, the cusp integral has the form
\begin{equation}
    u_{\rm cusp} = \sqrt{\frac{\nu}{2\pi}}\int\dd{t} \exp(i\nu\left[\frac{t^4}{4}+At^3 + Bt^2 + C t + D\right]),
\end{equation}
(we will not list explicit forms of all intermediate variables, as this section is simply to show that such a set of transformations can be done). Now we want to remove the $\nu$ dependence in a way which lets us analyse the fold contribution to this integral coming from $t=0$. Under the coordinate change $t=y/(3A\nu)^{1/3}$ the cubic term will become $y^3/3$ resembling the fold integral. The resulting integral is
\begin{align}\label{eqn:cusp_transformed_1}
    u_{\rm cusp} = \sqrt{\frac{\nu}{2\pi}}\frac{1}{(3A\nu)^{1/3}} &\int\dd{y} \exp(i\frac{\nu^{-1/3}}{4(3A)^{4/3}}y^4)\times \nonumber \\ 
    &\exp(i\left[\frac{y^3}{3} + \tilde{B}y^2 + \tilde{C} y + \tilde{D}\right]).
\end{align}

We then notice that the quartic term is suppressed in the $\nu\to \infty$ limit, and we can expand the exponential 
\begin{align}
    \exp(i\frac{\nu^{-1/3}}{4(3A)^{4/3}}y^4) \sim 1 + \order{\nu^{-1/3}}.
\end{align}
So to leading order in $\nu$ we can neglect the quartic term's contribution to the integral in equation~\eqref{eqn:cusp_transformed_1} at $y=0$. Note that this is the same trick used in standard stationary phase analysis, but here we retain the cubic term in the exponent, where in stationary phase we would expand all exponential terms of cubic or higher order and then perform the resulting Gaussian integrals. 

We can turn the resulting equation into the standard fold integral by taking $y = x - \tilde{B}$, which will transform this cubic into a depressed cubic (one without a quadratic term). This results in the contribution
\begin{align}
    \hat{u} = \frac{\nu^{1/6}}{\sqrt{2\pi}(3A)^{1/3}} \int\dd{x} \exp\left(i\left[\frac{x^3}{3} + Ex + F\right]\right),
\end{align}
to the cusp integral. We have written $\hat{u}$ here rather than $u_{\rm cusp}$ to remind us that this is an approximation to the contribution only around $s=s_2$, rather than a full analysis of the cusp integral. This fold contribution $\hat{u}$ is now in the form of the standard fold diffraction integral
\begin{align}
    \hat{u} &= \frac{e^{iF}\nu^{1/6}}{(3A)^{1/3}}u_{\rm fold }(E;\nu=1)\,, \\
    &= \sqrt{2\pi}\frac{e^{iF}}{(3A)^{1/3}}\nu^{1/6}\operatorname{Ai}(E)\,.
\end{align}
The explicit forms of $E$ and $F$ are 
\begin{align}
    E &= \nu^{2/3}\frac{3^{2/3} \left(12 C_{1} s_{2} - C_{2}^{2} + 3 s_{2}^{2} \left(2 C_{2} + s_{2}^{2}\right)\right)}{36 s_{2}^{4/3}}\,, \\
    F &= \nu \frac{ \left(- 18 C_{1} C_{2} s_{2} + C_{2}^{3} + 9 s_{2}^{2} \left(6 C_{1} s_{2} - C_{2}^{2} + C_{2} s_{2}^{2}\right)\right)}{108 s_{2}^{2}}\,.
\end{align}
Notice that $E\sim \nu^{2/3}$, correctly recovering the fringe index for the fold catastrophe.

One could perform a similar analysis on the full wavefunction with cosine initial conditions~\eqref{eqn:psi_for_catastrophe_expansion} on constant time slices post shell-crossing. Doing this would result in a set of coordinate relations between $(x,a)$ and the single control parameter of the standard fold integral. Doing this explicitly is significantly messier than the case here, requiring an expansion of cosine around the caustic line described in equation~\eqref{eqn:shell_cross_region}. Additionally, the wavefunction contribution cannot be written in closed form since the solutions  $s_2$ cannot be written down analytically (which was only possible here because $\zeta_{\rm cusp}'=0$ is a cubic equation).

\bibliography{refs}

\begin{thebibliography}{106}%
\makeatletter
\providecommand \@ifxundefined [1]{%
 \@ifx{#1\undefined}
}%
\providecommand \@ifnum [1]{%
 \ifnum #1\expandafter \@firstoftwo
 \else \expandafter \@secondoftwo
 \fi
}%
\providecommand \@ifx [1]{%
 \ifx #1\expandafter \@firstoftwo
 \else \expandafter \@secondoftwo
 \fi
}%
\providecommand \natexlab [1]{#1}%
\providecommand \enquote  [1]{``#1''}%
\providecommand \bibnamefont  [1]{#1}%
\providecommand \bibfnamefont [1]{#1}%
\providecommand \citenamefont [1]{#1}%
\providecommand \href@noop [0]{\@secondoftwo}%
\providecommand \href [0]{\begingroup \@sanitize@url \@href}%
\providecommand \@href[1]{\@@startlink{#1}\@@href}%
\providecommand \@@href[1]{\endgroup#1\@@endlink}%
\providecommand \@sanitize@url [0]{\catcode `\\12\catcode `\$12\catcode
  `\&12\catcode `\#12\catcode `\^12\catcode `\_12\catcode `\%12\relax}%
\providecommand \@@startlink[1]{}%
\providecommand \@@endlink[0]{}%
\providecommand \url  [0]{\begingroup\@sanitize@url \@url }%
\providecommand \@url [1]{\endgroup\@href {#1}{\urlprefix }}%
\providecommand \urlprefix  [0]{URL }%
\providecommand \Eprint [0]{\href }%
\providecommand \doibase [0]{http://dx.doi.org/}%
\providecommand \selectlanguage [0]{\@gobble}%
\providecommand \bibinfo  [0]{\@secondoftwo}%
\providecommand \bibfield  [0]{\@secondoftwo}%
\providecommand \translation [1]{[#1]}%
\providecommand \BibitemOpen [0]{}%
\providecommand \bibitemStop [0]{}%
\providecommand \bibitemNoStop [0]{.\EOS\space}%
\providecommand \EOS [0]{\spacefactor3000\relax}%
\providecommand \BibitemShut  [1]{\csname bibitem#1\endcsname}%
\let\auto@bib@innerbib\@empty
\bibitem [{\citenamefont {{Laureijs \textit{et al.}}}(2011)}]{Euclid_mission}%
  \BibitemOpen
  \bibfield  {author} {\bibinfo {author} {\bibfnamefont {R.}~\bibnamefont
  {{Laureijs \textit{et al.}}}},\ }\href@noop {} {\bibfield  {journal}
  {\bibinfo  {journal} {arXiv e-prints}\ ,\ \bibinfo {eid} {arXiv:1110.3193}}
  (\bibinfo {year} {2011})},\ \Eprint {http://arxiv.org/abs/1110.3193}
  {arXiv:1110.3193 [astro-ph.CO]} \BibitemShut {NoStop}%
\bibitem [{\citenamefont {{Ivezi{\'c}}}\ \emph {et~al.}(2019)\citenamefont
  {{Ivezi{\'c}}}, \citenamefont {{Kahn}},\ and\ \citenamefont {{Tyson
  \textit{et al.}}}}]{LSST_mission}%
  \BibitemOpen
  \bibfield  {author} {\bibinfo {author} {\bibfnamefont {{\v{Z}}.}~\bibnamefont
  {{Ivezi{\'c}}}}, \bibinfo {author} {\bibfnamefont {S.~M.}\ \bibnamefont
  {{Kahn}}}, \ and\ \bibinfo {author} {\bibfnamefont {J.~A.}\ \bibnamefont
  {{Tyson \textit{et al.}}}},\ }\href {\doibase 10.3847/1538-4357/ab042c}
  {\bibfield  {journal} {\bibinfo  {journal} {\apj}\ }\textbf {\bibinfo
  {volume} {873}},\ \bibinfo {eid} {111} (\bibinfo {year} {2019})},\ \Eprint
  {http://arxiv.org/abs/0805.2366} {arXiv:0805.2366 [astro-ph]} \BibitemShut
  {NoStop}%
\bibitem [{\citenamefont {{Levi \textit{et al.}}}(2013)}]{DESI_mission}%
  \BibitemOpen
  \bibfield  {author} {\bibinfo {author} {\bibfnamefont {M.}~\bibnamefont
  {{Levi \textit{et al.}}}},\ }\href@noop {} {\bibfield  {journal} {\bibinfo
  {journal} {arXiv e-prints}\ ,\ \bibinfo {eid} {arXiv:1308.0847}} (\bibinfo
  {year} {2013})},\ \Eprint {http://arxiv.org/abs/1308.0847} {arXiv:1308.0847
  [astro-ph.CO]} \BibitemShut {NoStop}%
\bibitem [{\citenamefont {{Bernardeau}}\ \emph {et~al.}(2002)\citenamefont
  {{Bernardeau}}, \citenamefont {{Colombi}}, \citenamefont {{Gazta{\~n}aga}},\
  and\ \citenamefont {{Scoccimarro}}}]{Bernardeau2002PhR}%
  \BibitemOpen
  \bibfield  {author} {\bibinfo {author} {\bibfnamefont {F.}~\bibnamefont
  {{Bernardeau}}}, \bibinfo {author} {\bibfnamefont {S.}~\bibnamefont
  {{Colombi}}}, \bibinfo {author} {\bibfnamefont {E.}~\bibnamefont
  {{Gazta{\~n}aga}}}, \ and\ \bibinfo {author} {\bibfnamefont {R.}~\bibnamefont
  {{Scoccimarro}}},\ }\href {\doibase 10.1016/S0370-1573(02)00135-7} {\bibfield
   {journal} {\bibinfo  {journal} {\physrep}\ }\textbf {\bibinfo {volume}
  {367}},\ \bibinfo {pages} {1} (\bibinfo {year} {2002})},\ \Eprint
  {http://arxiv.org/abs/astro-ph/0112551} {arXiv:astro-ph/0112551 [astro-ph]}
  \BibitemShut {NoStop}%
\bibitem [{\citenamefont {{Zel'dovich}}(1970)}]{Zeldovich1970}%
  \BibitemOpen
  \bibfield  {author} {\bibinfo {author} {\bibfnamefont {Y.~B.}\ \bibnamefont
  {{Zel'dovich}}},\ }\href@noop {} {\bibfield  {journal} {\bibinfo  {journal}
  {\aap}\ }\textbf {\bibinfo {volume} {5}},\ \bibinfo {pages} {84} (\bibinfo
  {year} {1970})}\BibitemShut {NoStop}%
\bibitem [{\citenamefont {{Buchert}}(1989)}]{Buchert1989A&A}%
  \BibitemOpen
  \bibfield  {author} {\bibinfo {author} {\bibfnamefont {T.}~\bibnamefont
  {{Buchert}}},\ }\href@noop {} {\bibfield  {journal} {\bibinfo  {journal}
  {\aap}\ }\textbf {\bibinfo {volume} {223}},\ \bibinfo {pages} {9} (\bibinfo
  {year} {1989})}\BibitemShut {NoStop}%
\bibitem [{\citenamefont {{Bouchet}}\ \emph {et~al.}(1992)\citenamefont
  {{Bouchet}}, \citenamefont {{Juszkiewicz}}, \citenamefont {{Colombi}},\ and\
  \citenamefont {{Pellat}}}]{Bouchet1992ApJL}%
  \BibitemOpen
  \bibfield  {author} {\bibinfo {author} {\bibfnamefont {F.~R.}\ \bibnamefont
  {{Bouchet}}}, \bibinfo {author} {\bibfnamefont {R.}~\bibnamefont
  {{Juszkiewicz}}}, \bibinfo {author} {\bibfnamefont {S.}~\bibnamefont
  {{Colombi}}}, \ and\ \bibinfo {author} {\bibfnamefont {R.}~\bibnamefont
  {{Pellat}}},\ }\href {\doibase 10.1086/186459} {\bibfield  {journal}
  {\bibinfo  {journal} {\apjl}\ }\textbf {\bibinfo {volume} {394}},\ \bibinfo
  {pages} {L5} (\bibinfo {year} {1992})}\BibitemShut {NoStop}%
\bibitem [{\citenamefont {{Bouchet}}\ \emph {et~al.}(1995)\citenamefont
  {{Bouchet}}, \citenamefont {{Colombi}}, \citenamefont {{Hivon}},\ and\
  \citenamefont {{Juszkiewicz}}}]{Bouchet1995A&A}%
  \BibitemOpen
  \bibfield  {author} {\bibinfo {author} {\bibfnamefont {F.~R.}\ \bibnamefont
  {{Bouchet}}}, \bibinfo {author} {\bibfnamefont {S.}~\bibnamefont
  {{Colombi}}}, \bibinfo {author} {\bibfnamefont {E.}~\bibnamefont {{Hivon}}},
  \ and\ \bibinfo {author} {\bibfnamefont {R.}~\bibnamefont {{Juszkiewicz}}},\
  }\href@noop {} {\bibfield  {journal} {\bibinfo  {journal} {\aap}\ }\textbf
  {\bibinfo {volume} {296}},\ \bibinfo {pages} {575} (\bibinfo {year}
  {1995})},\ \Eprint {http://arxiv.org/abs/astro-ph/9406013}
  {arXiv:astro-ph/9406013 [astro-ph]} \BibitemShut {NoStop}%
\bibitem [{\citenamefont {{Villone}}\ and\ \citenamefont
  {{Rampf}}(2017)}]{Villone2017EPJH}%
  \BibitemOpen
  \bibfield  {author} {\bibinfo {author} {\bibfnamefont {B.}~\bibnamefont
  {{Villone}}}\ and\ \bibinfo {author} {\bibfnamefont {C.}~\bibnamefont
  {{Rampf}}},\ }\href {\doibase 10.1140/epjh/e2017-80039-6} {\bibfield
  {journal} {\bibinfo  {journal} {European Physical Journal H}\ }\textbf
  {\bibinfo {volume} {42}} (\bibinfo {year} {2017}),\
  10.1140/epjh/e2017-80039-6},\ \Eprint {http://arxiv.org/abs/1707.01883}
  {arXiv:1707.01883 [math.HO]} \BibitemShut {NoStop}%
\bibitem [{\citenamefont {Uhlemann}\ \emph {et~al.}(2019)\citenamefont
  {Uhlemann}, \citenamefont {Rampf}, \citenamefont {Gosenca},\ and\
  \citenamefont {Hahn}}]{Uhlemann2019}%
  \BibitemOpen
  \bibfield  {author} {\bibinfo {author} {\bibfnamefont {C.}~\bibnamefont
  {Uhlemann}}, \bibinfo {author} {\bibfnamefont {C.}~\bibnamefont {Rampf}},
  \bibinfo {author} {\bibfnamefont {M.}~\bibnamefont {Gosenca}}, \ and\
  \bibinfo {author} {\bibfnamefont {O.}~\bibnamefont {Hahn}},\ }\href {\doibase
  10.1103/physrevd.99.083524} {\bibfield  {journal} {\bibinfo  {journal}
  {Physical Review D}\ }\textbf {\bibinfo {volume} {99}} (\bibinfo {year}
  {2019}),\ 10.1103/physrevd.99.083524}\BibitemShut {NoStop}%
\bibitem [{\citenamefont {{Rampf}}\ \emph {et~al.}(2021)\citenamefont
  {{Rampf}}, \citenamefont {{Uhlemann}},\ and\ \citenamefont
  {{Hahn}}}]{Rampf2021MNRAS}%
  \BibitemOpen
  \bibfield  {author} {\bibinfo {author} {\bibfnamefont {C.}~\bibnamefont
  {{Rampf}}}, \bibinfo {author} {\bibfnamefont {C.}~\bibnamefont {{Uhlemann}}},
  \ and\ \bibinfo {author} {\bibfnamefont {O.}~\bibnamefont {{Hahn}}},\ }\href
  {\doibase 10.1093/mnras/staa3605} {\bibfield  {journal} {\bibinfo  {journal}
  {\mnras}\ }\textbf {\bibinfo {volume} {503}},\ \bibinfo {pages} {406}
  (\bibinfo {year} {2021})},\ \Eprint {http://arxiv.org/abs/2008.09123}
  {arXiv:2008.09123 [astro-ph.CO]} \BibitemShut {NoStop}%
\bibitem [{\citenamefont {{Coles}}\ and\ \citenamefont
  {{Spencer}}(2003)}]{ColesSpencer2003}%
  \BibitemOpen
  \bibfield  {author} {\bibinfo {author} {\bibfnamefont {P.}~\bibnamefont
  {{Coles}}}\ and\ \bibinfo {author} {\bibfnamefont {K.}~\bibnamefont
  {{Spencer}}},\ }\href {\doibase 10.1046/j.1365-8711.2003.06529.x} {\bibfield
  {journal} {\bibinfo  {journal} {\mnras}\ }\textbf {\bibinfo {volume} {342}},\
  \bibinfo {pages} {176} (\bibinfo {year} {2003})},\ \Eprint
  {http://arxiv.org/abs/astro-ph/0212433} {arXiv:astro-ph/0212433 [astro-ph]}
  \BibitemShut {NoStop}%
\bibitem [{\citenamefont {{Short}}\ and\ \citenamefont
  {{Coles}}(2006)}]{ShortColes2006}%
  \BibitemOpen
  \bibfield  {author} {\bibinfo {author} {\bibfnamefont {C.~J.}\ \bibnamefont
  {{Short}}}\ and\ \bibinfo {author} {\bibfnamefont {P.}~\bibnamefont
  {{Coles}}},\ }\href {\doibase 10.1088/1475-7516/2006/12/016} {\bibfield
  {journal} {\bibinfo  {journal} {\jcap}\ }\textbf {\bibinfo {volume} {2006}},\
  \bibinfo {eid} {016} (\bibinfo {year} {2006})},\ \Eprint
  {http://arxiv.org/abs/astro-ph/0605013} {arXiv:astro-ph/0605013 [astro-ph]}
  \BibitemShut {NoStop}%
\bibitem [{\citenamefont {{Hui}}\ \emph {et~al.}(2017)\citenamefont {{Hui}},
  \citenamefont {{Ostriker}}, \citenamefont {{Tremaine}},\ and\ \citenamefont
  {{Witten}}}]{Hui2017}%
  \BibitemOpen
  \bibfield  {author} {\bibinfo {author} {\bibfnamefont {L.}~\bibnamefont
  {{Hui}}}, \bibinfo {author} {\bibfnamefont {J.~P.}\ \bibnamefont
  {{Ostriker}}}, \bibinfo {author} {\bibfnamefont {S.}~\bibnamefont
  {{Tremaine}}}, \ and\ \bibinfo {author} {\bibfnamefont {E.}~\bibnamefont
  {{Witten}}},\ }\href {\doibase 10.1103/PhysRevD.95.043541} {\bibfield
  {journal} {\bibinfo  {journal} {\prd}\ }\textbf {\bibinfo {volume} {95}},\
  \bibinfo {eid} {043541} (\bibinfo {year} {2017})},\ \Eprint
  {http://arxiv.org/abs/1610.08297} {arXiv:1610.08297 [astro-ph.CO]}
  \BibitemShut {NoStop}%
\bibitem [{\citenamefont {{Peccei}}\ and\ \citenamefont
  {{Quinn}}(1977)}]{Peccei1977PhRvL}%
  \BibitemOpen
  \bibfield  {author} {\bibinfo {author} {\bibfnamefont {R.~D.}\ \bibnamefont
  {{Peccei}}}\ and\ \bibinfo {author} {\bibfnamefont {H.~R.}\ \bibnamefont
  {{Quinn}}},\ }\href {\doibase 10.1103/PhysRevLett.38.1440} {\bibfield
  {journal} {\bibinfo  {journal} {\prl}\ }\textbf {\bibinfo {volume} {38}},\
  \bibinfo {pages} {1440} (\bibinfo {year} {1977})}\BibitemShut {NoStop}%
\bibitem [{\citenamefont {{Svrcek}}\ and\ \citenamefont
  {{Witten}}(2006)}]{Svrcek2006JHEP}%
  \BibitemOpen
  \bibfield  {author} {\bibinfo {author} {\bibfnamefont {P.}~\bibnamefont
  {{Svrcek}}}\ and\ \bibinfo {author} {\bibfnamefont {E.}~\bibnamefont
  {{Witten}}},\ }\href {\doibase 10.1088/1126-6708/2006/06/051} {\bibfield
  {journal} {\bibinfo  {journal} {Journal of High Energy Physics}\ }\textbf
  {\bibinfo {volume} {2006}},\ \bibinfo {eid} {051} (\bibinfo {year} {2006})},\
  \Eprint {http://arxiv.org/abs/hep-th/0605206} {arXiv:hep-th/0605206 [hep-th]}
  \BibitemShut {NoStop}%
\bibitem [{\citenamefont {{Arvanitaki}}\ \emph {et~al.}(2010)\citenamefont
  {{Arvanitaki}}, \citenamefont {{Dimopoulos}}, \citenamefont {{Dubovsky}},
  \citenamefont {{Kaloper}},\ and\ \citenamefont
  {{March-Russell}}}]{Arvanitaki2010PhRvD}%
  \BibitemOpen
  \bibfield  {author} {\bibinfo {author} {\bibfnamefont {A.}~\bibnamefont
  {{Arvanitaki}}}, \bibinfo {author} {\bibfnamefont {S.}~\bibnamefont
  {{Dimopoulos}}}, \bibinfo {author} {\bibfnamefont {S.}~\bibnamefont
  {{Dubovsky}}}, \bibinfo {author} {\bibfnamefont {N.}~\bibnamefont
  {{Kaloper}}}, \ and\ \bibinfo {author} {\bibfnamefont {J.}~\bibnamefont
  {{March-Russell}}},\ }\href {\doibase 10.1103/PhysRevD.81.123530} {\bibfield
  {journal} {\bibinfo  {journal} {\prd}\ }\textbf {\bibinfo {volume} {81}},\
  \bibinfo {eid} {123530} (\bibinfo {year} {2010})},\ \Eprint
  {http://arxiv.org/abs/0905.4720} {arXiv:0905.4720 [hep-th]} \BibitemShut
  {NoStop}%
\bibitem [{\citenamefont {{Jaeckel}}\ \emph {et~al.}(2022)\citenamefont
  {{Jaeckel}}, \citenamefont {{Rybka}},\ and\ \citenamefont
  {{Winslow}}}]{Jaeckel2022arXiv}%
  \BibitemOpen
  \bibfield  {author} {\bibinfo {author} {\bibfnamefont {J.}~\bibnamefont
  {{Jaeckel}}}, \bibinfo {author} {\bibfnamefont {G.}~\bibnamefont {{Rybka}}},
  \ and\ \bibinfo {author} {\bibfnamefont {L.}~\bibnamefont {{Winslow}}},\
  }\href@noop {} {\bibfield  {journal} {\bibinfo  {journal} {arXiv e-prints}\
  ,\ \bibinfo {eid} {arXiv:2203.14923}} (\bibinfo {year} {2022})},\ \Eprint
  {http://arxiv.org/abs/2203.14923} {arXiv:2203.14923 [hep-ex]} \BibitemShut
  {NoStop}%
\bibitem [{\citenamefont {{Niemeyer}}(2020)}]{Niemeyer2020}%
  \BibitemOpen
  \bibfield  {author} {\bibinfo {author} {\bibfnamefont {J.~C.}\ \bibnamefont
  {{Niemeyer}}},\ }\href {\doibase 10.1016/j.ppnp.2020.103787} {\bibfield
  {journal} {\bibinfo  {journal} {Progress in Particle and Nuclear Physics}\
  }\textbf {\bibinfo {volume} {113}},\ \bibinfo {eid} {103787} (\bibinfo {year}
  {2020})},\ \Eprint {http://arxiv.org/abs/1912.07064} {arXiv:1912.07064
  [astro-ph.CO]} \BibitemShut {NoStop}%
\bibitem [{\citenamefont {{Hui}}(2021)}]{Hui2021}%
  \BibitemOpen
  \bibfield  {author} {\bibinfo {author} {\bibfnamefont {L.}~\bibnamefont
  {{Hui}}},\ }\href {\doibase 10.1146/annurev-astro-120920-010024} {\bibfield
  {journal} {\bibinfo  {journal} {\araa}\ }\textbf {\bibinfo {volume} {59}}
  (\bibinfo {year} {2021}),\ 10.1146/annurev-astro-120920-010024},\ \Eprint
  {http://arxiv.org/abs/2101.11735} {arXiv:2101.11735 [astro-ph.CO]}
  \BibitemShut {NoStop}%
\bibitem [{\citenamefont {{Ferreira}}(2021)}]{Ferreira2021A&ARv}%
  \BibitemOpen
  \bibfield  {author} {\bibinfo {author} {\bibfnamefont {E.~G.~M.}\
  \bibnamefont {{Ferreira}}},\ }\href {\doibase 10.1007/s00159-021-00135-6}
  {\bibfield  {journal} {\bibinfo  {journal} {\aapr}\ }\textbf {\bibinfo
  {volume} {29}},\ \bibinfo {eid} {7} (\bibinfo {year} {2021})},\ \Eprint
  {http://arxiv.org/abs/2005.03254} {arXiv:2005.03254 [astro-ph.CO]}
  \BibitemShut {NoStop}%
\bibitem [{\citenamefont {{Widrow}}\ and\ \citenamefont
  {{Kaiser}}(1993)}]{WidrowKaiser1993}%
  \BibitemOpen
  \bibfield  {author} {\bibinfo {author} {\bibfnamefont {L.~M.}\ \bibnamefont
  {{Widrow}}}\ and\ \bibinfo {author} {\bibfnamefont {N.}~\bibnamefont
  {{Kaiser}}},\ }\href {\doibase 10.1086/187073} {\bibfield  {journal}
  {\bibinfo  {journal} {\apjl}\ }\textbf {\bibinfo {volume} {416}},\ \bibinfo
  {pages} {L71} (\bibinfo {year} {1993})}\BibitemShut {NoStop}%
\bibitem [{\citenamefont {{Uhlemann}}\ \emph {et~al.}(2014)\citenamefont
  {{Uhlemann}}, \citenamefont {{Kopp}},\ and\ \citenamefont
  {{Haugg}}}]{Uhlemann2014}%
  \BibitemOpen
  \bibfield  {author} {\bibinfo {author} {\bibfnamefont {C.}~\bibnamefont
  {{Uhlemann}}}, \bibinfo {author} {\bibfnamefont {M.}~\bibnamefont {{Kopp}}},
  \ and\ \bibinfo {author} {\bibfnamefont {T.}~\bibnamefont {{Haugg}}},\ }\href
  {\doibase 10.1103/PhysRevD.90.023517} {\bibfield  {journal} {\bibinfo
  {journal} {\prd}\ }\textbf {\bibinfo {volume} {90}},\ \bibinfo {eid} {023517}
  (\bibinfo {year} {2014})},\ \Eprint {http://arxiv.org/abs/1403.5567}
  {arXiv:1403.5567 [astro-ph.CO]} \BibitemShut {NoStop}%
\bibitem [{\citenamefont {{Kopp}}\ \emph {et~al.}(2017)\citenamefont {{Kopp}},
  \citenamefont {{Vattis}},\ and\ \citenamefont {{Skordis}}}]{Kopp2017}%
  \BibitemOpen
  \bibfield  {author} {\bibinfo {author} {\bibfnamefont {M.}~\bibnamefont
  {{Kopp}}}, \bibinfo {author} {\bibfnamefont {K.}~\bibnamefont {{Vattis}}}, \
  and\ \bibinfo {author} {\bibfnamefont {C.}~\bibnamefont {{Skordis}}},\ }\href
  {\doibase 10.1103/PhysRevD.96.123532} {\bibfield  {journal} {\bibinfo
  {journal} {\prd}\ }\textbf {\bibinfo {volume} {96}},\ \bibinfo {eid} {123532}
  (\bibinfo {year} {2017})},\ \Eprint {http://arxiv.org/abs/1711.00140}
  {arXiv:1711.00140 [astro-ph.CO]} \BibitemShut {NoStop}%
\bibitem [{\citenamefont {{Mocz}}\ \emph {et~al.}(2018)\citenamefont {{Mocz}},
  \citenamefont {{Lancaster}}, \citenamefont {{Fialkov}}, \citenamefont
  {{Becerra}},\ and\ \citenamefont {{Chavanis}}}]{Mocz2018}%
  \BibitemOpen
  \bibfield  {author} {\bibinfo {author} {\bibfnamefont {P.}~\bibnamefont
  {{Mocz}}}, \bibinfo {author} {\bibfnamefont {L.}~\bibnamefont {{Lancaster}}},
  \bibinfo {author} {\bibfnamefont {A.}~\bibnamefont {{Fialkov}}}, \bibinfo
  {author} {\bibfnamefont {F.}~\bibnamefont {{Becerra}}}, \ and\ \bibinfo
  {author} {\bibfnamefont {P.-H.}\ \bibnamefont {{Chavanis}}},\ }\href
  {\doibase 10.1103/PhysRevD.97.083519} {\bibfield  {journal} {\bibinfo
  {journal} {\prd}\ }\textbf {\bibinfo {volume} {97}},\ \bibinfo {eid} {083519}
  (\bibinfo {year} {2018})},\ \Eprint {http://arxiv.org/abs/1801.03507}
  {arXiv:1801.03507 [astro-ph.CO]} \BibitemShut {NoStop}%
\bibitem [{\citenamefont {{Garny}}\ \emph {et~al.}(2020)\citenamefont
  {{Garny}}, \citenamefont {{Konstandin}},\ and\ \citenamefont
  {{Rubira}}}]{Garny2020}%
  \BibitemOpen
  \bibfield  {author} {\bibinfo {author} {\bibfnamefont {M.}~\bibnamefont
  {{Garny}}}, \bibinfo {author} {\bibfnamefont {T.}~\bibnamefont
  {{Konstandin}}}, \ and\ \bibinfo {author} {\bibfnamefont {H.}~\bibnamefont
  {{Rubira}}},\ }\href {\doibase 10.1088/1475-7516/2020/04/003} {\bibfield
  {journal} {\bibinfo  {journal} {\jcap}\ }\textbf {\bibinfo {volume} {2020}},\
  \bibinfo {eid} {003} (\bibinfo {year} {2020})},\ \Eprint
  {http://arxiv.org/abs/1911.04505} {arXiv:1911.04505 [astro-ph.CO]}
  \BibitemShut {NoStop}%
\bibitem [{\citenamefont {{Eberhardt}}\ \emph {et~al.}(2020)\citenamefont
  {{Eberhardt}}, \citenamefont {{Banerjee}}, \citenamefont {{Kopp}},\ and\
  \citenamefont {{Abel}}}]{Eberhardt2020}%
  \BibitemOpen
  \bibfield  {author} {\bibinfo {author} {\bibfnamefont {A.}~\bibnamefont
  {{Eberhardt}}}, \bibinfo {author} {\bibfnamefont {A.}~\bibnamefont
  {{Banerjee}}}, \bibinfo {author} {\bibfnamefont {M.}~\bibnamefont {{Kopp}}},
  \ and\ \bibinfo {author} {\bibfnamefont {T.}~\bibnamefont {{Abel}}},\ }\href
  {\doibase 10.1103/PhysRevD.101.043011} {\bibfield  {journal} {\bibinfo
  {journal} {\prd}\ }\textbf {\bibinfo {volume} {101}},\ \bibinfo {eid}
  {043011} (\bibinfo {year} {2020})},\ \Eprint
  {http://arxiv.org/abs/2001.05791} {arXiv:2001.05791 [physics.comp-ph]}
  \BibitemShut {NoStop}%
\bibitem [{\citenamefont {{Rampf}}(2021)}]{Rampf2021arXiv}%
  \BibitemOpen
  \bibfield  {author} {\bibinfo {author} {\bibfnamefont {C.}~\bibnamefont
  {{Rampf}}},\ }\href@noop {} {\bibfield  {journal} {\bibinfo  {journal} {arXiv
  e-prints}\ ,\ \bibinfo {eid} {arXiv:2110.06265}} (\bibinfo {year} {2021})},\
  \Eprint {http://arxiv.org/abs/2110.06265} {arXiv:2110.06265 [astro-ph.CO]}
  \BibitemShut {NoStop}%
\bibitem [{\citenamefont {Uhlemann}(2018)}]{Uhlemann2018finitelygenerated}%
  \BibitemOpen
  \bibfield  {author} {\bibinfo {author} {\bibfnamefont {C.}~\bibnamefont
  {Uhlemann}},\ }\href {\doibase 10.1088/1475-7516/2018/10/030} {\bibfield
  {journal} {\bibinfo  {journal} {JCAP}\ }\textbf {\bibinfo {volume} {10}},\
  \bibinfo {pages} {030} (\bibinfo {year} {2018})},\ \Eprint
  {http://arxiv.org/abs/1807.07274} {arXiv:1807.07274 [astro-ph.CO]}
  \BibitemShut {NoStop}%
\bibitem [{\citenamefont {Schive}\ \emph {et~al.}(2014)\citenamefont {Schive},
  \citenamefont {Chiueh},\ and\ \citenamefont {Broadhurst}}]{Schive2014}%
  \BibitemOpen
  \bibfield  {author} {\bibinfo {author} {\bibfnamefont {H.-Y.}\ \bibnamefont
  {Schive}}, \bibinfo {author} {\bibfnamefont {T.}~\bibnamefont {Chiueh}}, \
  and\ \bibinfo {author} {\bibfnamefont {T.}~\bibnamefont {Broadhurst}},\
  }\href {\doibase 10.1038/nphys2996} {\bibfield  {journal} {\bibinfo
  {journal} {Nature Physics}\ }\textbf {\bibinfo {volume} {10}},\ \bibinfo
  {pages} {496} (\bibinfo {year} {2014})}\BibitemShut {NoStop}%
\bibitem [{\citenamefont {{Schwabe}}\ and\ \citenamefont
  {{Niemeyer}}(2022)}]{SchwabeNiemeyer2022}%
  \BibitemOpen
  \bibfield  {author} {\bibinfo {author} {\bibfnamefont {B.}~\bibnamefont
  {{Schwabe}}}\ and\ \bibinfo {author} {\bibfnamefont {J.~C.}\ \bibnamefont
  {{Niemeyer}}},\ }\href {\doibase 10.1103/PhysRevLett.128.181301} {\bibfield
  {journal} {\bibinfo  {journal} {\prl}\ }\textbf {\bibinfo {volume} {128}},\
  \bibinfo {eid} {181301} (\bibinfo {year} {2022})},\ \Eprint
  {http://arxiv.org/abs/2110.09145} {arXiv:2110.09145 [astro-ph.CO]}
  \BibitemShut {NoStop}%
\bibitem [{\citenamefont {Porqueres}\ \emph {et~al.}(2020)\citenamefont
  {Porqueres}, \citenamefont {Hahn}, \citenamefont {Jasche},\ and\
  \citenamefont {Lavaux}}]{Porqueres_2020}%
  \BibitemOpen
  \bibfield  {author} {\bibinfo {author} {\bibfnamefont {N.}~\bibnamefont
  {Porqueres}}, \bibinfo {author} {\bibfnamefont {O.}~\bibnamefont {Hahn}},
  \bibinfo {author} {\bibfnamefont {J.}~\bibnamefont {Jasche}}, \ and\ \bibinfo
  {author} {\bibfnamefont {G.}~\bibnamefont {Lavaux}},\ }\href {\doibase
  10.1051/0004-6361/202038482} {\bibfield  {journal} {\bibinfo  {journal}
  {Astronomy \& Astrophysics}\ }\textbf {\bibinfo {volume} {642}},\ \bibinfo
  {pages} {A139} (\bibinfo {year} {2020})}\BibitemShut {NoStop}%
\bibitem [{\citenamefont {{Mina}}\ \emph {et~al.}(2020)\citenamefont {{Mina}},
  \citenamefont {{Mota}},\ and\ \citenamefont {{Winther}}}]{Mina2020}%
  \BibitemOpen
  \bibfield  {author} {\bibinfo {author} {\bibfnamefont {M.}~\bibnamefont
  {{Mina}}}, \bibinfo {author} {\bibfnamefont {D.~F.}\ \bibnamefont {{Mota}}},
  \ and\ \bibinfo {author} {\bibfnamefont {H.~A.}\ \bibnamefont {{Winther}}},\
  }\href {\doibase 10.1051/0004-6361/201936272} {\bibfield  {journal} {\bibinfo
   {journal} {\aap}\ }\textbf {\bibinfo {volume} {641}},\ \bibinfo {eid} {A107}
  (\bibinfo {year} {2020})},\ \Eprint {http://arxiv.org/abs/1906.12160}
  {arXiv:1906.12160 [physics.comp-ph]} \BibitemShut {NoStop}%
\bibitem [{\citenamefont {{May}}\ and\ \citenamefont
  {{Springel}}(2021)}]{May2021}%
  \BibitemOpen
  \bibfield  {author} {\bibinfo {author} {\bibfnamefont {S.}~\bibnamefont
  {{May}}}\ and\ \bibinfo {author} {\bibfnamefont {V.}~\bibnamefont
  {{Springel}}},\ }\href {\doibase 10.1093/mnras/stab1764} {\bibfield
  {journal} {\bibinfo  {journal} {\mnras}\ }\textbf {\bibinfo {volume} {506}},\
  \bibinfo {pages} {2603} (\bibinfo {year} {2021})},\ \Eprint
  {http://arxiv.org/abs/2101.01828} {arXiv:2101.01828 [astro-ph.CO]}
  \BibitemShut {NoStop}%
\bibitem [{\citenamefont {{Rampf}}\ \emph {et~al.}(2015)\citenamefont
  {{Rampf}}, \citenamefont {{Villone}},\ and\ \citenamefont
  {{Frisch}}}]{Rampf2015}%
  \BibitemOpen
  \bibfield  {author} {\bibinfo {author} {\bibfnamefont {C.}~\bibnamefont
  {{Rampf}}}, \bibinfo {author} {\bibfnamefont {B.}~\bibnamefont {{Villone}}},
  \ and\ \bibinfo {author} {\bibfnamefont {U.}~\bibnamefont {{Frisch}}},\
  }\href {\doibase 10.1093/mnras/stv1365} {\bibfield  {journal} {\bibinfo
  {journal} {\mnras}\ }\textbf {\bibinfo {volume} {452}},\ \bibinfo {pages}
  {1421} (\bibinfo {year} {2015})},\ \Eprint {http://arxiv.org/abs/1504.00032}
  {arXiv:1504.00032 [astro-ph.CO]} \BibitemShut {NoStop}%
\bibitem [{\citenamefont {{Brenier}}\ \emph {et~al.}(2003)\citenamefont
  {{Brenier}}, \citenamefont {{Frisch}}, \citenamefont {{H{\'e}non}},
  \citenamefont {{Loeper}}, \citenamefont {{Matarrese}}, \citenamefont
  {{Mohayaee}},\ and\ \citenamefont {{Sobolevski{\u{i}}}}}]{Brenier2003}%
  \BibitemOpen
  \bibfield  {author} {\bibinfo {author} {\bibfnamefont {Y.}~\bibnamefont
  {{Brenier}}}, \bibinfo {author} {\bibfnamefont {U.}~\bibnamefont {{Frisch}}},
  \bibinfo {author} {\bibfnamefont {M.}~\bibnamefont {{H{\'e}non}}}, \bibinfo
  {author} {\bibfnamefont {G.}~\bibnamefont {{Loeper}}}, \bibinfo {author}
  {\bibfnamefont {S.}~\bibnamefont {{Matarrese}}}, \bibinfo {author}
  {\bibfnamefont {R.}~\bibnamefont {{Mohayaee}}}, \ and\ \bibinfo {author}
  {\bibfnamefont {A.}~\bibnamefont {{Sobolevski{\u{i}}}}},\ }\href {\doibase
  10.1046/j.1365-2966.2003.07106.x} {\bibfield  {journal} {\bibinfo  {journal}
  {\mnras}\ }\textbf {\bibinfo {volume} {346}},\ \bibinfo {pages} {501}
  (\bibinfo {year} {2003})},\ \Eprint {http://arxiv.org/abs/astro-ph/0304214}
  {arXiv:astro-ph/0304214 [astro-ph]} \BibitemShut {NoStop}%
\bibitem [{\citenamefont {{Gallagher}}\ and\ \citenamefont
  {{Coles}}(2022)}]{Gallagher_2022_SPvoids}%
  \BibitemOpen
  \bibfield  {author} {\bibinfo {author} {\bibfnamefont {A.}~\bibnamefont
  {{Gallagher}}}\ and\ \bibinfo {author} {\bibfnamefont {P.}~\bibnamefont
  {{Coles}}},\ }\href@noop {} {\bibfield  {journal} {\bibinfo  {journal} {arXiv
  e-prints}\ ,\ \bibinfo {eid} {arXiv:2208.13851}} (\bibinfo {year} {2022})},\
  \Eprint {http://arxiv.org/abs/2208.13851} {arXiv:2208.13851 [astro-ph.CO]}
  \BibitemShut {NoStop}%
\bibitem [{\citenamefont {{Madelung}}(1927)}]{Madelung1927}%
  \BibitemOpen
  \bibfield  {author} {\bibinfo {author} {\bibfnamefont {E.}~\bibnamefont
  {{Madelung}}},\ }\href {\doibase 10.1007/BF01400372} {\bibfield  {journal}
  {\bibinfo  {journal} {Zeitschrift fur Physik}\ }\textbf {\bibinfo {volume}
  {40}},\ \bibinfo {pages} {322} (\bibinfo {year} {1927})}\BibitemShut
  {NoStop}%
\bibitem [{\citenamefont {{Binney}}\ and\ \citenamefont
  {{Tremaine}}(2008)}]{BinneyTremaine2008}%
  \BibitemOpen
  \bibfield  {author} {\bibinfo {author} {\bibfnamefont {J.}~\bibnamefont
  {{Binney}}}\ and\ \bibinfo {author} {\bibfnamefont {S.}~\bibnamefont
  {{Tremaine}}},\ }\href@noop {} {\emph {\bibinfo {title} {{Galactic Dynamics:
  Second Edition}}}}\ (\bibinfo  {publisher} {Princeton University Press,
  Princeton, NJ USA},\ \bibinfo {year} {2008})\BibitemShut {NoStop}%
\bibitem [{\citenamefont {{Veltmaat}}\ and\ \citenamefont
  {{Niemeyer}}(2016)}]{Veltmaat2016PhRvD}%
  \BibitemOpen
  \bibfield  {author} {\bibinfo {author} {\bibfnamefont {J.}~\bibnamefont
  {{Veltmaat}}}\ and\ \bibinfo {author} {\bibfnamefont {J.~C.}\ \bibnamefont
  {{Niemeyer}}},\ }\href {\doibase 10.1103/PhysRevD.94.123523} {\bibfield
  {journal} {\bibinfo  {journal} {\prd}\ }\textbf {\bibinfo {volume} {94}},\
  \bibinfo {eid} {123523} (\bibinfo {year} {2016})},\ \Eprint
  {http://arxiv.org/abs/1608.00802} {arXiv:1608.00802 [astro-ph.CO]}
  \BibitemShut {NoStop}%
\bibitem [{\citenamefont {{Nori}}\ and\ \citenamefont
  {{Baldi}}(2018)}]{Nori_2018_AX-GADGET}%
  \BibitemOpen
  \bibfield  {author} {\bibinfo {author} {\bibfnamefont {M.}~\bibnamefont
  {{Nori}}}\ and\ \bibinfo {author} {\bibfnamefont {M.}~\bibnamefont
  {{Baldi}}},\ }\href {\doibase 10.1093/mnras/sty1224} {\bibfield  {journal}
  {\bibinfo  {journal} {\mnras}\ }\textbf {\bibinfo {volume} {478}},\ \bibinfo
  {pages} {3935} (\bibinfo {year} {2018})},\ \Eprint
  {http://arxiv.org/abs/1801.08144} {arXiv:1801.08144 [astro-ph.CO]}
  \BibitemShut {NoStop}%
\bibitem [{\citenamefont {Mocz}\ and\ \citenamefont
  {Succi}(2015)}]{Mocz_2015_SP_SPH}%
  \BibitemOpen
  \bibfield  {author} {\bibinfo {author} {\bibfnamefont {P.}~\bibnamefont
  {Mocz}}\ and\ \bibinfo {author} {\bibfnamefont {S.}~\bibnamefont {Succi}},\
  }\href {\doibase 10.1103/PhysRevE.91.053304} {\bibfield  {journal} {\bibinfo
  {journal} {Phys. Rev. E}\ }\textbf {\bibinfo {volume} {91}},\ \bibinfo
  {pages} {053304} (\bibinfo {year} {2015})}\BibitemShut {NoStop}%
\bibitem [{\citenamefont {{Hopkins}}(2019)}]{Hopkins_2019_numerics_FDM}%
  \BibitemOpen
  \bibfield  {author} {\bibinfo {author} {\bibfnamefont {P.~F.}\ \bibnamefont
  {{Hopkins}}},\ }\href {\doibase 10.1093/mnras/stz1922} {\bibfield  {journal}
  {\bibinfo  {journal} {\mnras}\ }\textbf {\bibinfo {volume} {489}},\ \bibinfo
  {pages} {2367} (\bibinfo {year} {2019})},\ \Eprint
  {http://arxiv.org/abs/1811.05583} {arXiv:1811.05583 [astro-ph.CO]}
  \BibitemShut {NoStop}%
\bibitem [{\citenamefont {{Wigner}}(1932)}]{Wigner1932}%
  \BibitemOpen
  \bibfield  {author} {\bibinfo {author} {\bibfnamefont {E.}~\bibnamefont
  {{Wigner}}},\ }\href {\doibase 10.1103/PhysRev.40.749} {\bibfield  {journal}
  {\bibinfo  {journal} {Physical Review}\ }\textbf {\bibinfo {volume} {40}},\
  \bibinfo {pages} {749} (\bibinfo {year} {1932})}\BibitemShut {NoStop}%
\bibitem [{\citenamefont {{Takahashi}}(1989)}]{Takahashi1989}%
  \BibitemOpen
  \bibfield  {author} {\bibinfo {author} {\bibfnamefont {K.}~\bibnamefont
  {{Takahashi}}},\ }\href {\doibase 10.1143/PTPS.98.109} {\bibfield  {journal}
  {\bibinfo  {journal} {Progress of Theoretical Physics Supplement}\ }\textbf
  {\bibinfo {volume} {98}},\ \bibinfo {pages} {109} (\bibinfo {year}
  {1989})}\BibitemShut {NoStop}%
\bibitem [{\citenamefont {Hudson}(1974)}]{Hudson1974}%
  \BibitemOpen
  \bibfield  {author} {\bibinfo {author} {\bibfnamefont {R.}~\bibnamefont
  {Hudson}},\ }\href {\doibase https://doi.org/10.1016/0034-4877(74)90007-X}
  {\bibfield  {journal} {\bibinfo  {journal} {Reports on Mathematical Physics}\
  }\textbf {\bibinfo {volume} {6}},\ \bibinfo {pages} {249} (\bibinfo {year}
  {1974})}\BibitemShut {NoStop}%
\bibitem [{\citenamefont {Husimi}(1940)}]{Husimi1940}%
  \BibitemOpen
  \bibfield  {author} {\bibinfo {author} {\bibfnamefont {K.}~\bibnamefont
  {Husimi}},\ }\href {\doibase 10.11429/ppmsj1919.22.4_264} {\bibfield
  {journal} {\bibinfo  {journal} {Proceedings of the Physico-Mathematical
  Society of Japan. 3rd Series}\ }\textbf {\bibinfo {volume} {22}},\ \bibinfo
  {pages} {264} (\bibinfo {year} {1940})}\BibitemShut {NoStop}%
\bibitem [{\citenamefont {{Lions}}\ and\ \citenamefont
  {{Paul}}(1993)}]{Lions1993}%
  \BibitemOpen
  \bibfield  {author} {\bibinfo {author} {\bibfnamefont {P.-L.}\ \bibnamefont
  {{Lions}}}\ and\ \bibinfo {author} {\bibfnamefont {T.}~\bibnamefont
  {{Paul}}},\ }\href {https://eudml.org/doc/39445} {\bibfield  {journal}
  {\bibinfo  {journal} {Revista Matemática Iberoamericana}\ }\textbf {\bibinfo
  {volume} {9}},\ \bibinfo {pages} {553} (\bibinfo {year} {1993})}\BibitemShut
  {NoStop}%
\bibitem [{\citenamefont {Gérard}\ \emph {et~al.}(1997)\citenamefont
  {Gérard}, \citenamefont {Markowich}, \citenamefont {Mauser},\ and\
  \citenamefont {Poupaud}}]{Gerard1997}%
  \BibitemOpen
  \bibfield  {author} {\bibinfo {author} {\bibfnamefont {P.}~\bibnamefont
  {Gérard}}, \bibinfo {author} {\bibfnamefont {P.~A.}\ \bibnamefont
  {Markowich}}, \bibinfo {author} {\bibfnamefont {N.~J.}\ \bibnamefont
  {Mauser}}, \ and\ \bibinfo {author} {\bibfnamefont {F.}~\bibnamefont
  {Poupaud}},\ }\href {\doibase
  10.1002/(SICI)1097-0312(199704)50:4<323::AID-CPA4>3.0.CO;2-C} {\bibfield
  {journal} {\bibinfo  {journal} {Communications on Pure and Applied
  Mathematics}\ }\textbf {\bibinfo {volume} {50}},\ \bibinfo {pages} {323}
  (\bibinfo {year} {1997})}\BibitemShut {NoStop}%
\bibitem [{\citenamefont {Zhang}\ \emph {et~al.}(2002)\citenamefont {Zhang},
  \citenamefont {Zheng},\ and\ \citenamefont {Mauser}}]{Zhang2002}%
  \BibitemOpen
  \bibfield  {author} {\bibinfo {author} {\bibfnamefont {P.}~\bibnamefont
  {Zhang}}, \bibinfo {author} {\bibfnamefont {Y.}~\bibnamefont {Zheng}}, \ and\
  \bibinfo {author} {\bibfnamefont {N.}~\bibnamefont {Mauser}},\ }\href
  {\doibase 10.1002/cpa.3017} {\bibfield  {journal} {\bibinfo  {journal}
  {Communications on Pure and Applied Mathematics}\ }\textbf {\bibinfo {volume}
  {55}},\ \bibinfo {pages} {582} (\bibinfo {year} {2002})}\BibitemShut
  {NoStop}%
\bibitem [{\citenamefont {Athanassoulis}\ \emph {et~al.}(2009)\citenamefont
  {Athanassoulis}, \citenamefont {Mauser},\ and\ \citenamefont
  {Paul}}]{Athanassoulis2009}%
  \BibitemOpen
  \bibfield  {author} {\bibinfo {author} {\bibfnamefont {A.~G.}\ \bibnamefont
  {Athanassoulis}}, \bibinfo {author} {\bibfnamefont {N.~J.}\ \bibnamefont
  {Mauser}}, \ and\ \bibinfo {author} {\bibfnamefont {T.}~\bibnamefont
  {Paul}},\ }\href {\doibase https://doi.org/10.1016/j.matpur.2009.01.001}
  {\bibfield  {journal} {\bibinfo  {journal} {Journal de Mathématiques Pures
  et Appliquées}\ }\textbf {\bibinfo {volume} {91}},\ \bibinfo {pages} {296}
  (\bibinfo {year} {2009})}\BibitemShut {NoStop}%
\bibitem [{\citenamefont {Athanassoulis}(2018)}]{Athanassoulis2018}%
  \BibitemOpen
  \bibfield  {author} {\bibinfo {author} {\bibfnamefont {A.}~\bibnamefont
  {Athanassoulis}},\ }\href {\doibase 10.1088/1361-6544/aa9a86} {\bibfield
  {journal} {\bibinfo  {journal} {Nonlinearity}\ }\textbf {\bibinfo {volume}
  {31}},\ \bibinfo {pages} {1045} (\bibinfo {year} {2018})}\BibitemShut
  {NoStop}%
\bibitem [{\citenamefont {{Galazo Garc{\'\i}a}}\ \emph
  {et~al.}(2022)\citenamefont {{Galazo Garc{\'\i}a}}, \citenamefont {{Brax}},\
  and\ \citenamefont {{Valageas}}}]{Galazo2022}%
  \BibitemOpen
  \bibfield  {author} {\bibinfo {author} {\bibfnamefont {R.}~\bibnamefont
  {{Galazo Garc{\'\i}a}}}, \bibinfo {author} {\bibfnamefont {P.}~\bibnamefont
  {{Brax}}}, \ and\ \bibinfo {author} {\bibfnamefont {P.}~\bibnamefont
  {{Valageas}}},\ }\href@noop {} {\bibfield  {journal} {\bibinfo  {journal}
  {arXiv e-prints}\ ,\ \bibinfo {eid} {arXiv:2203.05995}} (\bibinfo {year}
  {2022})},\ \Eprint {http://arxiv.org/abs/2203.05995} {arXiv:2203.05995
  [astro-ph.CO]} \BibitemShut {NoStop}%
\bibitem [{\citenamefont {{Bender}}\ and\ \citenamefont
  {{Orszag}}(1978)}]{BenderandOrszag}%
  \BibitemOpen
  \bibfield  {author} {\bibinfo {author} {\bibfnamefont {C.~M.}\ \bibnamefont
  {{Bender}}}\ and\ \bibinfo {author} {\bibfnamefont {S.~A.}\ \bibnamefont
  {{Orszag}}},\ }\href {\doibase 10.1007/978-1-4757-3069-2} {\emph {\bibinfo
  {title} {Advanced Mathematical Methods for Scientists and Engineers}}}\
  (\bibinfo  {publisher} {Springer New York, NY},\ \bibinfo {year}
  {1978})\BibitemShut {NoStop}%
\bibitem [{\citenamefont {{Wright}}(1980)}]{Wright1980}%
  \BibitemOpen
  \bibfield  {author} {\bibinfo {author} {\bibfnamefont {F.~J.}\ \bibnamefont
  {{Wright}}},\ }\href {\doibase 10.1088/0305-4470/13/9/018} {\bibfield
  {journal} {\bibinfo  {journal} {Journal of Physics A Mathematical General}\
  }\textbf {\bibinfo {volume} {13}},\ \bibinfo {pages} {2913} (\bibinfo {year}
  {1980})}\BibitemShut {NoStop}%
\bibitem [{\citenamefont {{Feldbrugge}}\ \emph {et~al.}(2019)\citenamefont
  {{Feldbrugge}}, \citenamefont {{Pen}},\ and\ \citenamefont
  {{Turok}}}]{Feldbrugge2019}%
  \BibitemOpen
  \bibfield  {author} {\bibinfo {author} {\bibfnamefont {J.}~\bibnamefont
  {{Feldbrugge}}}, \bibinfo {author} {\bibfnamefont {U.-L.}\ \bibnamefont
  {{Pen}}}, \ and\ \bibinfo {author} {\bibfnamefont {N.}~\bibnamefont
  {{Turok}}},\ }\href@noop {} {\bibfield  {journal} {\bibinfo  {journal} {arXiv
  e-prints}\ ,\ \bibinfo {eid} {arXiv:1909.04632}} (\bibinfo {year} {2019})},\
  \Eprint {http://arxiv.org/abs/1909.04632} {arXiv:1909.04632 [astro-ph.HE]}
  \BibitemShut {NoStop}%
\bibitem [{\citenamefont {{Eberhardt}}\ \emph {et~al.}(2021)\citenamefont
  {{Eberhardt}}, \citenamefont {{Kopp}}, \citenamefont {{Zamora}},\ and\
  \citenamefont {{Abel}}}]{Eberhardt_2021_CHiMES}%
  \BibitemOpen
  \bibfield  {author} {\bibinfo {author} {\bibfnamefont {A.}~\bibnamefont
  {{Eberhardt}}}, \bibinfo {author} {\bibfnamefont {M.}~\bibnamefont {{Kopp}}},
  \bibinfo {author} {\bibfnamefont {A.}~\bibnamefont {{Zamora}}}, \ and\
  \bibinfo {author} {\bibfnamefont {T.}~\bibnamefont {{Abel}}},\ }\href
  {\doibase 10.1103/PhysRevD.104.083007} {\bibfield  {journal} {\bibinfo
  {journal} {\prd}\ }\textbf {\bibinfo {volume} {104}},\ \bibinfo {eid}
  {083007} (\bibinfo {year} {2021})},\ \Eprint
  {http://arxiv.org/abs/2108.08849} {arXiv:2108.08849 [physics.comp-ph]}
  \BibitemShut {NoStop}%
\bibitem [{\citenamefont {Claasen}\ and\ \citenamefont
  {Mecklenbrauker}(1983)}]{Claasen1983}%
  \BibitemOpen
  \bibfield  {author} {\bibinfo {author} {\bibfnamefont {T.}~\bibnamefont
  {Claasen}}\ and\ \bibinfo {author} {\bibfnamefont {W.}~\bibnamefont
  {Mecklenbrauker}},\ }\href {\doibase 10.1109/TASSP.1983.1164212} {\bibfield
  {journal} {\bibinfo  {journal} {IEEE Transactions on Acoustics, Speech, and
  Signal Processing}\ }\textbf {\bibinfo {volume} {31}},\ \bibinfo {pages}
  {1067} (\bibinfo {year} {1983})}\BibitemShut {NoStop}%
\bibitem [{\citenamefont {Chassande-Mottin}\ and\ \citenamefont
  {Pai}(2005)}]{Chassande-Mottin2005}%
  \BibitemOpen
  \bibfield  {author} {\bibinfo {author} {\bibfnamefont {E.}~\bibnamefont
  {Chassande-Mottin}}\ and\ \bibinfo {author} {\bibfnamefont {A.}~\bibnamefont
  {Pai}},\ }\href {\doibase 10.1109/LSP.2005.849493} {\bibfield  {journal}
  {\bibinfo  {journal} {IEEE Signal Processing Letters}\ }\textbf {\bibinfo
  {volume} {12}},\ \bibinfo {pages} {508} (\bibinfo {year} {2005})}\BibitemShut
  {NoStop}%
\bibitem [{\citenamefont {{Veltmaat}}\ \emph {et~al.}(2018)\citenamefont
  {{Veltmaat}}, \citenamefont {{Niemeyer}},\ and\ \citenamefont
  {{Schwabe}}}]{Veltmaat_2018_FDM_halos}%
  \BibitemOpen
  \bibfield  {author} {\bibinfo {author} {\bibfnamefont {J.}~\bibnamefont
  {{Veltmaat}}}, \bibinfo {author} {\bibfnamefont {J.~C.}\ \bibnamefont
  {{Niemeyer}}}, \ and\ \bibinfo {author} {\bibfnamefont {B.}~\bibnamefont
  {{Schwabe}}},\ }\href {\doibase 10.1103/PhysRevD.98.043509} {\bibfield
  {journal} {\bibinfo  {journal} {\prd}\ }\textbf {\bibinfo {volume} {98}},\
  \bibinfo {eid} {043509} (\bibinfo {year} {2018})},\ \Eprint
  {http://arxiv.org/abs/1804.09647} {arXiv:1804.09647 [astro-ph.CO]}
  \BibitemShut {NoStop}%
\bibitem [{\citenamefont {Wyatt}\ and\ \citenamefont
  {Trahan}(2005)}]{Wyatt_2005_quantum_trajectories}%
  \BibitemOpen
  \bibfield  {author} {\bibinfo {author} {\bibfnamefont {R.~E.}\ \bibnamefont
  {Wyatt}}\ and\ \bibinfo {author} {\bibfnamefont {C.~J.}\ \bibnamefont
  {Trahan}},\ }\href {\doibase https://doi.org/10.1007/0-387-28145-2} {\emph
  {\bibinfo {title} {Quantum dynamics with trajectories: introduction to
  quantum hydrodynamics}}},\ \bibinfo {series} {Interdisciplinary applied
  mathematics}\ No.\ \bibinfo {number} {v. 28}\ (\bibinfo  {publisher}
  {Springer},\ \bibinfo {address} {New York},\ \bibinfo {year}
  {2005})\BibitemShut {NoStop}%
\bibitem [{\citenamefont {{Lagu{\"e}}}\ \emph {et~al.}(2021)\citenamefont
  {{Lagu{\"e}}}, \citenamefont {{Bond}}, \citenamefont {{Hlo{\v{z}}ek}},
  \citenamefont {{Marsh}},\ and\ \citenamefont
  {{S{\"o}ding}}}]{Lague_2021_FDM_LPT}%
  \BibitemOpen
  \bibfield  {author} {\bibinfo {author} {\bibfnamefont {A.}~\bibnamefont
  {{Lagu{\"e}}}}, \bibinfo {author} {\bibfnamefont {J.~R.}\ \bibnamefont
  {{Bond}}}, \bibinfo {author} {\bibfnamefont {R.}~\bibnamefont
  {{Hlo{\v{z}}ek}}}, \bibinfo {author} {\bibfnamefont {D.~J.~E.}\ \bibnamefont
  {{Marsh}}}, \ and\ \bibinfo {author} {\bibfnamefont {L.}~\bibnamefont
  {{S{\"o}ding}}},\ }\href {\doibase 10.1093/mnras/stab601} {\bibfield
  {journal} {\bibinfo  {journal} {\mnras}\ }\textbf {\bibinfo {volume} {504}},\
  \bibinfo {pages} {2391} (\bibinfo {year} {2021})},\ \Eprint
  {http://arxiv.org/abs/2004.08482} {arXiv:2004.08482 [astro-ph.CO]}
  \BibitemShut {NoStop}%
\bibitem [{\citenamefont {{Feynman}}(1958)}]{Feynman1958}%
  \BibitemOpen
  \bibfield  {author} {\bibinfo {author} {\bibfnamefont {R.~P.}\ \bibnamefont
  {{Feynman}}},\ }\href {\doibase 10.1016/S0031-8914(58)80495-4} {\bibfield
  {journal} {\bibinfo  {journal} {Physica}\ }\textbf {\bibinfo {volume} {24}},\
  \bibinfo {pages} {S18} (\bibinfo {year} {1958})}\BibitemShut {NoStop}%
\bibitem [{\citenamefont {{Gross}}(1961)}]{Gross1961}%
  \BibitemOpen
  \bibfield  {author} {\bibinfo {author} {\bibfnamefont {E.~P.}\ \bibnamefont
  {{Gross}}},\ }\href {\doibase 10.1007/BF02731494} {\bibfield  {journal}
  {\bibinfo  {journal} {Il Nuovo Cimento}\ }\textbf {\bibinfo {volume} {20}},\
  \bibinfo {pages} {454} (\bibinfo {year} {1961})}\BibitemShut {NoStop}%
\bibitem [{\citenamefont {Zinner}(2011)}]{hallock_vortex_2011}%
  \BibitemOpen
  \bibfield  {author} {\bibinfo {author} {\bibfnamefont {N.~T.}\ \bibnamefont
  {Zinner}},\ }\href {\doibase 10.1155/2011/734543} {\bibfield  {journal}
  {\bibinfo  {journal} {Physics Research International}\ }\textbf {\bibinfo
  {volume} {2011}},\ \bibinfo {pages} {734543} (\bibinfo {year} {2011})},\
  \bibinfo {note} {publisher: Hindawi Publishing Corporation}\BibitemShut
  {NoStop}%
\bibitem [{\citenamefont {{Rindler-Daller}}\ and\ \citenamefont
  {{Shapiro}}(2012)}]{Rindler-Daller2012MNRAS}%
  \BibitemOpen
  \bibfield  {author} {\bibinfo {author} {\bibfnamefont {T.}~\bibnamefont
  {{Rindler-Daller}}}\ and\ \bibinfo {author} {\bibfnamefont {P.~R.}\
  \bibnamefont {{Shapiro}}},\ }\href {\doibase
  10.1111/j.1365-2966.2012.20588.x} {\bibfield  {journal} {\bibinfo  {journal}
  {\mnras}\ }\textbf {\bibinfo {volume} {422}},\ \bibinfo {pages} {135}
  (\bibinfo {year} {2012})},\ \Eprint {http://arxiv.org/abs/1106.1256}
  {arXiv:1106.1256 [astro-ph.CO]} \BibitemShut {NoStop}%
\bibitem [{\citenamefont {{Hui}}\ \emph {et~al.}(2021)\citenamefont {{Hui}},
  \citenamefont {{Joyce}}, \citenamefont {{Landry}},\ and\ \citenamefont
  {{Li}}}]{Hui2021JCAP}%
  \BibitemOpen
  \bibfield  {author} {\bibinfo {author} {\bibfnamefont {L.}~\bibnamefont
  {{Hui}}}, \bibinfo {author} {\bibfnamefont {A.}~\bibnamefont {{Joyce}}},
  \bibinfo {author} {\bibfnamefont {M.~J.}\ \bibnamefont {{Landry}}}, \ and\
  \bibinfo {author} {\bibfnamefont {X.}~\bibnamefont {{Li}}},\ }\href {\doibase
  10.1088/1475-7516/2021/01/011} {\bibfield  {journal} {\bibinfo  {journal}
  {\jcap}\ }\textbf {\bibinfo {volume} {2021}},\ \bibinfo {eid} {011} (\bibinfo
  {year} {2021})},\ \Eprint {http://arxiv.org/abs/2004.01188} {arXiv:2004.01188
  [astro-ph.CO]} \BibitemShut {NoStop}%
\bibitem [{\citenamefont {{Schobesberger}}\ \emph {et~al.}(2021)\citenamefont
  {{Schobesberger}}, \citenamefont {{Rindler-Daller}},\ and\ \citenamefont
  {{Shapiro}}}]{Schobesberger2021MNRAS}%
  \BibitemOpen
  \bibfield  {author} {\bibinfo {author} {\bibfnamefont {S.~O.}\ \bibnamefont
  {{Schobesberger}}}, \bibinfo {author} {\bibfnamefont {T.}~\bibnamefont
  {{Rindler-Daller}}}, \ and\ \bibinfo {author} {\bibfnamefont {P.~R.}\
  \bibnamefont {{Shapiro}}},\ }\href {\doibase 10.1093/mnras/stab1153}
  {\bibfield  {journal} {\bibinfo  {journal} {\mnras}\ }\textbf {\bibinfo
  {volume} {505}},\ \bibinfo {pages} {802} (\bibinfo {year} {2021})},\ \Eprint
  {http://arxiv.org/abs/2101.04958} {arXiv:2101.04958 [astro-ph.GA]}
  \BibitemShut {NoStop}%
\bibitem [{\citenamefont {{Alexander}}\ \emph {et~al.}(2021)\citenamefont
  {{Alexander}}, \citenamefont {{Capanelli}}, \citenamefont {{Ferreira}},\ and\
  \citenamefont {{McDonough}}}]{Alexander2021arXiv}%
  \BibitemOpen
  \bibfield  {author} {\bibinfo {author} {\bibfnamefont {S.}~\bibnamefont
  {{Alexander}}}, \bibinfo {author} {\bibfnamefont {C.}~\bibnamefont
  {{Capanelli}}}, \bibinfo {author} {\bibfnamefont {E.~G.~M.}\ \bibnamefont
  {{Ferreira}}}, \ and\ \bibinfo {author} {\bibfnamefont {E.}~\bibnamefont
  {{McDonough}}},\ }\href@noop {} {\bibfield  {journal} {\bibinfo  {journal}
  {arXiv e-prints}\ ,\ \bibinfo {eid} {arXiv:2111.03061}} (\bibinfo {year}
  {2021})},\ \Eprint {http://arxiv.org/abs/2111.03061} {arXiv:2111.03061
  [astro-ph.CO]} \BibitemShut {NoStop}%
\bibitem [{\citenamefont {Helmholtz}(1858)}]{Helmholtz1858}%
  \BibitemOpen
  \bibfield  {author} {\bibinfo {author} {\bibfnamefont {H.}~\bibnamefont
  {Helmholtz}},\ }\href {\doibase doi:10.1515/crll.1858.55.25} {\bibfield
  {journal} {\bibinfo  {journal} {Journal für die reine und angewandte
  Mathematik}\ }\textbf {\bibinfo {volume} {1858}},\ \bibinfo {pages} {25}
  (\bibinfo {year} {1858})}\BibitemShut {NoStop}%
\bibitem [{\citenamefont {Thomson}(1869)}]{Kelvin1869}%
  \BibitemOpen
  \bibfield  {author} {\bibinfo {author} {\bibfnamefont {W.~L.~K.}\
  \bibnamefont {Thomson}},\ }\href@noop {} {\bibfield  {journal} {\bibinfo
  {journal} {Trans. Roy. Soc. Edinburgh}\ }\textbf {\bibinfo {volume} {25}},\
  \bibinfo {pages} {217} (\bibinfo {year} {1869})}\BibitemShut {NoStop}%
\bibitem [{\citenamefont {{Damski}}\ and\ \citenamefont
  {{Sacha}}(2003)}]{Damski2003}%
  \BibitemOpen
  \bibfield  {author} {\bibinfo {author} {\bibfnamefont {B.}~\bibnamefont
  {{Damski}}}\ and\ \bibinfo {author} {\bibfnamefont {K.}~\bibnamefont
  {{Sacha}}},\ }\href {\doibase 10.1088/0305-4470/36/9/311} {\bibfield
  {journal} {\bibinfo  {journal} {Journal of Physics A Mathematical General}\
  }\textbf {\bibinfo {volume} {36}},\ \bibinfo {pages} {2339} (\bibinfo {year}
  {2003})},\ \Eprint {http://arxiv.org/abs/quant-ph/0202137}
  {arXiv:quant-ph/0202137 [quant-ph]} \BibitemShut {NoStop}%
\bibitem [{\citenamefont {{Savchenko}}\ and\ \citenamefont
  {{Zel'Dovich}}(1999)}]{Savchenko1999}%
  \BibitemOpen
  \bibfield  {author} {\bibinfo {author} {\bibfnamefont {A.~Y.}\ \bibnamefont
  {{Savchenko}}}\ and\ \bibinfo {author} {\bibfnamefont {B.~Y.}\ \bibnamefont
  {{Zel'Dovich}}},\ }\href {\doibase 10.1364/JOSAA.16.001665} {\bibfield
  {journal} {\bibinfo  {journal} {Journal of the Optical Society of America A}\
  }\textbf {\bibinfo {volume} {16}},\ \bibinfo {pages} {1665} (\bibinfo {year}
  {1999})}\BibitemShut {NoStop}%
\bibitem [{\citenamefont {{Arnold}}(1978)}]{Arnold1978_classicalmechanics}%
  \BibitemOpen
  \bibfield  {author} {\bibinfo {author} {\bibfnamefont {V.~I.}\ \bibnamefont
  {{Arnold}}},\ }\href {\doibase https://doi.org/10.1007/978-1-4757-1693-1}
  {\emph {\bibinfo {title} {{Mathematical methods of classical mechanics}}}}\
  (\bibinfo  {publisher} {Springer New York, NY},\ \bibinfo {year}
  {1978})\BibitemShut {NoStop}%
\bibitem [{\citenamefont {Heller}(2018)}]{heller_semiclassical_2018}%
  \BibitemOpen
  \bibfield  {author} {\bibinfo {author} {\bibfnamefont {E.~J.}\ \bibnamefont
  {Heller}},\ }\href {\doibase 10.2307/j.ctvc77gwd} {\emph {\bibinfo {title}
  {The {Semiclassical} {Way} to {Dynamics} and {Spectroscopy}}}}\ (\bibinfo
  {publisher} {Princeton University Press},\ \bibinfo {year}
  {2018})\BibitemShut {NoStop}%
\bibitem [{\citenamefont {{Fried}}(1998)}]{Fried1998}%
  \BibitemOpen
  \bibfield  {author} {\bibinfo {author} {\bibfnamefont {D.~L.}\ \bibnamefont
  {{Fried}}},\ }\href {\doibase 10.1364/JOSAA.15.002759} {\bibfield  {journal}
  {\bibinfo  {journal} {Journal of the Optical Society of America A}\ }\textbf
  {\bibinfo {volume} {15}},\ \bibinfo {pages} {2759} (\bibinfo {year}
  {1998})}\BibitemShut {NoStop}%
\bibitem [{\citenamefont {{Buehlmann}}\ and\ \citenamefont
  {{Hahn}}(2019)}]{BuehlmannHahn2019}%
  \BibitemOpen
  \bibfield  {author} {\bibinfo {author} {\bibfnamefont {M.}~\bibnamefont
  {{Buehlmann}}}\ and\ \bibinfo {author} {\bibfnamefont {O.}~\bibnamefont
  {{Hahn}}},\ }\href {\doibase 10.1093/mnras/stz1243} {\bibfield  {journal}
  {\bibinfo  {journal} {\mnras}\ }\textbf {\bibinfo {volume} {487}},\ \bibinfo
  {pages} {228} (\bibinfo {year} {2019})},\ \Eprint
  {http://arxiv.org/abs/1812.07489} {arXiv:1812.07489 [astro-ph.CO]}
  \BibitemShut {NoStop}%
\bibitem [{\citenamefont {{Yahalom}}(2018)}]{Yahalom2018MolPh}%
  \BibitemOpen
  \bibfield  {author} {\bibinfo {author} {\bibfnamefont {A.}~\bibnamefont
  {{Yahalom}}},\ }\href {\doibase 10.1080/00268976.2018.1457808} {\bibfield
  {journal} {\bibinfo  {journal} {Molecular Physics}\ }\textbf {\bibinfo
  {volume} {116}},\ \bibinfo {pages} {2698} (\bibinfo {year}
  {2018})}\BibitemShut {NoStop}%
\bibitem [{\citenamefont {Thom}(1994)}]{Thom1994}%
  \BibitemOpen
  \bibfield  {author} {\bibinfo {author} {\bibfnamefont {R.}~\bibnamefont
  {Thom}},\ }\href {https://books.google.ca/books?id=YVNPDwAAQBAJ} {\emph
  {\bibinfo {title} {{Structural Stability and Morphogenesis}}}}\ (\bibinfo
  {publisher} {Avalon Publishing},\ \bibinfo {year} {1994})\BibitemShut
  {NoStop}%
\bibitem [{\citenamefont {Arnol'd}(1973)}]{Arnold1973}%
  \BibitemOpen
  \bibfield  {author} {\bibinfo {author} {\bibfnamefont {V.~I.}\ \bibnamefont
  {Arnol'd}},\ }\href {\doibase 10.1070/RM1973v028n05ABEH001609} {\bibfield
  {journal} {\bibinfo  {journal} {Russian Mathematical Surveys}\ }\textbf
  {\bibinfo {volume} {28}},\ \bibinfo {pages} {19} (\bibinfo {year} {1973})},\
  \bibinfo {note} {publisher: IOP Publishing}\BibitemShut {NoStop}%
\bibitem [{\citenamefont {Arnol'd}(1975)}]{Arnold1975}%
  \BibitemOpen
  \bibfield  {author} {\bibinfo {author} {\bibfnamefont {V.~I.}\ \bibnamefont
  {Arnol'd}},\ }\href {\doibase 10.1070/RM1975v030n05ABEH001521} {\bibfield
  {journal} {\bibinfo  {journal} {Russian Mathematical Surveys}\ }\textbf
  {\bibinfo {volume} {30}},\ \bibinfo {pages} {1} (\bibinfo {year} {1975})},\
  \bibinfo {note} {publisher: IOP Publishing}\BibitemShut {NoStop}%
\bibitem [{\citenamefont {{Arnold}}\ \emph {et~al.}(1982)\citenamefont
  {{Arnold}}, \citenamefont {{Shandarin}},\ and\ \citenamefont
  {{Zeldovich}}}]{ArnoldShandarinZeldovich1982}%
  \BibitemOpen
  \bibfield  {author} {\bibinfo {author} {\bibfnamefont {V.~I.}\ \bibnamefont
  {{Arnold}}}, \bibinfo {author} {\bibfnamefont {S.~F.}\ \bibnamefont
  {{Shandarin}}}, \ and\ \bibinfo {author} {\bibfnamefont {I.~B.}\ \bibnamefont
  {{Zeldovich}}},\ }\href {\doibase 10.1080/03091928208209001} {\bibfield
  {journal} {\bibinfo  {journal} {Geophysical and Astrophysical Fluid
  Dynamics}\ }\textbf {\bibinfo {volume} {20}},\ \bibinfo {pages} {111}
  (\bibinfo {year} {1982})}\BibitemShut {NoStop}%
\bibitem [{\citenamefont {{Hidding}}\ \emph {et~al.}(2014)\citenamefont
  {{Hidding}}, \citenamefont {{Shandarin}},\ and\ \citenamefont {{van de
  Weygaert}}}]{Hidding2014}%
  \BibitemOpen
  \bibfield  {author} {\bibinfo {author} {\bibfnamefont {J.}~\bibnamefont
  {{Hidding}}}, \bibinfo {author} {\bibfnamefont {S.~F.}\ \bibnamefont
  {{Shandarin}}}, \ and\ \bibinfo {author} {\bibfnamefont {R.}~\bibnamefont
  {{van de Weygaert}}},\ }\href {\doibase 10.1093/mnras/stt2142} {\bibfield
  {journal} {\bibinfo  {journal} {\mnras}\ }\textbf {\bibinfo {volume} {437}},\
  \bibinfo {pages} {3442} (\bibinfo {year} {2014})},\ \Eprint
  {http://arxiv.org/abs/1311.7134} {arXiv:1311.7134 [astro-ph.CO]} \BibitemShut
  {NoStop}%
\bibitem [{\citenamefont {{Feldbrugge}}\ \emph {et~al.}(2014)\citenamefont
  {{Feldbrugge}}, \citenamefont {{Hidding}},\ and\ \citenamefont {{van de
  Weygaert}}}]{Feldbrugge2014}%
  \BibitemOpen
  \bibfield  {author} {\bibinfo {author} {\bibfnamefont {J.}~\bibnamefont
  {{Feldbrugge}}}, \bibinfo {author} {\bibfnamefont {J.}~\bibnamefont
  {{Hidding}}}, \ and\ \bibinfo {author} {\bibfnamefont {R.}~\bibnamefont {{van
  de Weygaert}}},\ }\href@noop {} {\bibfield  {journal} {\bibinfo  {journal}
  {arXiv e-prints}\ ,\ \bibinfo {eid} {arXiv:1412.5121}} (\bibinfo {year}
  {2014})},\ \Eprint {http://arxiv.org/abs/1412.5121} {arXiv:1412.5121
  [astro-ph.CO]} \BibitemShut {NoStop}%
\bibitem [{\citenamefont {{Feldbrugge}}\ \emph {et~al.}(2018)\citenamefont
  {{Feldbrugge}}, \citenamefont {{van de Weygaert}}, \citenamefont
  {{Hidding}},\ and\ \citenamefont {{Feldbrugge}}}]{Feldbrugge2018}%
  \BibitemOpen
  \bibfield  {author} {\bibinfo {author} {\bibfnamefont {J.}~\bibnamefont
  {{Feldbrugge}}}, \bibinfo {author} {\bibfnamefont {R.}~\bibnamefont {{van de
  Weygaert}}}, \bibinfo {author} {\bibfnamefont {J.}~\bibnamefont {{Hidding}}},
  \ and\ \bibinfo {author} {\bibfnamefont {J.}~\bibnamefont {{Feldbrugge}}},\
  }\href {\doibase 10.1088/1475-7516/2018/05/027} {\bibfield  {journal}
  {\bibinfo  {journal} {\jcap}\ }\textbf {\bibinfo {volume} {2018}},\ \bibinfo
  {eid} {027} (\bibinfo {year} {2018})},\ \Eprint
  {http://arxiv.org/abs/1703.09598} {arXiv:1703.09598 [astro-ph.CO]}
  \BibitemShut {NoStop}%
\bibitem [{\citenamefont {Berry}(1977)}]{Berry1977_focusing}%
  \BibitemOpen
  \bibfield  {author} {\bibinfo {author} {\bibfnamefont {M.~V.}\ \bibnamefont
  {Berry}},\ }\href {\doibase 10.1088/0305-4470/10/12/015} {\bibfield
  {journal} {\bibinfo  {journal} {Journal of Physics A: Mathematical and
  General}\ }\textbf {\bibinfo {volume} {10}},\ \bibinfo {pages} {2061}
  (\bibinfo {year} {1977})},\ \bibinfo {note} {publisher: IOP
  Publishing}\BibitemShut {NoStop}%
\bibitem [{\citenamefont {Berry}\ and\ \citenamefont
  {Nye}(1977)}]{Berry1977_finestructure}%
  \BibitemOpen
  \bibfield  {author} {\bibinfo {author} {\bibfnamefont {M.~V.}\ \bibnamefont
  {Berry}}\ and\ \bibinfo {author} {\bibfnamefont {J.~F.}\ \bibnamefont
  {Nye}},\ }\href {\doibase 10.1038/267034a0} {\bibfield  {journal} {\bibinfo
  {journal} {Nature}\ }\textbf {\bibinfo {volume} {267}},\ \bibinfo {pages}
  {34} (\bibinfo {year} {1977})},\ \bibinfo {note} {number: 5606 Publisher:
  Nature Publishing Group}\BibitemShut {NoStop}%
\bibitem [{\citenamefont {{Berry}}\ \emph {et~al.}(1979)\citenamefont
  {{Berry}}, \citenamefont {{Nye}},\ and\ \citenamefont
  {{Wright}}}]{BerryNyeWright1979}%
  \BibitemOpen
  \bibfield  {author} {\bibinfo {author} {\bibfnamefont {M.~V.}\ \bibnamefont
  {{Berry}}}, \bibinfo {author} {\bibfnamefont {J.~F.}\ \bibnamefont {{Nye}}},
  \ and\ \bibinfo {author} {\bibfnamefont {F.~J.}\ \bibnamefont {{Wright}}},\
  }\href {\doibase 10.1098/rsta.1979.0039} {\bibfield  {journal} {\bibinfo
  {journal} {Philosophical Transactions of the Royal Society of London Series
  A}\ }\textbf {\bibinfo {volume} {291}},\ \bibinfo {pages} {453} (\bibinfo
  {year} {1979})}\BibitemShut {NoStop}%
\bibitem [{\citenamefont {{Berry}}\ and\ \citenamefont
  {{Upstill}}(1980)}]{BerryUpstill1980}%
  \BibitemOpen
  \bibfield  {author} {\bibinfo {author} {\bibfnamefont {M.~V.}\ \bibnamefont
  {{Berry}}}\ and\ \bibinfo {author} {\bibfnamefont {C.}~\bibnamefont
  {{Upstill}}},\ }\href {\doibase 10.1016/S0079-6638(08)70215-4} {\bibfield
  {journal} {\bibinfo  {journal} {Progess in Optics}\ }\textbf {\bibinfo
  {volume} {18}},\ \bibinfo {pages} {257} (\bibinfo {year} {1980})}\BibitemShut
  {NoStop}%
\bibitem [{\citenamefont {Arnold}\ \emph {et~al.}(2012)\citenamefont {Arnold},
  \citenamefont {Gusein-Zade},\ and\ \citenamefont
  {Varchenko}}]{Arnold2012_catastrophebook}%
  \BibitemOpen
  \bibfield  {author} {\bibinfo {author} {\bibfnamefont {V.}~\bibnamefont
  {Arnold}}, \bibinfo {author} {\bibfnamefont {S.}~\bibnamefont {Gusein-Zade}},
  \ and\ \bibinfo {author} {\bibfnamefont {A.}~\bibnamefont {Varchenko}},\
  }\href {\doibase 10.1007/978-0-8176-8340-5} {\emph {\bibinfo {title}
  {{Singularities of {Differentiable} {Maps}, {Volume} 1}}}}\ (\bibinfo
  {publisher} {Birkhäuser},\ \bibinfo {address} {Boston},\ \bibinfo {year}
  {2012})\BibitemShut {NoStop}%
\bibitem [{\citenamefont {Pearcey}(1946)}]{Pearcey1946}%
  \BibitemOpen
  \bibfield  {author} {\bibinfo {author} {\bibfnamefont {T.}~\bibnamefont
  {Pearcey}},\ }\href {\doibase 10.1080/14786444608561335} {\bibfield
  {journal} {\bibinfo  {journal} {The London, Edinburgh, and Dublin
  Philosophical Magazine and Journal of Science}\ }\textbf {\bibinfo {volume}
  {37}},\ \bibinfo {pages} {311} (\bibinfo {year} {1946})},\ \bibinfo {note}
  {publisher: Taylor \& Francis \_eprint:
  https://doi.org/10.1080/14786444608561335}\BibitemShut {NoStop}%
\bibitem [{\citenamefont {{Dalal}}\ \emph {et~al.}(2021)\citenamefont
  {{Dalal}}, \citenamefont {{Bovy}}, \citenamefont {{Hui}},\ and\ \citenamefont
  {{Li}}}]{Dalal2021JCAP}%
  \BibitemOpen
  \bibfield  {author} {\bibinfo {author} {\bibfnamefont {N.}~\bibnamefont
  {{Dalal}}}, \bibinfo {author} {\bibfnamefont {J.}~\bibnamefont {{Bovy}}},
  \bibinfo {author} {\bibfnamefont {L.}~\bibnamefont {{Hui}}}, \ and\ \bibinfo
  {author} {\bibfnamefont {X.}~\bibnamefont {{Li}}},\ }\href {\doibase
  10.1088/1475-7516/2021/03/076} {\bibfield  {journal} {\bibinfo  {journal}
  {\jcap}\ }\textbf {\bibinfo {volume} {2021}},\ \bibinfo {eid} {076} (\bibinfo
  {year} {2021})},\ \Eprint {http://arxiv.org/abs/2011.13141} {arXiv:2011.13141
  [astro-ph.CO]} \BibitemShut {NoStop}%
\bibitem [{\citenamefont {{Colombi}}(2015)}]{Colombi2015}%
  \BibitemOpen
  \bibfield  {author} {\bibinfo {author} {\bibfnamefont {S.}~\bibnamefont
  {{Colombi}}},\ }\href {\doibase 10.1093/mnras/stu2308} {\bibfield  {journal}
  {\bibinfo  {journal} {\mnras}\ }\textbf {\bibinfo {volume} {446}},\ \bibinfo
  {pages} {2902} (\bibinfo {year} {2015})},\ \Eprint
  {http://arxiv.org/abs/1411.4165} {arXiv:1411.4165 [astro-ph.CO]} \BibitemShut
  {NoStop}%
\bibitem [{\citenamefont {{Taruya}}\ and\ \citenamefont
  {{Colombi}}(2017)}]{TaruyaColombi2017}%
  \BibitemOpen
  \bibfield  {author} {\bibinfo {author} {\bibfnamefont {A.}~\bibnamefont
  {{Taruya}}}\ and\ \bibinfo {author} {\bibfnamefont {S.}~\bibnamefont
  {{Colombi}}},\ }\href {\doibase 10.1093/mnras/stx1501} {\bibfield  {journal}
  {\bibinfo  {journal} {\mnras}\ }\textbf {\bibinfo {volume} {470}},\ \bibinfo
  {pages} {4858} (\bibinfo {year} {2017})},\ \Eprint
  {http://arxiv.org/abs/1701.09088} {arXiv:1701.09088 [astro-ph.CO]}
  \BibitemShut {NoStop}%
\bibitem [{\citenamefont {Pietroni}(2018)}]{Pietroni2018}%
  \BibitemOpen
  \bibfield  {author} {\bibinfo {author} {\bibfnamefont {M.}~\bibnamefont
  {Pietroni}},\ }\href {\doibase 10.1088/1475-7516/2018/06/028} {\bibfield
  {journal} {\bibinfo  {journal} {Journal of Cosmology and Astroparticle
  Physics}\ }\textbf {\bibinfo {volume} {2018}},\ \bibinfo {pages} {028}
  (\bibinfo {year} {2018})}\BibitemShut {NoStop}%
\bibitem [{\citenamefont {{Saga}}\ \emph {et~al.}(2022)\citenamefont {{Saga}},
  \citenamefont {{Taruya}},\ and\ \citenamefont
  {{Colombi}}}]{Saga_2022_LPT_Vlasov}%
  \BibitemOpen
  \bibfield  {author} {\bibinfo {author} {\bibfnamefont {S.}~\bibnamefont
  {{Saga}}}, \bibinfo {author} {\bibfnamefont {A.}~\bibnamefont {{Taruya}}}, \
  and\ \bibinfo {author} {\bibfnamefont {S.}~\bibnamefont {{Colombi}}},\ }\href
  {\doibase 10.1051/0004-6361/202142756} {\bibfield  {journal} {\bibinfo
  {journal} {\aap}\ }\textbf {\bibinfo {volume} {664}},\ \bibinfo {eid} {A3}
  (\bibinfo {year} {2022})},\ \Eprint {http://arxiv.org/abs/2111.08836}
  {arXiv:2111.08836 [astro-ph.CO]} \BibitemShut {NoStop}%
\bibitem [{\citenamefont {{Zimmermann}}\ \emph {et~al.}(2021)\citenamefont
  {{Zimmermann}}, \citenamefont {{Schwersenz}}, \citenamefont {{Pietroni}},\
  and\ \citenamefont {{Wimberger}}}]{Zimmermann2021PhRvD}%
  \BibitemOpen
  \bibfield  {author} {\bibinfo {author} {\bibfnamefont {T.}~\bibnamefont
  {{Zimmermann}}}, \bibinfo {author} {\bibfnamefont {N.}~\bibnamefont
  {{Schwersenz}}}, \bibinfo {author} {\bibfnamefont {M.}~\bibnamefont
  {{Pietroni}}}, \ and\ \bibinfo {author} {\bibfnamefont {S.}~\bibnamefont
  {{Wimberger}}},\ }\href {\doibase 10.1103/PhysRevD.103.083018} {\bibfield
  {journal} {\bibinfo  {journal} {\prd}\ }\textbf {\bibinfo {volume} {103}},\
  \bibinfo {eid} {083018} (\bibinfo {year} {2021})},\ \Eprint
  {http://arxiv.org/abs/2102.13619} {arXiv:2102.13619 [astro-ph.CO]}
  \BibitemShut {NoStop}%
\bibitem [{\citenamefont {{Zagorac}}\ \emph {et~al.}(2022)\citenamefont
  {{Zagorac}}, \citenamefont {{Sands}}, \citenamefont {{Padmanabhan}},\ and\
  \citenamefont {{Easther}}}]{Zagorac2022PhRvD}%
  \BibitemOpen
  \bibfield  {author} {\bibinfo {author} {\bibfnamefont {J.~L.}\ \bibnamefont
  {{Zagorac}}}, \bibinfo {author} {\bibfnamefont {I.}~\bibnamefont {{Sands}}},
  \bibinfo {author} {\bibfnamefont {N.}~\bibnamefont {{Padmanabhan}}}, \ and\
  \bibinfo {author} {\bibfnamefont {R.}~\bibnamefont {{Easther}}},\ }\href
  {\doibase 10.1103/PhysRevD.105.103506} {\bibfield  {journal} {\bibinfo
  {journal} {\prd}\ }\textbf {\bibinfo {volume} {105}},\ \bibinfo {eid}
  {103506} (\bibinfo {year} {2022})},\ \Eprint
  {http://arxiv.org/abs/2109.01920} {arXiv:2109.01920 [astro-ph.CO]}
  \BibitemShut {NoStop}%
\bibitem [{\citenamefont {{Taruya}}\ and\ \citenamefont
  {{Saga}}(2022)}]{Taruya_2022_corehalo}%
  \BibitemOpen
  \bibfield  {author} {\bibinfo {author} {\bibfnamefont {A.}~\bibnamefont
  {{Taruya}}}\ and\ \bibinfo {author} {\bibfnamefont {S.}~\bibnamefont
  {{Saga}}},\ }\href@noop {} {\bibfield  {journal} {\bibinfo  {journal} {arXiv
  e-prints}\ ,\ \bibinfo {eid} {arXiv:2208.06562}} (\bibinfo {year} {2022})},\
  \Eprint {http://arxiv.org/abs/2208.06562} {arXiv:2208.06562 [astro-ph.CO]}
  \BibitemShut {NoStop}%
\bibitem [{\citenamefont {{Angulo}}\ and\ \citenamefont
  {{Hahn}}(2022)}]{AnguloHahn2022}%
  \BibitemOpen
  \bibfield  {author} {\bibinfo {author} {\bibfnamefont {R.~E.}\ \bibnamefont
  {{Angulo}}}\ and\ \bibinfo {author} {\bibfnamefont {O.}~\bibnamefont
  {{Hahn}}},\ }\href {\doibase 10.1007/s41115-021-00013-z} {\bibfield
  {journal} {\bibinfo  {journal} {Living Reviews in Computational
  Astrophysics}\ }\textbf {\bibinfo {volume} {8}},\ \bibinfo {eid} {1}
  (\bibinfo {year} {2022})},\ \Eprint {http://arxiv.org/abs/2112.05165}
  {arXiv:2112.05165 [astro-ph.CO]} \BibitemShut {NoStop}%
\bibitem [{\citenamefont {{Hahn}}\ \emph {et~al.}(2021)\citenamefont {{Hahn}},
  \citenamefont {{Rampf}},\ and\ \citenamefont
  {{Uhlemann}}}]{Hahn_2021_ICs_2fluid}%
  \BibitemOpen
  \bibfield  {author} {\bibinfo {author} {\bibfnamefont {O.}~\bibnamefont
  {{Hahn}}}, \bibinfo {author} {\bibfnamefont {C.}~\bibnamefont {{Rampf}}}, \
  and\ \bibinfo {author} {\bibfnamefont {C.}~\bibnamefont {{Uhlemann}}},\
  }\href {\doibase 10.1093/mnras/staa3773} {\bibfield  {journal} {\bibinfo
  {journal} {\mnras}\ }\textbf {\bibinfo {volume} {503}},\ \bibinfo {pages}
  {426} (\bibinfo {year} {2021})},\ \Eprint {http://arxiv.org/abs/2008.09124}
  {arXiv:2008.09124 [astro-ph.CO]} \BibitemShut {NoStop}%
\bibitem [{\citenamefont {{Hunter \textit{et al.}}}(2007)}]{matplotlib}%
  \BibitemOpen
  \bibfield  {author} {\bibinfo {author} {\bibfnamefont {J.~D.}\ \bibnamefont
  {{Hunter \textit{et al.}}}},\ }\href {\doibase 10.1109/MCSE.2007.55}
  {\bibfield  {journal} {\bibinfo  {journal} {Computing in Science \&
  Engineering}\ }\textbf {\bibinfo {volume} {9}},\ \bibinfo {pages} {90}
  (\bibinfo {year} {2007})}\BibitemShut {NoStop}%
\bibitem [{\citenamefont {{Harris \textit{et al.}}}(2020)}]{numpy}%
  \BibitemOpen
  \bibfield  {author} {\bibinfo {author} {\bibfnamefont {C.~R.}\ \bibnamefont
  {{Harris \textit{et al.}}}},\ }\href {\doibase 10.1038/s41586-020-2649-2}
  {\bibfield  {journal} {\bibinfo  {journal} {Nature}\ }\textbf {\bibinfo
  {volume} {585}},\ \bibinfo {pages} {357} (\bibinfo {year}
  {2020})}\BibitemShut {NoStop}%
\bibitem [{\citenamefont {{Virtanen \textit{et al.}}}(2020)}]{2020SciPy-NMeth}%
  \BibitemOpen
  \bibfield  {author} {\bibinfo {author} {\bibfnamefont {P.}~\bibnamefont
  {{Virtanen \textit{et al.}}}},\ }\href {\doibase 10.1038/s41592-019-0686-2}
  {\bibfield  {journal} {\bibinfo  {journal} {Nature Methods}\ }\textbf
  {\bibinfo {volume} {17}},\ \bibinfo {pages} {261} (\bibinfo {year}
  {2020})}\BibitemShut {NoStop}%
\bibitem [{\citenamefont {{van der Walt \textit{et
  al.}}}(2014)}]{scikit-image}%
  \BibitemOpen
  \bibfield  {author} {\bibinfo {author} {\bibfnamefont {S.}~\bibnamefont {{van
  der Walt \textit{et al.}}}},\ }\href {\doibase 10.7717/peerj.453} {\bibfield
  {journal} {\bibinfo  {journal} {PeerJ}\ }\textbf {\bibinfo {volume} {2}},\
  \bibinfo {pages} {e453} (\bibinfo {year} {2014})}\BibitemShut {NoStop}%
\bibitem [{\citenamefont {{Wegert}}(2012)}]{Wegert2012}%
  \BibitemOpen
  \bibfield  {author} {\bibinfo {author} {\bibfnamefont {E.}~\bibnamefont
  {{Wegert}}},\ }\href {\doibase 10.1007/978-3-0348-0180-5} {\emph {\bibinfo
  {title} {{Visual Complex Functions}}}}\ (\bibinfo  {publisher} {Birkhäuser,
  Basel},\ \bibinfo {year} {2012})\BibitemShut {NoStop}%
\end{thebibliography}%

\end{document}